\documentclass[prx,aps,twocolumn,floatfix,superscriptaddress,nofootinbib]{revtex4-2}
\usepackage[utf8]{inputenc}
\usepackage{booktabs}
\usepackage{graphicx}% Include figure files
\usepackage{braket}
\usepackage{stix}
\usepackage{xcolor}

\usepackage{amsmath}
\usepackage{physics}

\usepackage{calc}

\definecolor{reviewblue}{RGB}{0,85,160}
\definecolor{reviewgreen}{RGB}{0,150,0}

\usepackage[hidelinks]{hyperref}
\usepackage[nameinlink,noabbrev]{cleveref}
\usepackage{tasks}
\usepackage{adjustbox} % in preamble

\makeatletter
\usepackage{dcolumn}% Align table columns on the decimal point
\usepackage{soul}
\usepackage{bm}% bold math
\usepackage{hyperref}
\usepackage{amsthm}
\usepackage{enumitem}
\usepackage{amsmath}

\usepackage[utf8]{inputenc}
\usepackage[T1]{fontenc}
\usepackage{xr-hyper}
\usepackage{soul}
\usepackage[caption=false]{subfig}
\hypersetup{colorlinks,
	citecolor=blue
}

\usepackage{mathtools}
\usepackage{apptools}
\usepackage{nameref}
\AtAppendix{\counterwithin{lemma}{section}}

\makeatletter
\newcommand*{\addFileDependency}[1]{% argument=file name and extension
  \typeout{(#1)}
  \@addtofilelist{#1}
  \IfFileExists{#1}{}{\typeout{No file #1.}}
}
\makeatother

\crefname{section}{Appendix}{Appendices}
\Crefname{section}{Appendix}{Appendices}

\begin{document}

\preprint{APS/123-QED}

\title{\textbf{Emergence of a Luttinger Liquid Phase in an Array of Chiral Molecules} 
}% 

\author{Muhammad Arsalan Ali Akbar}
\affiliation{Department of Electrical and Computer Engineering, North Carolina State University, Raleigh, NC 27606}
\affiliation{ Department of Chemistry, North Carolina State University, Raleigh, NC 27695}

\author{Bretislav Friedrich}
\affiliation{Fritz-Haber-Institut der Max-Planck-Gesellschaft, Faradayweg 4-6, D-14195 Berlin, Germany}

\author{Sabre Kais}
\email{skais@ncsu.edu}
\affiliation{Department of Electrical and Computer Engineering, North Carolina State University, Raleigh, NC 27606}
\affiliation{ Department of Chemistry, North Carolina State University, Raleigh, NC 27695}

\begin{abstract}
We propose a robust platform for simulating chiral quantum magnetism using linear arrays of trapped asymmetric top molecules, specifically 1,2-propanediol ($\mathrm{C_{3}H_{8}O_{2}}$). By mapping the Stark-dressed rotational states onto an effective spin-$1/2$ subspace, we rigorously derive a generalized $XXZ$ Heisenberg Hamiltonian governing the underlying many-body dynamics. Unlike standard solid-state models where the topological Dzyaloshinskii-Moriya Interaction (DMI) is introduced phenomenologically, we demonstrate that DMI emerges \textit{ab initio} from the molecular stereochemistry. Specifically, the interference between the transition dipole moments of heterochiral enantiomer pairs (L-R), which breaks inversion symmetry, generates a tunable DMI that stabilizes a Chiral Luttinger Liquid phase. Through a comprehensive phase-diagram analysis, we identify an optimal experimental regime characterized by intermolecular separations of \( r \approx 1.5~\mathrm{nm} \) and intermediate electric-field strengths \( d\varepsilon/B \approx 2.5 \). In this window, the system is protected from trivial field-polarized phases and exhibits a robust gapless spin-spiral texture. Our results establish 1,2-propanediol arrays as a versatile quantum simulator, providing a direct microscopic link between molecular chirality and topological many-body phases.
\end{abstract}

\maketitle

%\tableofcontents
\begin{figure*}[t] % two-column APS layout
\centering

% --- Top row: two side-by-side boxes, no captions --- 
\begin{minipage}[t]{1.0\textwidth}
    \centering
    \fbox{\includegraphics[width=\linewidth]{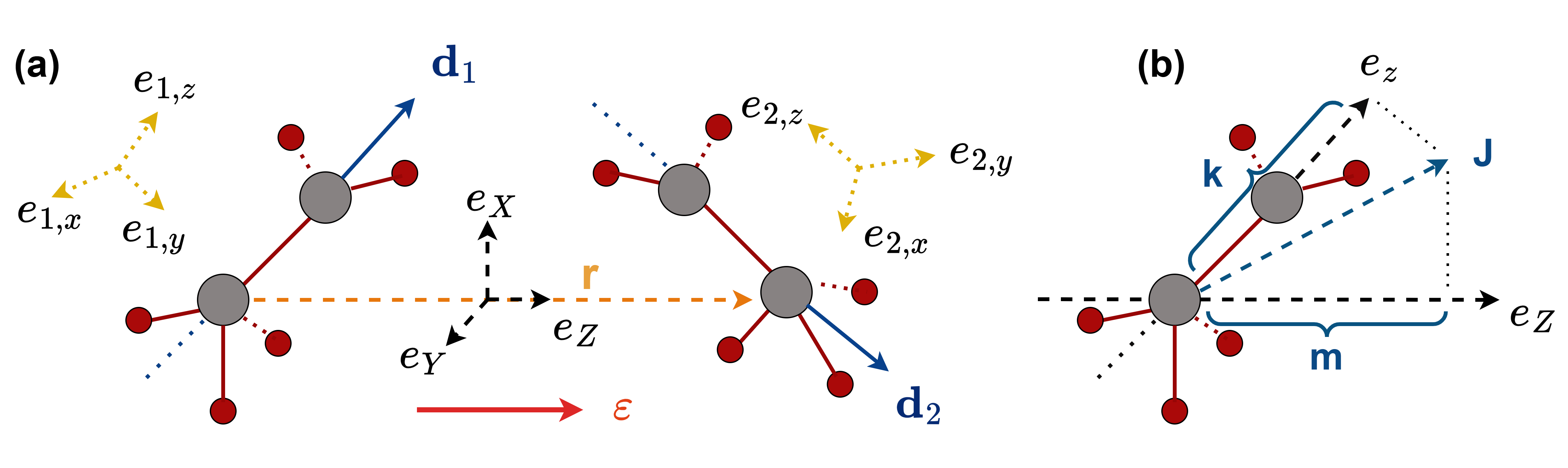}}
\end{minipage}
\vspace{0.5em} % space between rows

% % --- Bottom row: single wide box with the only caption ---

% \begin{minipage}[t]{1.0\textwidth}
%     \centering
%     \fbox{\includegraphics[width=\linewidth]{Figure 2b.png}}

\caption{(a) Schematic representation of two rotating prolate asymmetric-top molecules (1,2-propanediol) with permanent electric dipole moments $\mathbf{d}_1$ and $\mathbf{d}_2$, interacting with an external dc electric field $\boldsymbol{\varepsilon}$. The laboratory-fixed reference frame $(X, Y, Z)$ is chosen such that the intermolecular separation vector satisfies $\mathbf{r}=r\,\mathbf{e}_Z$. Each molecule is associated with its own molecule-fixed frame $(x_i,y_i,z_i)$. Due to molecular chirality, the dipole moments $\mathbf{d}_i$ have opposite orientations with respect to their corresponding body-fixed frames for the two enantiomers.(b) Schematic representation of the total angular momentum $\mathbf{J}$ with respect to the laboratory-fixed axis $\mathbf{e}_Z$ and the molecule-fixed principal axis $\mathbf{e}_z$.
}

    \label{fig:01}
% \end{minipage}
\end{figure*}
\section{Introduction}
\label{sec:level1}
An object is said to be chiral if it cannot be superposed on its mirror image -- like the right hand cannot be superposed on the left hand \cite{feringa1999absolute,bonner1991origin,mislow1999molecular}. The chirality -- or handedness -- of molecules was first identified in 1848 by Louis Pasteur \cite{pasteur1848memoires,pasteur1848relations}, who observed that the left (L) and right (R) handed isomers (termed enantiomers) of a given molecule interact differently with polarized light, rotating its plane of polarization either clockwise or counterclockwise. %This showed that chirality gives rise to an observable effect with direct physical consequences in electromagnetic fields. 
In 1927, Friedrich Hund analyzed molecular chirality quantum-mechanically and introduced, in the process, the concept of tunneling through the potential barrier separating the L- and R-enantiomers \cite{hund1927,FriedrichHerschbach1996}. 

Biological systems are homochiral, meaning that life uses only a single handedness, such as right-handed sugars and left-handed amino acids \cite{bada1995biomolecules,keszthelyi1995origin,blackmond2010origin,hein2012origin}. However, in the absence of a chiral bias, chemistry produces equal proportions of left- and right-handed enantiomers, known as racemic mixtures, which are achiral \cite{whitesell1985new,strauss1999biocatalytic}. Chirality dictates function through a lock-and-key-like mechanism: a molecule must have the correct handedness in order to bind to a chiral receptor, as the interaction between two chiral molecules depends strongly on their handedness and relative orientation. It has been proposed that homochirality may involve magnetic effects. Indeed, experimental studies suggest that magnetic surfaces could create a bias for the formation of one enantiomer over another, which could have played a role in the early stages of life on Earth \cite{ozturk2023chirality,ozturk2022origins,yashima2016supramolecular,liu2015supramolecular,herbst2009complex}.

Chirality-induced spin selectivity (CISS) describes the phenomenon where chiral molecules induce spin-dependent electron transport even in the absence of circularly polarized light or an external magnetic field \cite{evers2022theory,bloom2024chiral,yang2025quantification}. So far, CISS has been studied in three experimental domains: (i) Electron transmission across chiral monolayers, such as DNA or peptides, which may exhibit spin polarization of up to 60\% \cite{ray1999asymmetric,kettner2011spin,kettner2018chirality}. (ii) Even greater spin selectivity (80-90\%) is produced by electron transport in chiral perovskite films and molecular junctions \cite{long2020chiral,wei2021chiral,waldeck2021spin}. (iii) Spin-polarized electron-mediated enantioselectivity in chemical processes, connecting CISS to spin-dependent catalysis and enantioseparation \cite{naaman2019chiral}. Spin-orbit coupling (SOC) is the primary mechanism proposed for CISS, but theoretical calculations based solely on SOC predict spin polarizations several orders of magnitude smaller than the experimentally determined values. The primary reason for this discrepancy is a lack of a unified theory of CISS that would integrate spin-orbit coupling, electron-phonon interactions, molecular structure, and environmental effects \cite{gobel2025chirality,mohtashim2025rinq}. In this paper, we aim to develop an effective many-body model for an asymmetric top molecule (1,2-propanediole) rather than a simple, single-particle scattering model for a chiral molecular configuration.

Recently, several research groups have studied the many-body dynamics of linear, spherical, and symmetric-top molecules by mapping their configurations onto the $XXZ$ spin-1/2 model \cite{yue2022realization,wall2015realizing,wall2013simulating}. The realization of the $XXZ$ model in a molecular system offers a key advantage: the ability to implement an effective spin degree of freedom without any fine-tuning of the field. Additionally, the study of ground-state phase diagrams is significantly easier.  In this paper, we chose a chiral asymmetric top molecule (1,2-propanediole), as its energy spectrum is particularly rich due to the absence of inversion symmetry \cite{isaule2024rotational}. For each fixed total angular momentum $\mathbf{J}$, the Hilbert space scales with the basis states $j$ as $(2j+1)^{2}$. The absence of a conserved molecular-fixed projection quantum number leads to strong mixing among the rotational states; therefore, the eigenstates cannot be labelled by a single value of the projection $k$ on the figure axis of the molecule. In addition, unless the electric field is oriented along the laboratory  $Z$ axis, mixing among different $m$ states further increases the complexity of the field-induced state. Together, these features result in highly coupled rotations and amide folds, making the theoretical description of 1,2-propanediol significantly more challenging than that of linear or symmetric-top molecules. Herein, we select the two lowest-energy states, $\lvert j=0,k=0,m=0\rangle$ and $\lvert j=1,k=-1,m=\pm1\rangle$, which are identified as the pseudo-spin states $\lvert\uparrow\rangle$ (dashed blue curve) and $\lvert\downarrow\rangle$ (solid red curve) shown, respectively, in Fig. \ref{fig:02}. We then demonstrate how an effective spin-$1/2$ $XXZ$ model can be realized directly from these states without invoking any approximations. 

The standard $XXZ$ model has been derived from achiral molecules, where the interactions are symmetric. In other spin models, the antisymmetric exchange term, $\vec{D}\cdot\left(\vec{S}_i \times \vec{S}_j\right)$, is introduced as a phenomenological fitting parameter; in our work,  chiral coupling is derived \emph{ab initio} from the molecular stereochemistry. We explicitly demonstrate how the biological handedness of enantiomers is encoded in dipole moments and aptly translated into a topological spin-spin interaction. This approach provides a detailed recipe for engineering Hamiltonian parameters through chemical synthesis and establishes a direct link between molecular chirality and the emergence of chiral spin textures. Our phase-diagram analysis reveals a broad sweet spot at intermediate electric fields $(d\varepsilon/B \approx 2.0)$ and a lattice spacing of $r = 1.5\,\mathrm{nm}$. By mapping the complex rotational dynamics onto the universal phase diagram of the $XXZ$ chain, we demonstrate that at the many-body level, the antisymmetric interactions qualitatively change the low-energy physics of the spin chain. Instead of forming conventional Luttinger liquid phases of achiral $XXZ$ models, the system can stabilize chiral quantum phases, such as the Chiral Luttinger Liquid.\\

The outline of this paper is as follows: In Sec.~\ref{sec 1}, we calculate the eigenenergies of the asymmetric-top molecule $(\ensuremath{\mathrm{C_{3}H_{8}O_{2}}})$ in an external electrostatic (dc) field as a function of the dimensionless parameter $d\varepsilon/B$, where $d$ is the permanent electric dipole moment, $\varepsilon$ is the electric field strength, and $B$ is the rotational constant. We then construct the matrix elements of the dipole-dipole interaction in the pseudospin basis $\{\lvert\downarrow\downarrow\rangle,\,\lvert\downarrow\uparrow\rangle,\,\lvert\uparrow\downarrow\rangle,\,\lvert\uparrow\uparrow\rangle\}$ and finally realize the $XXZ$ spin-$1/2$ model by projection onto an effective two-qubit subspace. In Sec.~\ref{sec 3}, we present and analyze the coupling constants as functions of $d\varepsilon/B$ for several intermolecular separations. In Sec.~\ref{sec 4}, we examine the resulting phase diagram of the array of asymmetric-top molecules. In Sec.~\ref{sec 5}, we discuss the experimental implementation of a quantum simulator of a Luttinger liquid based on an array of asymmetric top molecules. Finally, in Sec.~\ref{sec 6} we present our conclusions.

\section{Model}
\label{sec 1}
\subsection{Single-Molecule Hamiltonian ($\hat{H}_{\mathrm{rot}} + \hat{H}_{\mathrm{dc}}$)}
\begin{figure}[t!]
\centering  
\includegraphics[width=1.0\linewidth]{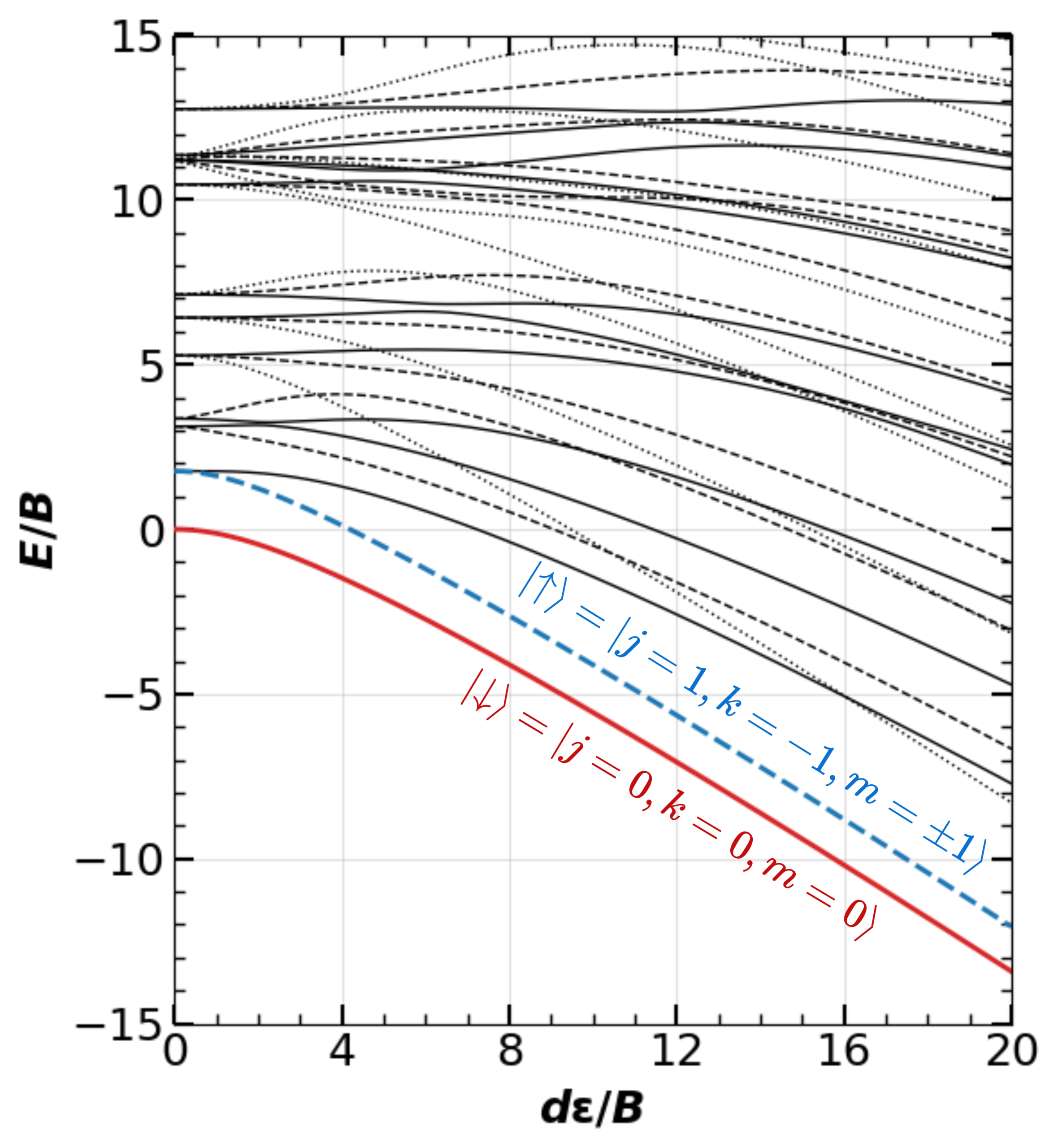} 
\caption{Eigenenergies of the asymmetric-top molecule 1,2-propanediol $(\ensuremath{\mathrm{C_{3}H_{8}O_{2}}})$ in an external dc electric field, plotted as a function of $d\varepsilon/B$, where $d$ is the permanent dipole moment, $\varepsilon$ is the electric-field strength, and $B$ is the rotational constant. In the absence of the electric field, the states $\lvert j=0,k=0,m=0\rangle$ and $\lvert j=1,k=-1,m=\pm1\rangle$ are identified as the pseudo-spin states $\lvert\downarrow\rangle$ (solid red curve) and $\lvert\uparrow\rangle$(dotted blue curve), respectively.}
\label{fig:02}
\end{figure}
We consider two asymmetric-top molecules (1,2-propanediol) in their
vibrational ground state, separated by a distance $r$ as illustrated
in the Fig. \ref{fig:01}. Each molecule is treated as a rigid rotor with a permanent
dipole moment $d_{i}(i=1,2)$, leading to dipole--dipole interactions,
subject to a dc electric field $\varepsilon$ applied along the laboratory
$Z-\text{axis }$ at fixed intermolecular separation. The total Hamiltonian
for the system reads 
\begin{equation}
\hat{H}_{total}=\hat{H}_{rot}+\hat{H}_{dc}+\hat{H}_{dd}
\label{Eq1}
\end{equation}
where $\hat{H}_{rot}$ describes the rotation of both molecules relative
to each other, $\hat{H}_{dc}$ captures the interaction of the molecules
with an external electric field, and $\hat{H}_{dd}$ gives the diole
dipole interaction. We have ignored nuclear spins of the molecules.
For a general asymmetric top molecule (1,2-propanediol), the rotation
of the molecules under rigid rotor approximation is given as \cite{zare1988angular,gordy1984microwave} 
\begin{equation}
\hat{H}_{rot}=A\hat{J}_{a}^{2}+B\hat{J}_{b}^{2}+C\hat{J}_{c}^{2}
\label{Eq2}
\end{equation}
Here \(A=\frac{\hbar^{2}}{2I_{a}},\ B=\frac{\hbar^{2}}{2I_{b}},\)
and \(C=\frac{\hbar^{2}}{2I_{c}}\) are the rotational constants
associated with the principal moments of inertia
\(I_{a}, I_{b}, I_{c}\), respectively, with
\(I_{a}\neq I_{b}\neq I_{c}\). The operators \(\hat{J}_{a}^{2}\), \(\hat{J}_{b}^{2}\), and
\(\hat{J}_{c}^{2}\) denote the squared components of the angular
momentum along the corresponding principal axes. For asymmetric-top
molecules, where \(I_{a}\neq I_{b}\neq I_{c}\), none of the angular
momentum components \(\hat{J}_{a}\), \(\hat{J}_{b}\), and
\(\hat{J}_{c}\) commute with the rotational Hamiltonian.
\[
[\hat{H}_{rot},\hat{J}_{a}]\neq0\,\,\,[\hat{H}_{rot},\hat{J}_{b}]\neq0\,\,\,[\hat{H}_{rot},\hat{J}_{c}]\neq0
\]
Consequently, no molecule-fixed projection of the total angular momentum
$\vec{J}$ is conserved. Based on microwave
spectroscopy measurements \cite{patterson2013enantiomer,lovas2009microwave}, the rotational constants for a conformer
of 1,2-propanediol ($\mathrm{C_{3}H_{8}O_{2}}$) are $A=8572.05\,\mathrm{MHz}$,
$B=3640.11\,\mathrm{MHz}$, and $C=2790.97\,\mathrm{MHz}$. The permanent
electric dipole moment $\boldsymbol{d}$ possesses nonzero components
along all three principal axes, with $d_{a}=1.201\,\mathrm{D}$, $d_{b}=1.916\,\mathrm{D}$,
and $d_{c}=\pm0.365\,\mathrm{D}$.

Since the rotational constants of 1,2-propanediol ($\mathrm{C_{3}H_{8}O_{2}}$)
satisfy $A>B>C$, the molecule is classified as a prolate asymmetric
top. For a prolate asymmetric top, we define the molecule-fixed frame
as $(x_{i},y_{i},z_{i})=(b_{i},c_{i},a_{i})$ for each molecule $i$
while the laboratory-fixed frame $(X,Y,Z)$ is chosen such that the
intermolecular separation vector satisfies $\boldsymbol{r}=r\,\boldsymbol{e}_{Z}$ \cite{gordy1984microwave}.

The primary objective of this work is to investigate the interaction
between two enantiomers of 1,2-propanediol ($\mathrm{C_{3}H_{8}O_{2}}$).
The two enantiomers are identical in all spectroscopic properties,
including their rotational constants and energy spectra; however,
they differ in the orientation of one component of the permanent dipole
moment in the molecule-fixed frame. Specifically, for the $L$ enantiomer
the dipole moment components are $d_{a}=1.201\,\mathrm{D}$, $d_{b}=1.916\,\mathrm{D}$,
and $d_{c}=+0.365\,\mathrm{D}$, whereas for the $R$ enantiomer the
components $d_{a}$ and $d_{b}$ remain unchanged while the $c$-axis
component reverses sign, $d_{c}=-0.365\,\mathrm{D}$. Our goal is
to encode this enantiomer-dependent dipole structure directly into
the Hamiltonian in order to study its impact on molecular chirality
and, indirectly, on the resulting phase diagrams. 

For any rotating molecule -- linear, symmetric-top, or asymmetric-top -- the
total angular momentum operator \(\boldsymbol{J}\) satisfies
\(\boldsymbol{J}^{2}=\hbar^{2}j(j+1)\), where \(j=0,1,2,\ldots\) is always
a good quantum number. For 1,2-propanediol $(\ensuremath{\mathrm{C_{3}H_{8}O_{2}}})$, each
fixed value of $j$ corresponds to $(2j+1)^{2}$ rotational basis
states $\lvert jkm\rangle$. The quantum number $k$ represents the
projection of $\boldsymbol{J}$ on the molecule-fixed $z$-axis,
which coincides with the $a$-axis for prolate asymmetric tops. For
a given $j$, $k$ can take $2j+1$ values in the range $k=-j,-j+1,\ldots,j-1,j$.
Since $k$ is not a conserved quantum number, the true eigenstates
are labeled by $\tilde{k}$. Similarly, $m$ denotes the projection
of $\boldsymbol{J}$ on the laboratory $Z$-axis and, for a given
$j$, can take $2j+1$ values in the range $m=-j,-j+1,\ldots,j-1,j$.
The quantum number $m$ remains conserved if
the system possesses cylindrical symmetry about the $Z$-axis, i.e.,
if the applied external electric field $\boldsymbol{\varepsilon}$
is oriented along $Z$. If the field is not oriented along the
$Z$-axis, different $m$ states are mixed \cite{townes1955microwave}. Overall, for a fixed value
of $j$, all combinations of $k$ and $m$ satisfying $|k|\leq j$
and $|m|\leq j$ are allowed.

The interaction between the permanent dipole and the external dc electric
field is given by 
\begin{equation}
\hat{H}_{\mathrm{dc}}=-\boldsymbol{d}\cdot\boldsymbol{\varepsilon}
\label{Eq3}
\end{equation}
where $\boldsymbol{d}$ is the dipole moment operator and $\boldsymbol{\varepsilon}$
is the external dc electric field. Both $\boldsymbol{d}$ and $\boldsymbol{\varepsilon}$
can be expressed as rank-1 spherical tensors, leading to 
\begin{equation}
\boldsymbol{d}\cdot\boldsymbol{\varepsilon}=\sum_{q=-1}^{1}(-1)^{q}d_{q}\,\varepsilon_{-q}
\end{equation}
where $d_{q}$ and $\varepsilon_{-q}$ denote the spherical components
in the laboratory frame. Consequently, 
\begin{equation}
\hat{H}_{\mathrm{dc}}=-\sum_{q=-1}^{1}(-1)^{q}d_{q}\,\varepsilon_{-q}
\label{Eq5}
\end{equation}

Since the molecular dipole moment is naturally defined in the molecule-fixed
frame through the components $d_{r}^{(\mathrm{mol})}$, whereas $d_{q}$
denotes the corresponding components in the laboratory frame, it is
necessary to transform $d_{q}$ into the molecule-fixed frame. This
transformation is given by 
\begin{equation}
d_{q}=\sum_{r=-1}^{1}D_{q,r}^{1*}(\Omega)\,d_{r}^{(\mathrm{mol})}
\end{equation}
where $d_{r}^{(\mathrm{mol})}$ are fixed molecular constants and
$\Omega$ specifies the molecular orientation. Substituting this expression
into Eq. (\ref{Eq5}) gives
\begin{equation}
\hat{H}_{\mathrm{dc}}=-\sum_{q=-1}^{1}\sum_{r=-1}^{1}(-1)^{q}D_{q,r}^{1*}(\Omega)\,d_{r}^{(\mathrm{mol})}\,\varepsilon_{-q}.
\end{equation}

The corresponding matrix elements in the $\lvert jkm\rangle$ basis
become
\begin{equation}
\begin{aligned}
\langle jkm \lvert \hat{H}_{\mathrm{dc}} \lvert j'k'm' \rangle
&= - \sum_{q=-1}^{1} \sum_{r=-1}^{1} (-1)^{q}
\langle jkm \lvert D^{1*}_{q,r}(\Omega) \lvert j'k'm' \rangle \\
&\quad \times d_{r}^{(\mathrm{mol})}\,\varepsilon_{-q}.
\end{aligned}
\end{equation}

Since the external electric field has only a single component along
the laboratory $z$-axis $(\boldsymbol{\varepsilon}=\varepsilon\,\boldsymbol{e}_{z})$,
its spherical components satisfy 
\begin{equation}
\varepsilon_{0}=\varepsilon_{z},\qquad\varepsilon_{\pm1}=0.
\end{equation}
As a result, only the $(q=0)$ term contributes to the dc Stark Hamiltonian.
The matrix elements, therefore, reduce to 
\begin{equation}
\langle jkm\vert\hat{H}_{\mathrm{dc}}\vert j'k'm'\rangle=-\varepsilon_{Z}\sum_{r=-1}^{1}\langle jkm\vert D_{0,r}^{1*}(\Omega)\vert j'k'm'\rangle\,d_{r}^{(\mathrm{mol})}
\end{equation}

The detailed derivation of the matrix elements for $\hat{H}_{\mathrm{rot}}$
and $\hat{H}_{\mathrm{dc}}$ is provided in the Appendix \ref{app:B}. For a single
molecule interacting with an external electric field, we compute
the energy spectrum of the asymmetric-top molecule (1,2-propanediol)
and plot it as a function of $d\varepsilon/B$ in Fig. \ref{fig:02}. The results show
that increasing the electric field leads to enhanced mixing of states
with different $j$ and $k$ quantum numbers. For higher rotational
quantum numbers $j\geq2$, we clearly observe the onset of level crossings,
reflecting the fact that asymmetric-top molecules exhibit mixing between
states with different $k$. As a result, neither $j$ nor $k$ remains
a good quantum number, and we therefore denote them by $\tilde{j}$
and $\tilde{k}$. In contrast, the quantum number $m$ remains conserved,
since the external electric field has only a single component along
the laboratory $z$-axis, $(\boldsymbol{\varepsilon}=\varepsilon\,\boldsymbol{e}_{z})$,
with spherical components $\varepsilon_{0}=\varepsilon_{z}$ and $\varepsilon_{\pm1}=0$.

We select two rotational states -- the ground state $\lvert j=0,k=0,m=0\rangle$
(red solid line) and the first excited state $\lvert j=1,k=-1,m=\pm1\rangle$
(blue dotted line) -- to define effective pseudo-spin states $\lvert\downarrow\rangle$
and $\lvert\uparrow\rangle$, respectively, as illustrated in Fig. \ref{fig:02}.
An external electric field is then used to couple these two states,
forming an effective two-level system. Owing to field-induced mixing,
the resulting pseudo-spin states are linear combinations of basis
states characterized by different $\tilde{j}$ and $\tilde{k}$:
\begin{align}
\lvert\downarrow\rangle & =\sum_{\tilde{j},\tilde{k}}c_{\tilde{j},\tilde{k}}^{(\uparrow)}(x=d\varepsilon/B)\lvert\tilde{j},\tilde{k},m=0\rangle\\
\lvert\uparrow\rangle & =\sum_{\tilde{j},\tilde{k}}c_{\tilde{j},\tilde{k}}^{(\downarrow)}(x=d\varepsilon/B)\lvert\tilde{j},\tilde{k},m=1\rangle
\end{align}

\begin{figure}[t!]
  \centering
  % left plot - width adjusted to fit half the column
  \includegraphics[width=0.50\columnwidth]{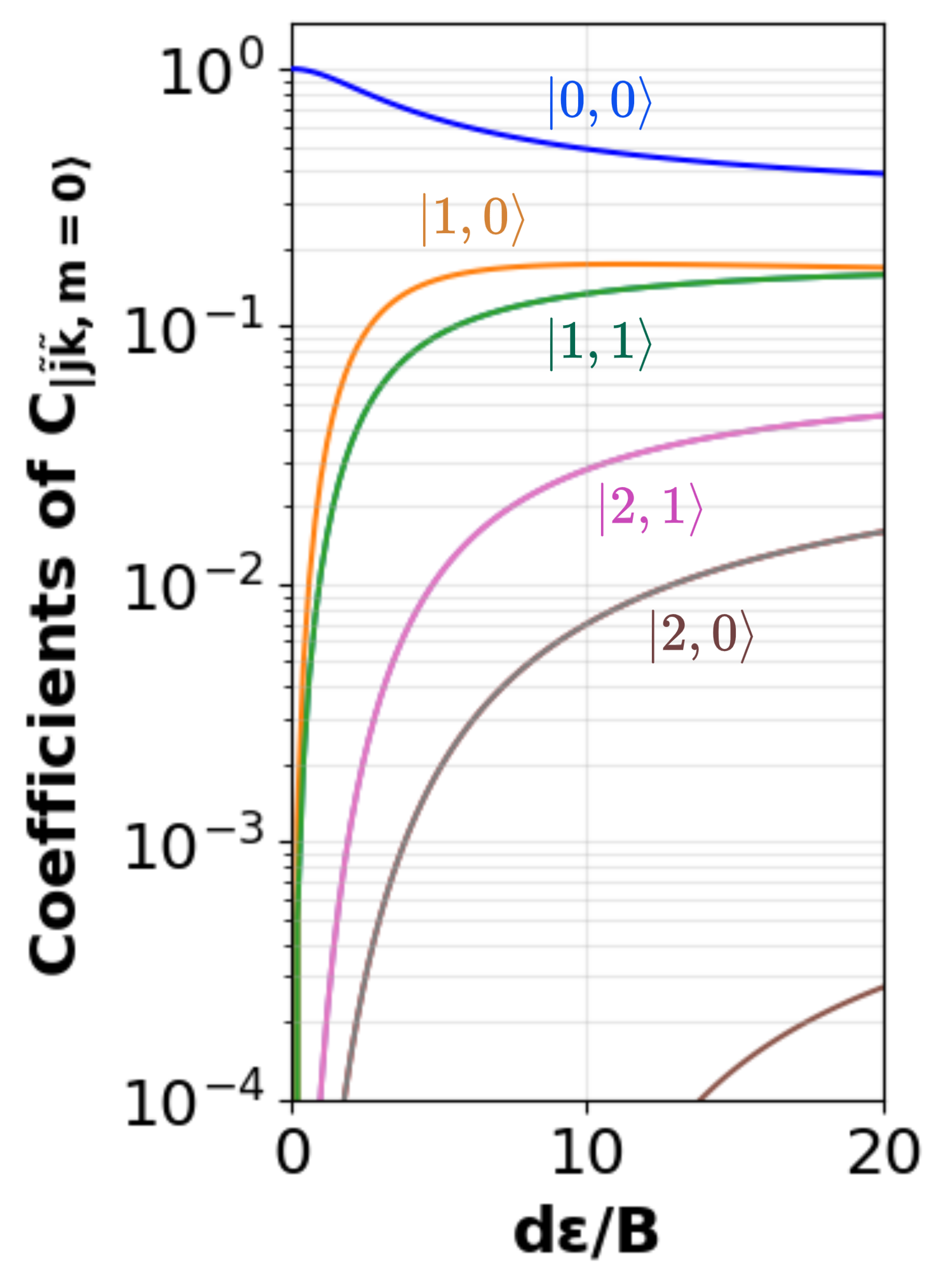}\hfill
  % right plot
  \includegraphics[width=0.50\columnwidth]{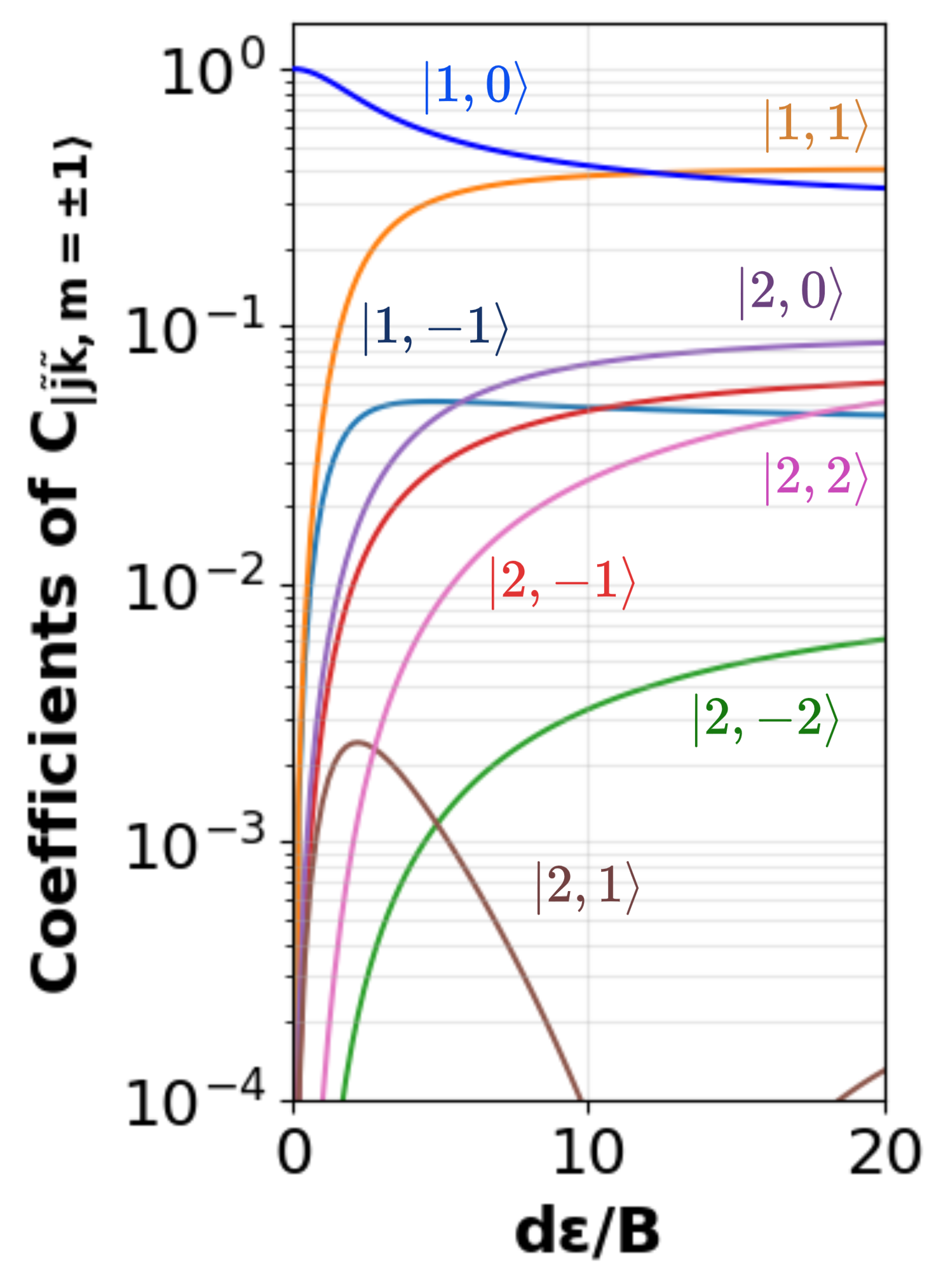}
  
  \caption{Field dependence of the expansion coefficients $c_{\tilde{j},\tilde{k}}(x=d\varepsilon/B)$
  for the pseudo-spin states $\lvert\downarrow\rangle=\lvert\tilde{j},\tilde{k},m=0\rangle$ (left panel) 
  and $\lvert\uparrow\rangle=\lvert\tilde{j},\tilde{k},m=\pm1\rangle$ (right panel); see Eq. \ref{Eq3}.}
  \label{fig:03}
\end{figure}

Fig. \ref{fig:03} shows the expansion coefficients as functions
of $d\varepsilon/B$. At $d\varepsilon/B=0$, both $\lvert\downarrow\rangle$
and $\lvert\uparrow\rangle$ correspond to pure rotational states,
consisting solely of the components $\lvert j=0,k=0\rangle$ and $\lvert j=1,k=-1\rangle$,
respectively. As $d\varepsilon/B$ increases, additional $j$ and
$k$ components are progressively admixed, while the initially dominant
contributions decrease accordingly. For $\lvert\uparrow\rangle$,
the dominant component $\lvert j=0,k=0\rangle$ (shown in blue) decreases
slowly and saturates along the component $\lvert j=1,k=1\rangle$
for $d\varepsilon/B\gtrsim5.0$. In contrast, for $\lvert\downarrow\rangle$
the initial dominant component $\lvert j=1,k=-1\rangle$ (also shown
in blue) decreases more gradually, but is eventually overtaken by
the $\lvert j=1,k=1\rangle$ component at sufficiently large $d\varepsilon/B$.

% \begin{figure}
% 
% \begin{minipage}[t]{0.45\columnwidth}%
% \includegraphics[scale=0.75]{\Figure 03a.png}%
% \end{minipage}\enskip{}\enskip{}\enskip{}\enskip{}\enskip{}\enskip{}\enskip{}\enskip{}\enskip{}%
% \begin{minipage}[t]{0.45\columnwidth}%
% \includegraphics[scale=0.75]{Figure 3b.png}%
% \end{minipage}

% \caption{Field dependence of the expansion coefficients $c_{\tilde{j},\tilde{k}}(x=d\varepsilon/B)$
% for the pseudo-spin states $\lvert\downarrow\rangle=\lvert\tilde{j},\tilde{k},m=0\rangle$(left
% panel) and$\lvert\uparrow\rangle=\lvert\tilde{j},\tilde{k},m=\pm1\rangle$(right
% panel); see Eq.(3).}\label{fig: 03}
% \selectlanguage{english}%
% \end{figure}
\subsection{Dipole Dipole Interaction}

The dipole--dipole interaction is described by the Hamiltonian
\begin{equation}
\hat{H}_{\mathrm{dd}}=-\frac{(\hat{\mathbf{d}}_{1}\cdot\hat{\mathbf{d}}_{2})-3(\hat{\mathbf{d}}_{1}\cdot\mathbf{e}_{r})(\hat{\mathbf{d}}_{2}\cdot\mathbf{e}_{r})}{r^{3}}
\end{equation}
where $\mathbf{e}_{r}=\mathbf{e}_{Z}$ in our chosen frame. For simplicity,
in evaluating the matrix element of $\hat{H}_{dd}$ we set $1/(4\pi\varepsilon_{0})=1$,
where $\varepsilon_{0}$ is the vacuum permittivity. We rewrite the
dipole--dipole interaction in terms of spherical tensors \cite{zare1988angular,man2013cartesian}, which allows
$\hat{H}_{\mathrm{dd}}$ to be written as \cite{krems2018molecules}
\begin{equation}
\hat{H}_{\mathrm{dd}}=-\frac{\sqrt{6}}{r^{3}}\sum_{q=-2}^{2}(-1)^{q}C_{-q}^{(2)}(\Omega_{r})\left[\hat{\mathbf{d}}_{1}\otimes\hat{\mathbf{d}}_{2}\right]_{q}^{(2)}
\end{equation}
Here \(\Omega_{r}=(\theta_{r},\phi_{r})\) denotes the polar and azimuthal
angles specifying the orientation of the intermolecular separation
vector \(\mathbf{r}\) in the laboratory-fixed frame, and
\(C_{q}^{(2)}(\Omega_{r})\) denotes an unnormalized spherical harmonic.
The molecular orientation with respect to the laboratory frame is
described separately by the Euler angles
\(\Omega=(\alpha,\beta,\gamma)\).
The term in the bracket,
\begin{equation}
\left[d_{1}\otimes d_{2}\right]_{q}^{(2)}=\sum_{q'=-1}^{1}\langle1\,q',\,1\,q-q'|2\,q\rangle\;d_{1,q'}\,d_{2,q-q'}
\end{equation}
represent a rank-2 tensor product of the two dipole moments. Following
the formulation detailed in Appendix \ref{app:A}, the dipole operators are expressed
in spherical tensor form as
\begin{equation}
d_{0}=d_{Z},\qquad d_{\pm1}=\mp\frac{1}{\sqrt{2}}(d_{X}\pm id_{Y})
\end{equation}

For molecules separated by a distance $r$ with the electric field
oriented along the laboratory $z-\text{axis,}$ the laboratory-frame
orientation is $\Omega_{r}=(\theta_{r},\phi_{r})=(0,0)$,
and  the $q=0$ term is the only nonvanishing contribution. Hence, the expression
reduces to \cite{dawid2018two,wall2015realizing}
\begin{equation}
\hat{H}_{\mathrm{dd}}=-\frac{1}{r^{3}}\left(2\,d_{1,0}d_{2,0}+d_{1,-1}d_{2,1}+d_{1,1}d_{2,-1}\right)
\end{equation}

In this equation, the dipole components are expressed in the laboratory
frame and must be transformed to the molecule-fixed frame. Accordingly,

\begin{equation}
d_{i,q}=\sum_{r_{i}=-1}^{1}D_{q,r_{i}}^{1*}(\Omega_{i})\,d_{i,r_{i}}^{(\mathrm{mol})},
\end{equation}
where $q=-1,0,1$ and $r=-1,0,1$ label the laboratory-frame and molecule-fixed
spherical components, respectively; $d_{i,r_{i}}^{(mol)}$ are the
known dipole components of the molecule $i$ and $D_{q,r_{i}}^{1*}(\Omega_{i})$
is the Winger $D-\text{matrix}$ associated with the orientation $\Omega_{i}.$
Expressing the dipole moment components in the molecule-fixed frame
gives
\begin{equation}
\begin{aligned}
\hat{H}_{\mathrm{dd}}
&= -\frac{1}{r^{3}}
\sum_{r_{1}=-1}^{1}\sum_{r_{2}=-1}^{1}
\Big[
2\,D_{0,r_{1}}^{1*}(\Omega_{1})\,d_{1,r_{1}}^{(\mathrm{mol})}
D_{0,r_{2}}^{1*}(\Omega_{2})\,d_{2,r_{2}}^{(\mathrm{mol})}
\\[4pt]
&\qquad
+ D_{-1,r_{1}}^{1*}(\Omega_{1})\,d_{1,r_{1}}^{(\mathrm{mol})}
D_{1,r_{2}}^{1*}(\Omega_{2})\,d_{2,r_{2}}^{(\mathrm{mol})}
\\[4pt]
&\qquad
+ D_{1,r_{1}}^{1*}(\Omega_{1})\,d_{1,r_{1}}^{(\mathrm{mol})}
D_{-1,r_{2}}^{1*}(\Omega_{2})\,d_{2,r_{2}}^{(\mathrm{mol})}
\Big].
\end{aligned}
\end{equation}

The expression inside the brackets shows that the dipole--dipole
interaction is projected onto the laboratory frame with the intermolecular
axis oriented along the $z-\text{axis}$. Physically,
this implies the following contributions: 

\begin{enumerate}
\renewcommand{\labelenumi}{(\roman{enumi})}
\item The first term corresponds to both dipoles having a laboratory-frame
projections \(m_{1}=0\) and \(m_{2}=0\), i.e., both dipoles are oriented
along the laboratory \(Z\) axis.

\item The second term describes a configuration in which dipole~1 has
\(m_{1}=-1\), while dipole~2 has \(m_{2}=+1\).

\item The third term corresponds to dipole~1 having \(m_{1}=+1\) and
dipole~2 having \(m_{2}=-1\).
\end{enumerate}
These are the only physically allowed combinations, as they conserve
the total laboratory-frame projection quantum number $(m=m_{1}+m_{2})$, when
the intermolecular axis is oriented along the $z-\text{axis}.$ Consequently, the only nonzero matrix elements of $\hat{H}_{dd}$ are
given by Eq. (\ref{EqC14}).

We now construct the matrix elements of the dipole--dipole interaction
in the dressed-state basis
\begin{align*}
\lvert\downarrow\rangle & =\sum_{\tilde{j},\tilde{k}}c_{\tilde{j},\tilde{k}}^{(\uparrow)}(x=d\varepsilon/B)\lvert\tilde{j},\tilde{k},m=0\rangle\\
\lvert\uparrow\rangle & =\sum_{\tilde{j},\tilde{k}}c_{\tilde{j},\tilde{k}}^{(\downarrow)}(x=d\varepsilon/B)\lvert\tilde{j},\tilde{k},m=1\rangle
\end{align*}

When the dipole--dipole Hamiltonian $\hat{H}_{dd}$ is projected
onto the pseudospin basis $\{\lvert\downarrow\downarrow\rangle,\,\lvert\downarrow\uparrow\rangle,\,\lvert\uparrow\downarrow\rangle,\,\lvert\uparrow\uparrow\rangle\}$,
its structure naturally separates into two distinct physical contributions:
a longitudinal (static) interaction, corresponding to diagonal
matrix elements, and a transverse (resonant) exchange interaction,
corresponding to off-diagonal matrix elements. The diagonal terms arise
from the selection rules 
\[
\Delta m_{1}=0\;\Rightarrow\;m_{1}=m_{1}',\qquad\Delta m_{2}=0\;\Rightarrow\;m_{2}=m_{2}',
\]
and physically represent a classical-like interaction between the
time-averaged static dipole moments induced by the external electric
field. Although each molecule occupies a rotational superposition,
the strong DC field polarizes the molecules, resulting in a nonzero
expectation value $\langle d_{z}\rangle$ along the quantization axis.
Consequently, the dipole--dipole Hamiltonian reduces, by virtue of the
selection rules discussed above, to purely diagonal contributions in
the dressed pseudo-spin basis. These correspond to the matrix elements
\(\langle\downarrow\downarrow\lvert\hat{H}_{\mathrm{dd}}\lvert\downarrow\downarrow\rangle\),
\(\langle\downarrow\uparrow\lvert\hat{H}_{\mathrm{dd}}\lvert\downarrow\uparrow\rangle\),
\(\langle\uparrow\downarrow\lvert\hat{H}_{\mathrm{dd}}\lvert\uparrow\downarrow\rangle\),
and
\(\langle\uparrow\uparrow\lvert\hat{H}_{\mathrm{dd}}\lvert\uparrow\uparrow\rangle\),
which define the coefficients \(C_{1}\)–\(C_{4}\), respectively:

\begin{equation}
\begin{aligned}
C_{1}
&= -\sum_{\substack{\tilde{j}_{1},\tilde{k}_{1}\\ \tilde{j}'_{1},\tilde{k}'_{1}}}
\sum_{\substack{\tilde{j}_{2},\tilde{k}_{2}\\ \tilde{j}'_{2},\tilde{k}'_{2}}}
c_{\tilde{j}_{1},\tilde{k}_{1}}^{(\downarrow)*}
c_{\tilde{j}'_{1},\tilde{k}'_{1}}^{(\downarrow)}
c_{\tilde{j}_{2},\tilde{k}_{2}}^{(\downarrow)*}
c_{\tilde{j}'_{2},\tilde{k}'_{2}}^{(\downarrow)}
\nonumber\\[2pt]
&\quad\times
\Big[
2\,
\langle \tilde{j}_{1}\tilde{k}_{1}1 \lvert D^{1}_{0,\tilde{k}_{1}-\tilde{k}'_{1}} \lvert
\tilde{j}'_{1}\tilde{k}'_{1}1 \rangle\,
\langle \tilde{j}_{2}\tilde{k}_{2}1 \lvert D^{1}_{0,\tilde{k}_{2}-\tilde{k}'_{2}} \lvert
\tilde{j}'_{2}\tilde{k}'_{2}1 \rangle
\Big]
\nonumber\\[2pt]
&\quad\times
d_{1,\tilde{k}_{1}-\tilde{k}'_{1}}\,
d_{2,\tilde{k}_{2}-\tilde{k}'_{2}} ,
\end{aligned}
\end{equation}

\begin{equation}
\begin{aligned}
C_{2}
&= -\sum_{\substack{\tilde{j}_{1},\tilde{k}_{1}\\ \tilde{j}'_{1},\tilde{k}'_{1}}}
\sum_{\substack{\tilde{j}_{2},\tilde{k}_{2}\\ \tilde{j}'_{2},\tilde{k}'_{2}}}
c_{\tilde{j}_{1},\tilde{k}_{1}}^{(\downarrow)*}
c_{\tilde{j}'_{1},\tilde{k}'_{1}}^{(\downarrow)}
c_{\tilde{j}_{2},\tilde{k}_{2}}^{(\uparrow)*}
c_{\tilde{j}'_{2},\tilde{k}'_{2}}^{(\uparrow)}
\nonumber\\[2pt]
&\quad\times
\Big[
2\,
\langle \tilde{j}_{1}\tilde{k}_{1}1 \lvert D^{1}_{0,\tilde{k}_{1}-\tilde{k}'_{1}} \lvert
\tilde{j}'_{1}\tilde{k}'_{1}1 \rangle\,
\langle \tilde{j}_{2}\tilde{k}_{2}0 \lvert D^{1}_{0,\tilde{k}_{2}-\tilde{k}'_{2}} \lvert
\tilde{j}'_{2}\tilde{k}'_{2}0 \rangle
\Big]
\nonumber\\[2pt]
&\quad\times
d_{1,\tilde{k}_{1}-\tilde{k}'_{1}}\,
d_{2,\tilde{k}_{2}-\tilde{k}'_{2}} ,
\end{aligned}
\end{equation}

\begin{equation}
\begin{aligned}
C_{3}
&= -\sum_{\substack{\tilde{j}_{1},\tilde{k}_{1}\\ \tilde{j}'_{1},\tilde{k}'_{1}}}
\sum_{\substack{\tilde{j}_{2},\tilde{k}_{2}\\ \tilde{j}'_{2},\tilde{k}'_{2}}}
c_{\tilde{j}_{1},\tilde{k}_{1}}^{(\uparrow)*}
c_{\tilde{j}'_{1},\tilde{k}'_{1}}^{(\uparrow)}
c_{\tilde{j}_{2},\tilde{k}_{2}}^{(\downarrow)*}
c_{\tilde{j}'_{2},\tilde{k}'_{2}}^{(\downarrow)}
\nonumber\\[2pt]
&\quad\times
\Big[
2\,
\langle \tilde{j}_{1}\tilde{k}_{1}0 \lvert D^{1}_{0,\tilde{k}_{1}-\tilde{k}'_{1}} \lvert
\tilde{j}'_{1}\tilde{k}'_{1}0 \rangle\,
\langle \tilde{j}_{2}\tilde{k}_{2}1 \lvert D^{1}_{0,\tilde{k}_{2}-\tilde{k}'_{2}} \lvert
\tilde{j}'_{2}\tilde{k}'_{2}1 \rangle
\Big]
\nonumber\\[2pt]
&\quad\times
d_{1,\tilde{k}_{1}-\tilde{k}'_{1}}\,
d_{2,\tilde{k}_{2}-\tilde{k}'_{2}} ,
\end{aligned}
\end{equation}

and

\begin{equation}
\begin{aligned}
C_{4}
&= -\sum_{\substack{\tilde{j}_{1},\tilde{k}_{1}\\ \tilde{j}'_{1},\tilde{k}'_{1}}}
\sum_{\substack{\tilde{j}_{2},\tilde{k}_{2}\\ \tilde{j}'_{2},\tilde{k}'_{2}}}
c_{\tilde{j}_{1},\tilde{k}_{1}}^{(\uparrow)*}
c_{\tilde{j}'_{1},\tilde{k}'_{1}}^{(\uparrow)}
c_{\tilde{j}_{2},\tilde{k}_{2}}^{(\uparrow)*}
c_{\tilde{j}'_{2},\tilde{k}'_{2}}^{(\uparrow)}
\nonumber\\[2pt]
&\quad\times
\Big[
2\,
\langle \tilde{j}_{1}\tilde{k}_{1}0 \lvert D^{1}_{0,\tilde{k}_{1}-\tilde{k}'_{1}} \lvert
\tilde{j}'_{1}\tilde{k}'_{1}0 \rangle\,
\langle \tilde{j}_{2}\tilde{k}_{2}0 \lvert D^{1}_{0,\tilde{k}_{2}-\tilde{k}'_{2}} \lvert
\tilde{j}'_{2}\tilde{k}'_{2}0 \rangle
\Big]
\nonumber\\[2pt]
&\quad\times
d_{1,\tilde{k}_{1}-\tilde{k}'_{1}}\,
d_{2,\tilde{k}_{2}-\tilde{k}'_{2}} .
\end{aligned}
\end{equation}

\begin{figure}[t!]
\centering  
\includegraphics[width=1.0\linewidth]{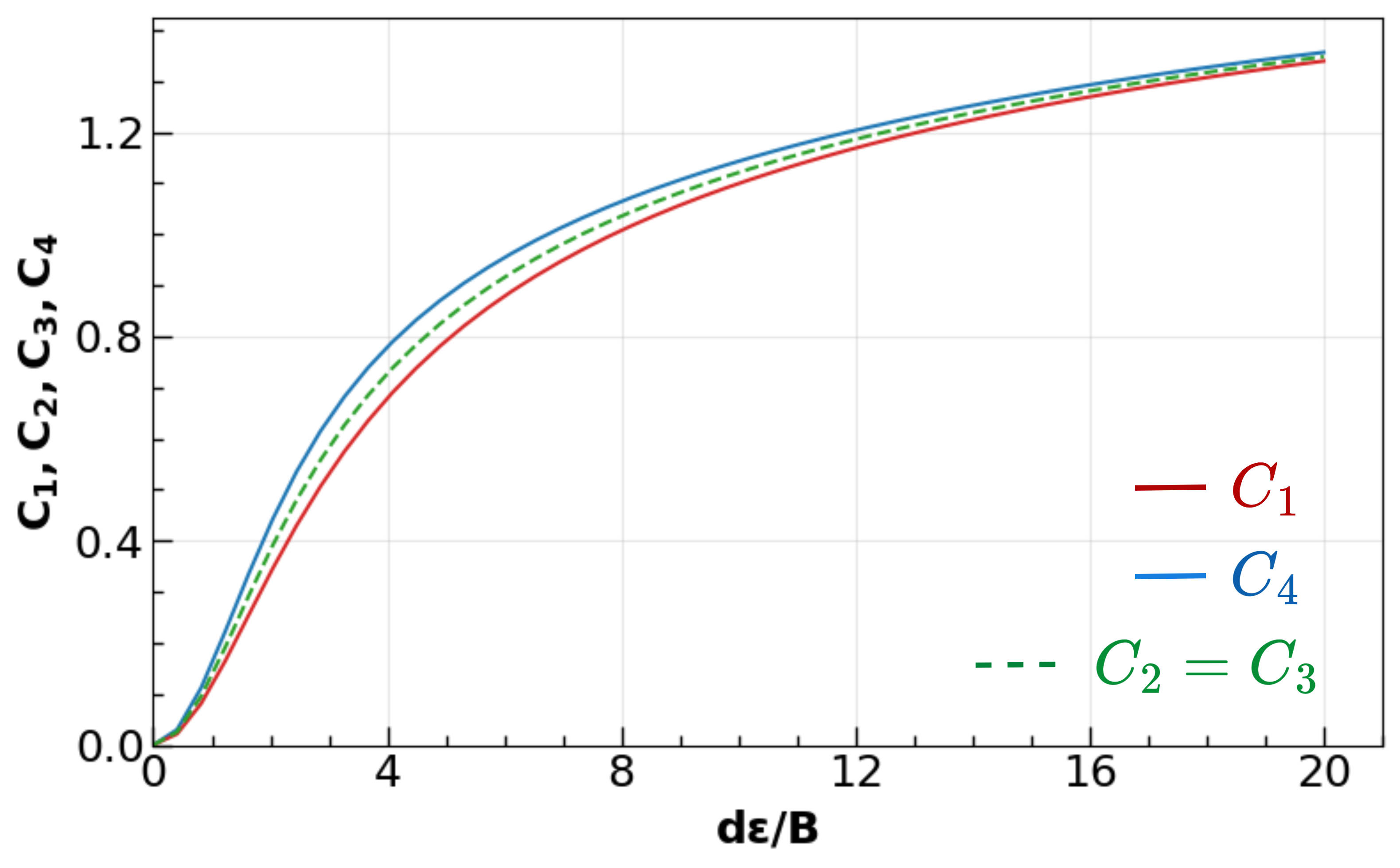}
\caption{Matrix elements of $C_{1}$, $C_{2}$, $C_{3}$, and $C_{4}$ as functions of the dimensionless electric-field interaction parameter $d\varepsilon/B$. The dotted green curves correspond to $C_{2}$ and $C_{3}$, which are equal in magnitude for the (1,2-propanediol) molecule configuration. In contrast, $C_{1}$ (solid red) and $C_{4}$ (solid blue) are initially distinct but gradually converge at larger field strengths.}
\label{fig:04}
\end{figure}

The like-state interactions, $C_{1}$ and $C_{4}$, correspond to the
energy cost when both molecules occupy the same pseudo-spin manifold,
either $\lvert\downarrow\rangle$ or $\lvert\uparrow\rangle$. In
contrast, the unlike-state interactions, $C_{2}$ and $C_{3}$, describe
the interaction energy when the two molecules occupy different pseudo-spin
states. These terms are particularly important in the context of the
effective spin-$\tfrac{1}{2}$ description, as the longitudinal interactions
give rise to the Ising anisotropy $J_{z}$ and the effective longitudinal
field $h$. Collectively, they play a central role in orienting the
pseudo-spins along the $z$-axis, favoring either ferromagnetic or
anti-ferromagnetic order depending on the sign and relative magnitude
of the interaction.

To quantify the longitudinal (static) interaction, we numerically
evaluate the diagonal matrix elements $C_{1}$--$C_{4}$ in the dressed-state
basis as a function of the applied dc electric field. The results
are shown in Fig. \ref{fig:04}. At $\varepsilon=0$, all coefficients vanish. Although chiral molecules possess indefinite parity due to the localization of enantiomers, the spatial isotropy of the field-free Hamiltonian ensures that the time-averaged dipole moment along any fixed laboratory axis satisfies $\langle\hat{d}_{z}\rangle=0$, and thus no static dipole--dipole coupling is present.

As the electric field increases, the rotational states of opposite parity
are mixed via the Stark effect, inducing a finite dipole moment oriented along the field direction. Since the diagonal interaction scales with
the product of these induced moments, $C_{ij}\propto\langle\hat{d}_{z}\rangle_{i}\langle\hat{d}_{z}\rangle_{j}$,
the coefficients increase monotonically with field strength. At larger
fields ($d\varepsilon/B\gtrsim6$), the growth slows down and the coefficients
begin to saturate, signaling the strong-field regime in which the
molecules form highly oriented pendular states pinned along the field
direction. In this limit, the  dipole moment induced in the laboratory frame approaches the
magnitude of the permanent molecular dipole, yielding a maximal static
interaction strength.

While all four coefficients exhibit the same overall behavior, they
separate into two distinct groups. The interactions between identical
pseudospin states ($C_{1}$ for $\lvert\downarrow\downarrow\rangle$
and $C_{4}$ for $\lvert\uparrow\uparrow\rangle$, solid lines) are
slightly stronger than the cross-state interactions ($C_{2}$ for
$\lvert\downarrow\uparrow\rangle$ and $C_{3}$ for $\lvert\uparrow\downarrow\rangle$,
dashed lines). This splitting arises from the different degrees of orientation
of the dressed states $\lvert\uparrow\rangle$ and $\lvert\downarrow\rangle$,
which lead to unequal induced dipole moments, $\langle\hat{d}_{z}\rangle_{\uparrow}\neq\langle\hat{d}_{z}\rangle_{\downarrow}$.

\begin{figure}[t!]
\centering  
\includegraphics[width=1.0\linewidth]{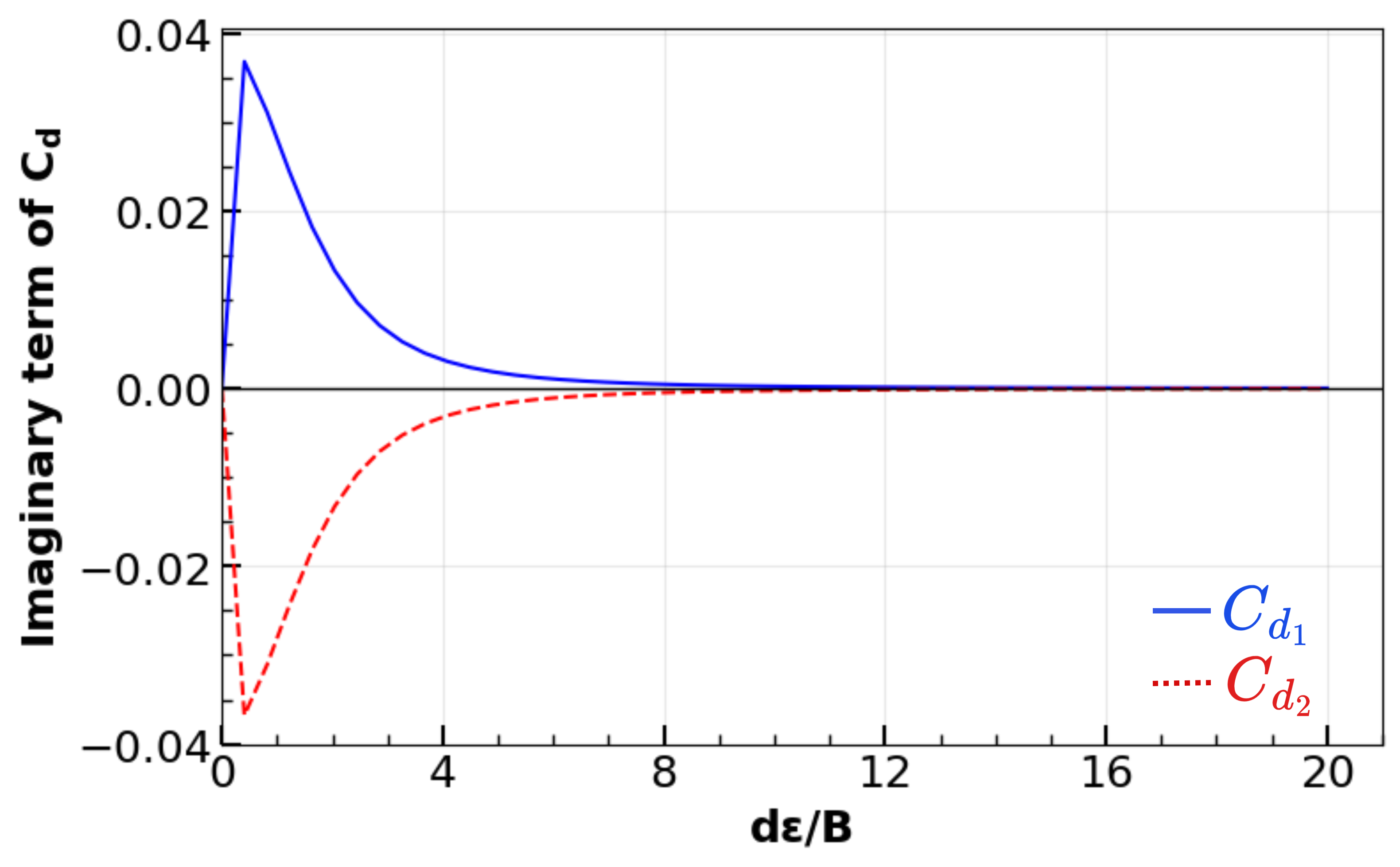}
\caption{Imaginary parts of the dipolar coefficients $C_{d_1}$ and $C_{d_2}$ as functions of the electric-field parameter $d\varepsilon/B$. The two coefficients are equal in magnitude and opposite in sign ($C_{d_1} = -C_{d_2}$); this antisymmetry constitutes the microscopic origin of the effective Dzyaloshinskii--Moriya interaction, which causes neighboring spins to be slightly twisted relative to each other, rather than strictly parallel or antiparallel.
}
\label{fig:05}
\end{figure}

The off-diagonal coefficients \(C_{d_{1}}\) and \(C_{d_{2}}\) originate
from the dipole--dipole selection rules
\[
(m_{1}-m'_{1},\,m_{2}-m'_{2}) = (-1,+1)\ \text{or}\ (+1,-1),
\]
which couple the pseudo-spin states
\(\langle\downarrow\uparrow\lvert\hat{H}_{\mathrm{dd}}\lvert\uparrow\downarrow\rangle\)
and
\(\langle\uparrow\downarrow\lvert\hat{H}_{\mathrm{dd}}\lvert\downarrow\uparrow\rangle\),
respectively.

\begin{equation}
\begin{aligned}
C_{d_{1}}
&= -\sum_{\substack{\tilde{j}_{1},\tilde{k}_{1}\\ \tilde{j}'_{1},\tilde{k}'_{1}}}
\sum_{\substack{\tilde{j}_{2},\tilde{k}_{2}\\ \tilde{j}'_{2},\tilde{k}'_{2}}}
c_{\tilde{j}_{1},\tilde{k}_{1}}^{(\downarrow)*}
c_{\tilde{j}'_{1},\tilde{k}'_{1}}^{(\uparrow)}
c_{\tilde{j}_{2},\tilde{k}_{2}}^{(\uparrow)*}
c_{\tilde{j}'_{2},\tilde{k}'_{2}}^{(\downarrow)}
\nonumber\\[2pt]
&\quad\times
\langle \tilde{j}_{1}\tilde{k}_{1}1 \lvert D^{1}_{+1,\tilde{k}_{1}-\tilde{k}'_{1}} \lvert
\tilde{j}'_{1}\tilde{k}'_{1}0 \rangle\,
\langle \tilde{j}_{2}\tilde{k}_{2}0 \lvert D^{1}_{-1,\tilde{k}_{2}-\tilde{k}'_{2}} \lvert
\tilde{j}'_{2}\tilde{k}'_{2}1 \rangle
\nonumber\\[2pt]
&\quad\times
d_{1,\tilde{k}_{1}-\tilde{k}'_{1}}\,
d_{2,\tilde{k}_{2}-\tilde{k}'_{2}} .
\end{aligned}
\end{equation}

and 

\noindent\quad{}$\langle\uparrow\downarrow\lvert H_{\mathrm{dd}}\lvert\downarrow\uparrow\rangle$

\begin{equation}
\begin{aligned}
C_{d_{2}}
&= -\sum_{\substack{\tilde{j}_{1},\tilde{k}_{1}\\ \tilde{j}'_{1},\tilde{k}'_{1}}}
\sum_{\substack{\tilde{j}_{2},\tilde{k}_{2}\\ \tilde{j}'_{2},\tilde{k}'_{2}}}
c_{\tilde{j}_{1},\tilde{k}_{1}}^{(\uparrow)*}
c_{\tilde{j}'_{1},\tilde{k}'_{1}}^{(\downarrow)}
c_{\tilde{j}_{2},\tilde{k}_{2}}^{(\downarrow)*}
c_{\tilde{j}'_{2},\tilde{k}'_{2}}^{(\uparrow)}
\nonumber\\[2pt]
&\quad\times
\langle \tilde{j}_{1}\tilde{k}_{1}0 \lvert D^{1}_{-1,\tilde{k}_{1}-\tilde{k}'_{1}} \lvert
\tilde{j}'_{1}\tilde{k}'_{1}1 \rangle\,
\langle \tilde{j}_{2}\tilde{k}_{2}1 \lvert D^{1}_{+1,\tilde{k}_{2}-\tilde{k}'_{2}} \lvert
\tilde{j}'_{2}\tilde{k}'_{2}0 \rangle
\nonumber\\[2pt]
&\quad\times
d_{1,\tilde{k}_{1}-\tilde{k}'_{1}}\,
d_{2,\tilde{k}_{2}-\tilde{k}'_{2}} .
\end{aligned}
\end{equation}

These terms describe a quantum-coherent exchange, or flip--flop,
process mediated by the transverse components of the dipole operator
$(d_{+},d_{-})$. In this process, one molecule undergoes a transition
from $\lvert\uparrow\rangle\rightarrow\lvert\downarrow\rangle$ while
its neighbor is simultaneously excited from $\lvert\downarrow\rangle\rightarrow\lvert\uparrow\rangle$.
This resonant energy transfer is driven by transition dipole moments
that rotate in the plane perpendicular to the applied electric field.

Collecting all non-zero terms, the dipole--dipole Hamiltonian in
the dressed basis $\{\lvert\downarrow\downarrow\rangle,\,\lvert\downarrow\uparrow\rangle,\,\lvert\uparrow\downarrow\rangle,\,\lvert\uparrow\uparrow\rangle\}$
becomes 
\begin{equation}
H_{\mathrm{dd}}=-\frac{1}{r^{3}}\begin{pmatrix}C_{1} & 0 & 0 & 0\\
0 & C_{2} & C_{d_{1}} & 0\\
0 & C_{d_{2}} & C_{3} & 0\\
0 & 0 & 0 & C_{4}
\end{pmatrix}
\end{equation}

Since we consider interactions between two non-identical molecules---specifically,
the $L$ and $R$ enantiomers of 1,2-propanediol $(\ensuremath{\mathrm{C_{3}H_{8}O_{2}}})$---the
off-diagonal matrix elements$C_{d_{1}}$ and $C_{d_{2}}$ are related
by Hermitian conjugation. These terms generally carry complex phases
originating from the transverse dipole components $d_{\pm}$, whose
molecule-fixed-frame expressions involve the combinations $\mp(d_{b}\pm id_{c})/\sqrt{2}$.
The presence of these complex phases constitutes the microscopic origin
of the chirality-dependent interaction in the effective Hamiltonian given in Eq. (\ref{Eq29}).
The coefficients $C_{d_{1}}$ and $C_{d_{2}}$ are directly
proportional to the products $d_{1,\tilde{k}_{1}-\tilde{k}'_{1}}\,d_{2,\tilde{k}_{2}-\tilde{k}'_{2}}$,
indicating that the interaction strength depends on which dipole-moment
components couple the initial $\tilde{k}$ state to the final $\tilde{k}'$
state for each molecule. For the specific transition contributing
to the $C_{d_{1}}$ term, molecule 1 undergoes a transition $\lvert\downarrow\rangle\rightarrow\lvert\uparrow\rangle$,
corresponding to a change in the laboratory-frame projection $m:1\rightarrow0$.
Conservation of angular momentum, therefore, requires the dipole operator
to carry a laboratory-frame component $q=+1$, which maps in the molecule-fixed
frame to the transverse dipole component $d_{+}^{(1)}$. Similarly,
molecule 2 undergoes the transition $\lvert\uparrow\rangle\rightarrow\lvert\downarrow\rangle$,
corresponding to $m:0\rightarrow1$. In this case, angular momentum
conservation requires a laboratory-frame component $q=-1$, which
in the molecule-fixed frame corresponds to the transverse dipole component
$d_{-}^{(2)}$. For these specific off-diagonal
matrix elements, the generic indices $\tilde{k}-\tilde{k}'$ are fixed
by the character of the transition, such that 
\[
d_{1,\tilde{k}_{1}-\tilde{k}'_{1}}\rightarrow d_{+}^{(1)},\qquad d_{2,\tilde{k}_{2}-\tilde{k}'_{2}}\rightarrow d_{-}^{(2)}.
\]
As a result, the matrix element $C_{d_{1}}$ is given by the product
of these dipole components multiplied by a real Wigner coefficient
$W$, 
\[
C_{d_{1}}\propto W\times\bigl(d_{+}^{(1)}d_{-}^{(2)}\bigr).
\]

Here $d_{\pm}=\mp\frac{d_{b}\pm id_{c}}{\sqrt{2}}$, are the spherical
tensor components of the dipole operator in the molecule-fixed frame.
The chirality of 1,2-propanediol ($\mathrm{C_{3}H_{8}O_{2}}$) is
encoded in the sign of the $c$-axis dipole component $d_{c}$. Consider
a heterochiral pair in which molecule~1 is left-handed $(+d_{c})$
and molecule~2 is right-handed $(-d_{c})$. For the left-handed $(L)$
enantiomer, 
\[
d_{+}^{(L)}=-(d_{b}+id_{c}),\qquad d_{-}^{(L)}=(d_{b}-id_{c}),
\]
while for the right-handed $(R)$ enantiomer, 
\[
d_{+}^{(R)}=-(d_{b}-id_{c}),\qquad d_{-}^{(R)}=(d_{b}+id_{c}).
\]
The product of the transition dipoles for the $L$-- $R$ pair is
therefore 
\begin{equation}
d_{+}^{(1)}d_{-}^{(2)}=d_{+}^{(L)}d_{-}^{(R)}=-(d_{b}+id_{c})(d_{b}+id_{c}),
\end{equation}
which evaluates to 
\begin{equation}
d_{+}^{(L)}d_{-}^{(R)}=-(d_{b}^{2}-d_{c}^{2})-2id_{b}d_{c}.
\label{Eq22}
\end{equation}
The nonzero imaginary part constitutes the microscopic origin of the
Dzyaloshinskii--Moriya interaction (DMI), reflecting constructive
interference between the handedness of the two molecules, leading
to a chiral twisting of excitation transfer that breaks inversion
symmetry. For the opposite ordering, where molecule~1 is right-handed
and molecule~2 is left-handed, one finds 
\begin{equation}
d_{+}^{(1)}d_{-}^{(2)}=d_{+}^{(R)}d_{-}^{(L)}=-(d_{b}-id_{c})(d_{b}-id_{c}),
\end{equation}
which yields 
\begin{equation}
d_{+}^{(R)}d_{-}^{(L)}=-(d_{b}^{2}-d_{c}^{2})+2id_{b}d_{c}.
\label{Eq24}
\end{equation}
Thus, exchanging the order of the enantiomers from an $L$-- $R$
pair to an $R$-- $L$ pair reverses the sign of the imaginary contribution,
confirming that the interaction possesses a vector chirality determined
by the spatial arrangement of the enantiomers. In contrast, for a
homo-chiral pair (LL or RR), the transition dipoles carry conjugate
phases that cancel exactly: 

\[
d_{+}^{(1)}d_{-}^{(2)}=d_{+}^{(L)}d_{-}^{(L)}=-(d_{b}+id_{c})(d_{b}-id_{c})
\]

\begin{equation}
d_{+}^{(L)}d_{-}^{(L)}=(d_{b}+d_{c})^{2}
\end{equation}

This product is purely real. Consequently, the imaginary term vanishes
$(D=0)$ and the interaction reduces to a pure symmetric exchange
$(J_{xy}).$ In order to validate the above analytic derivation of the origin of
the Dzyaloshinskii--Moriya interaction (DMI), we numerically evaluated
the imaginary part of the off-diagonal coupling coefficient $C_{d}$
as a function of the applied electric field. The results are shown
in Fig. \ref{fig:05}, providing a direct visualization of the symmetry breaking
discussed in Eq. (\ref{Eq22}) and (\ref{Eq24}). The solid blue curve corresponds to the right--left
(RL) heterochiral pair and exhibits a positive DMI strength ($D>0$),
while the dashed red curve corresponds to the left--right (LR) configuration
and shows an equal-magnitude but opposite sign ($D<0$). The mirror
symmetry of the two curves is in perfect agreement with the analytical
prediction that exchanging the spatial order of the enantiomers reverses
the sign of the interference term, 
\[
\Im\!\left[d_{+}^{(L)}d_{-}^{(R)}\right]=-\Im\!\left[d_{+}^{(R)}d_{-}^{(L)}\right],
\]
confirming that the interaction possesses a vector chirality determined
by the structural handedness of the molecules.

A further key feature apparent in Fig. \ref{fig:05} is that the DMI is
not a static property, but is widely tunable by the external electric
field. At $\varepsilon=0$, the interaction vanishes because the rotational
eigenstates have well-defined parity, which forbids the simultaneous
$\Delta m=\pm1$ transitions required for the flip--flop process.
At intermediate fields, $d\varepsilon/B\approx0.5$, the DMI reaches
a pronounced maximum, corresponding to an optimal regime in which
the Stark effect sufficiently mixes states of opposite parity to activate
the transverse transition dipoles $d_{\pm1}$ without locking
the rotors so tightly that transverse fluctuations are suppressed.
At larger fields, the DMI decreases as the molecules become strongly
polarized along the $z$-axis, approaching the Ising limit in which
transverse flip--flop exchange processes are quenched in favor of
static longitudinal interactions.

\subsection{Realization of the $XXZ$  spin-1/2 model from asymmetric top molecules}

As we are considering a pair of interacting 1,2-propanediol molecules
($\mathrm{C_{3}H_{8}O_{2}}$), prepared in opposite enantiomer configurations,
one left-handed (L) and one right-handed (R). Restricting each molecule
to an effective two-level dressed rotational manifold, the composite
system is described in the pseudo-spin basis $\{\lvert\!\downarrow\downarrow\rangle,\lvert\!\downarrow\uparrow\rangle,\lvert\!\uparrow\downarrow\rangle,\lvert\!\uparrow\uparrow\rangle\}$.

The total Hamiltonian consists of single-molecule rotational and Stark
contributions together with the intermolecular dipole--dipole interaction,
\begin{equation}
\hat{H}_{\mathrm{tot}}=\hat{H}_{\mathrm{rot}}^{(L)}+\hat{H}_{\mathrm{rot}}^{(R)}+\hat{H}_{\mathrm{dc}}^{(L)}+\hat{H}_{\mathrm{dc}}^{(R)}+\hat{H}_{\mathrm{dd}}^{(LR)}.
\end{equation}

In the pseudo-spin basis, the Hamiltonian takes the matrix form 
\begin{equation}
\setlength{\arraycolsep}{1.5pt} % Drastically reduces space between columns
\small % Slightly reduces font size
\hat{H}_{\mathrm{tot}}
=
\begin{pmatrix}
2E_{\downarrow}-\dfrac{C_{1}}{r^{3}} & 0 & 0 & 0\\[6pt]
0 & E_{\downarrow}+E_{\uparrow}-\dfrac{C_{2}}{r^{3}} & \dfrac{C_{d_{1}}}{r^{3}} & 0\\[6pt]
0 & \dfrac{C_{d_{2}}}{r^{3}} & E_{\uparrow}+E_{\downarrow}-\dfrac{C_{3}}{r^{3}} & 0\\[6pt]
0 & 0 & 0 & 2E_{\uparrow}-\dfrac{C_{4}}{r^{3}}
\end{pmatrix}
\end{equation}
 \begin{figure*}[t!] % spans both columns, top of page
  \centering
  % left plot
  \includegraphics[width=0.50\textwidth]{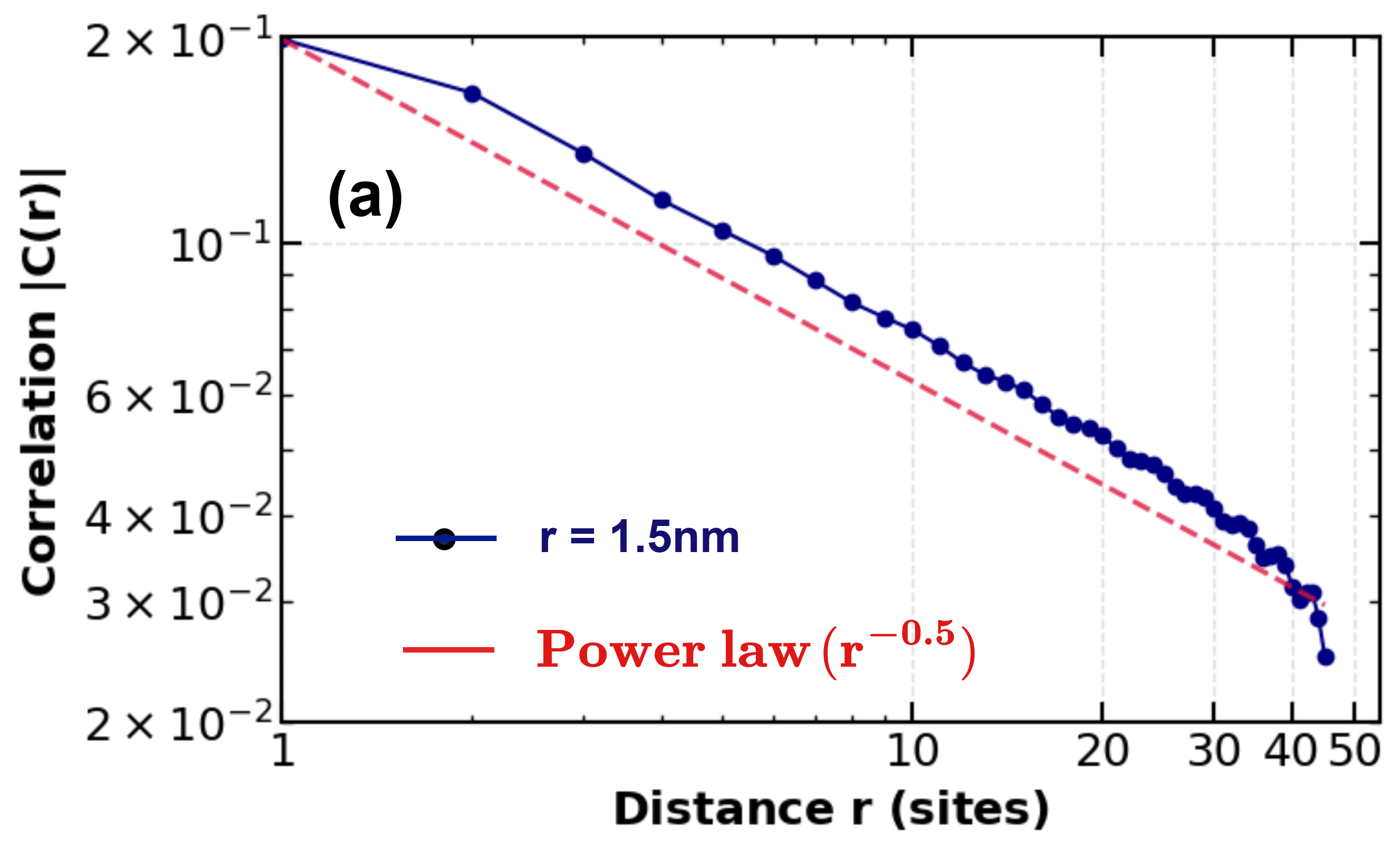}\hfill
  % right plot
  \includegraphics[width=0.49\textwidth]{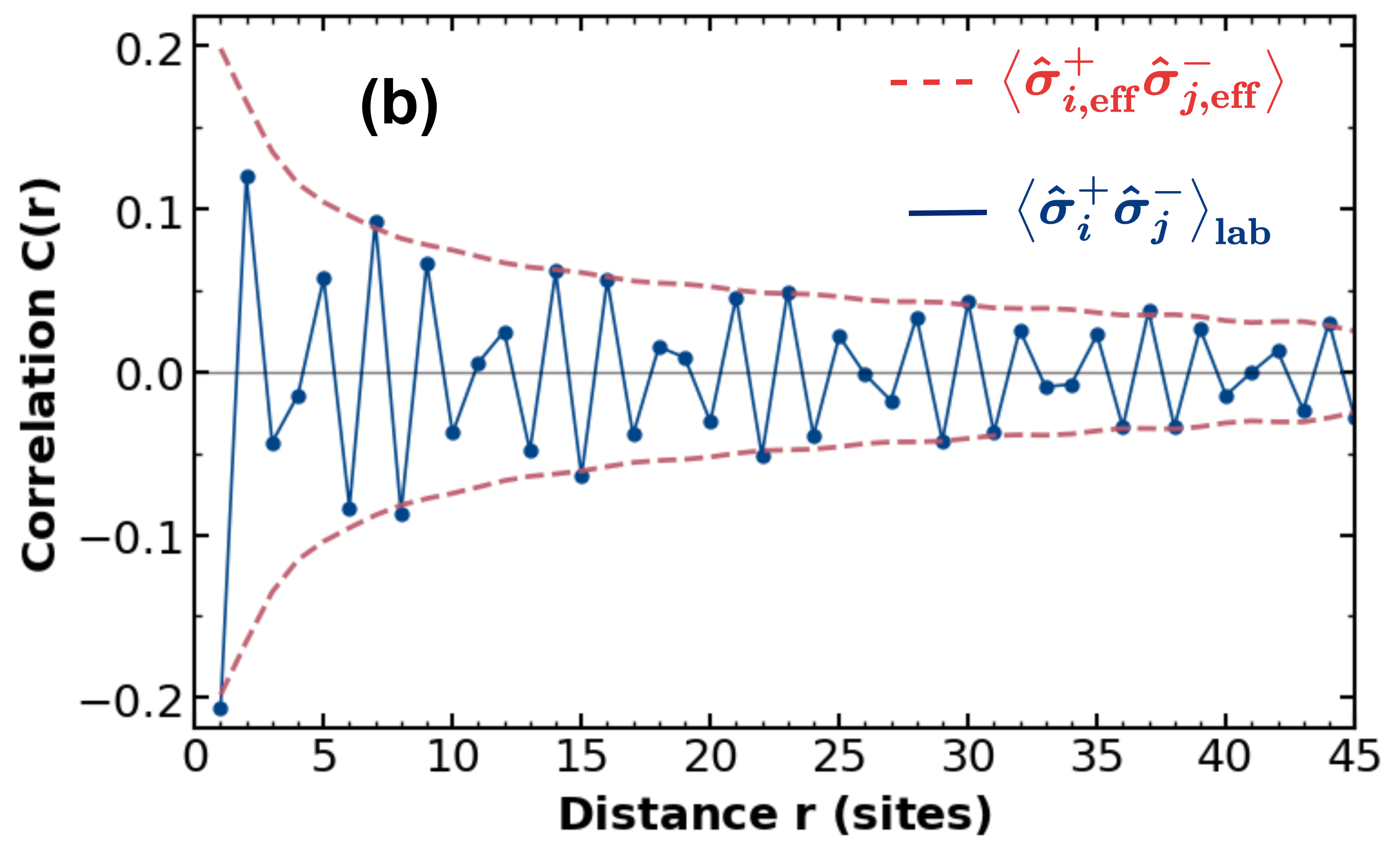}
  \caption{(a) The red dotted line represents the reference power-law decay \(r^{-0.5}\), while the solid blue line shows the standard Luttinger-liquid behavior after the gauge transformation, in which \(J_{xy}\) and \(D\) combine into the effective coupling 
\(\tilde{J}_{xy} = \sqrt{J_{xy}^{2} + D^{2}}\).
  (b)The red dotted curve represents the standard Luttinger-liquid correlation plotted on a linear scale and serves as an envelope for the correlations in the presence of the Dzyaloshinskii–Moriya interaction. 
The oscillations of the laboratory-frame correlations (solid blue curve) within this envelope arise from the phase factor \(e^{i\theta(i-j)}\), providing clear evidence of the twisting induced by the DMI. 
The corresponding twist angle is given by \(\theta = \tan^{-1}(D/J_{xy})\).
}
  \label{fig:06}
\end{figure*}
\begin{figure*}[t!]
  \centering

\end{figure*}

Here $E_{\uparrow}$ and $E_{\downarrow}$ denote the single-molecule
dressed energies, while the coefficients $C_{1}$--$C_{4}$ arise
from diagonal dipole--dipole interactions. The off-diagonal coefficients
$C_{d_{1}}$ and $C_{d_{2}}$ originate from resonant dipolar exchange
processes between the two molecules. A crucial step in constructing
the effective spin model is the unification of physical units. The
single-molecule energies $(E_{\uparrow},E_{\downarrow})$ are naturally
expressed in frequency units set by the rotational constants, whereas
the dipole--dipole interaction coefficients initially carry geometric
units of electric dipole density (e.g., $\mathrm{Debye}^2/\mathrm{nm}^3$). To restore dimensional consistency,
the interaction energy is converted to frequency units by normalizing
with Planck’s constant $h$. We define a distance-dependent scaling
factor 
\begin{equation}
\varOmega(r)=\frac{1}{h}\left(\frac{\lvert\boldsymbol{d}\lvert^{2}}{4\pi\epsilon_{0}r^{3}}\right),
\end{equation}
which sets the maximal interaction strength between two dipoles of
magnitude $\lvert\boldsymbol{d}\lvert$ separated by a distance $r$. The effective
coupling constants are obtained as the product of this physical scale
and the dimensionless geometric coefficients $C_{i}$.

Projecting onto the effective two-qubit subspace (see the details in Appendix \ref{app:D}), the interacting
molecular dipoles map onto an anisotropic spin-$\tfrac{1}{2}$ $XXZ$
Heisenberg model, 
\begin{equation}
\begin{split}
\hat{H}_{\mathrm{spin}} &= \sum_{j=1}^{N-1} \Big[ 
J_{xy}\left(\hat{\sigma}_{j}^{x}\hat{\sigma}_{j+1}^{x}+\hat{\sigma}_{j}^{y}\hat{\sigma}_{j+1}^{y}\right)
-D\left(\hat{\sigma}_{j}^{x}\hat{\sigma}_{j+1}^{y}-\hat{\sigma}_{j}^{y}\hat{\sigma}_{j+1}^{x}\right)
\\
&\quad +J_{z}\,\hat{\sigma}_{j}^{z}\hat{\sigma}_{j+1}^{z} \Big] 
+ \mathfrak{h}\sum_{j=1}^{N}\hat{\sigma}_{j}^{z}.
\end{split}
\label{Eq29}
\end{equation}

The effective coupling constants are given by 
\begin{equation}
\begin{aligned}
J_{xy} &= -\frac{\varOmega(r)}{2}\Re\!\left(C_{d_{1}}\right),\\
D      &= \phantom{-}\frac{\varOmega(r)}{2}\Im\!\left(C_{d_{1}}\right),\\
J_{z}  &= \frac{\varOmega(r)}{4}\left[\left(C_{2}+C_{3}\right)-\left(C_{1}+C_{4}\right)\right],\\
\mathfrak{h}      &= \frac{2(E_{\uparrow}-E_{\downarrow})+\varOmega(r)(C_{1}-C_{4})}{4}.
\end{aligned}
\label{coupling parameters}
\end{equation}

The transverse couplings $J_{xy}$ and $D$ originate from the off-diagonal
dipole--dipole matrix elements and describe coherent resonant exchange
of rotational excitation between the two molecules, corresponding
to flip--flop processes $\lvert\downarrow\uparrow\rangle\leftrightarrow\lvert\uparrow\downarrow\rangle$.
The imaginary component of this exchange gives rise to the chiral
Dzyaloshinskii--Moriya interaction $D$. In contrast, the longitudinal parameters $J_{z}$ and $\mathfrak{h}$ arise from
diagonal dipole--dipole interactions and the single-molecule Stark
splitting. These terms encode the static interaction landscape generated
by field-induced permanent dipoles aligned along the laboratory axis.

To determine the ground-state phase diagram, we simplify the Hamiltonian by applying a gauge transformation that absorbs the Dzyaloshinskii–Moriya interaction into the transverse exchange coupling $J_{xy}$\cite{kaplan1983single,shekhtman1992moriya}. This is achieved by performing a site-dependent rotation about the $z$ axis, defined by the unitary operator (see Appendix \ref{D3} for details),

\[
U = \prod_{i=1}^{N} e^{-i \phi_i \hat{\sigma}_i^{z}} .
\]

where the rotation angle $\phi_{i}$ increase along the chain as $\phi_{i}=j.\theta$.
If we choose the twist angle $\theta=\tan^{-1}(D/J_{xy})$, the antisymmetric
DMI term and the symmetric exchange term will combine, and a renomorlized
planar interaction will form \cite{alcaraz1990heisenberg}:

\[
\tilde{J}_{xy}=\sqrt{J_{xy}^{2}+D^{2}}
\]

The transformed Hamiltonian takes the form of a conventional $XXZ$ spin model, characterized by transverse and longitudinal couplings $\tilde{J}_{xy}$ and $J_z$, respectively.
\[
\hat{H}_{\mathrm{spin}}=
\sum_{j=1}^{N-1}
\Big[
\tilde{J}_{xy}
\left(
\hat{\sigma}_{j}^{x}\hat{\sigma}_{j+1}^{x}
+
\hat{\sigma}_{j}^{y}\hat{\sigma}_{j+1}^{y}
\right)
+
J_{z}\,
\hat{\sigma}_{j}^{z}\hat{\sigma}_{j+1}^{z}
\Big]
+
\mathfrak{h}\sum_{j=1}^{N} \hat{\sigma}_{j}^{z}
\label{}
\]

% ===== Figure 09 (Four panels, 2x2 grid) =====
\begin{figure*}[t!]
  \centering

  % --- First Row ---
  % Top-Left Plot
  \begin{minipage}[t]{0.48\textwidth}
    \centering
    \includegraphics[width=\linewidth]{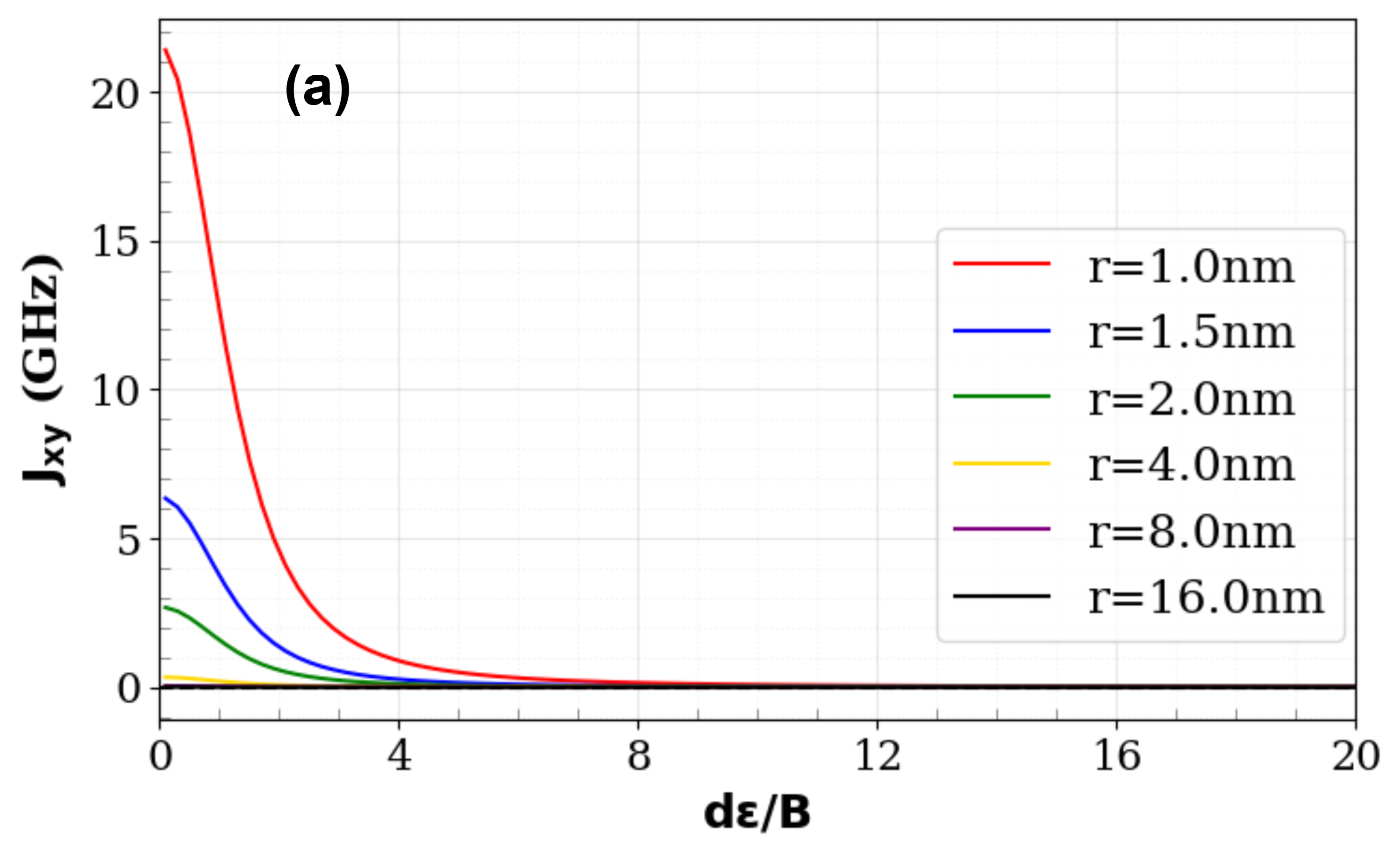}
  \end{minipage}\hfill
  % Top-Right Plot
  \begin{minipage}[t]{0.50\textwidth}
    \centering
    \includegraphics[width=\linewidth]{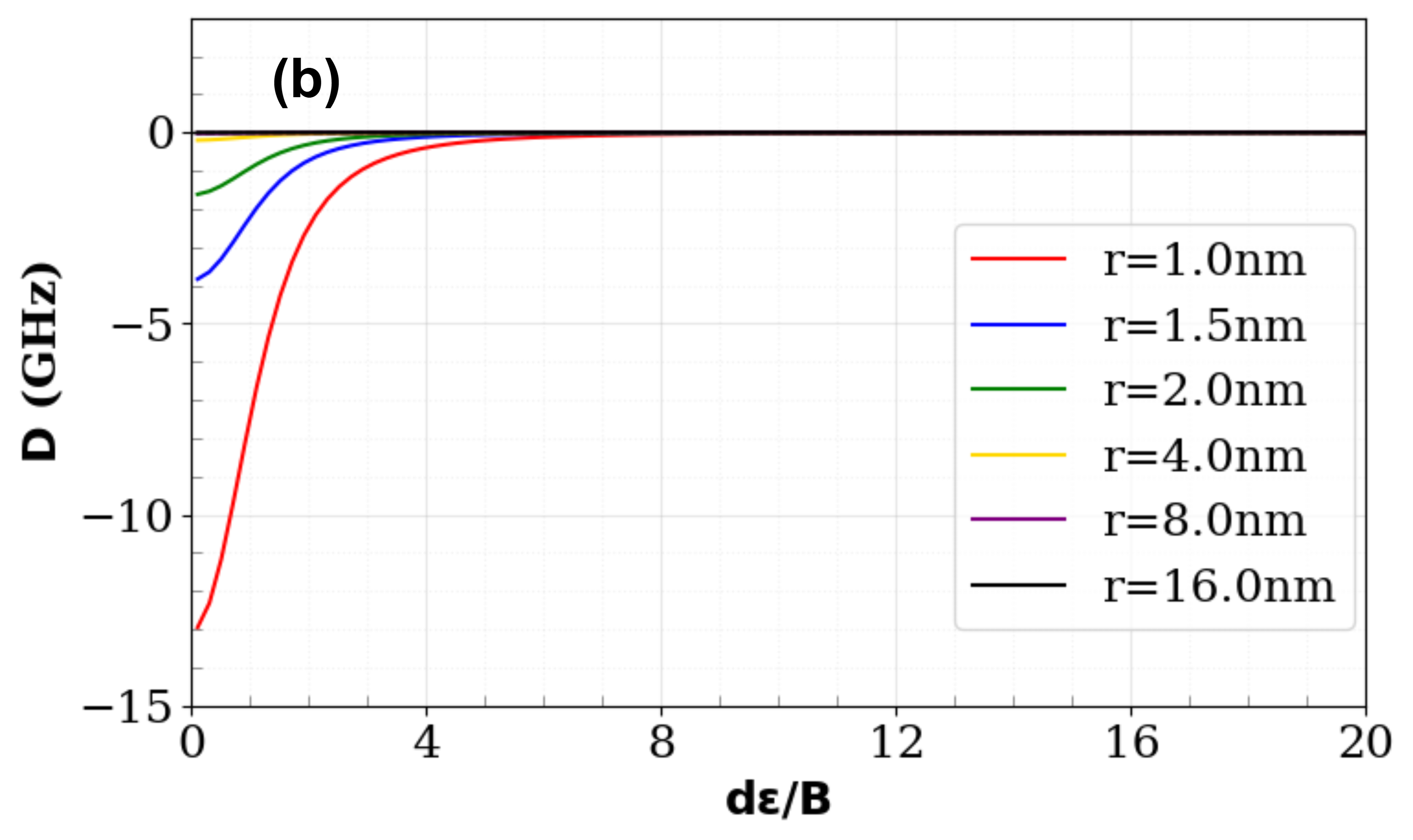}
  \end{minipage}

  \vspace{0.3cm} % Vertical space between rows

  % --- Second Row ---
  % Bottom-Left Plot
  \begin{minipage}[t]{0.50\textwidth}
    \centering
    \includegraphics[width=\linewidth]{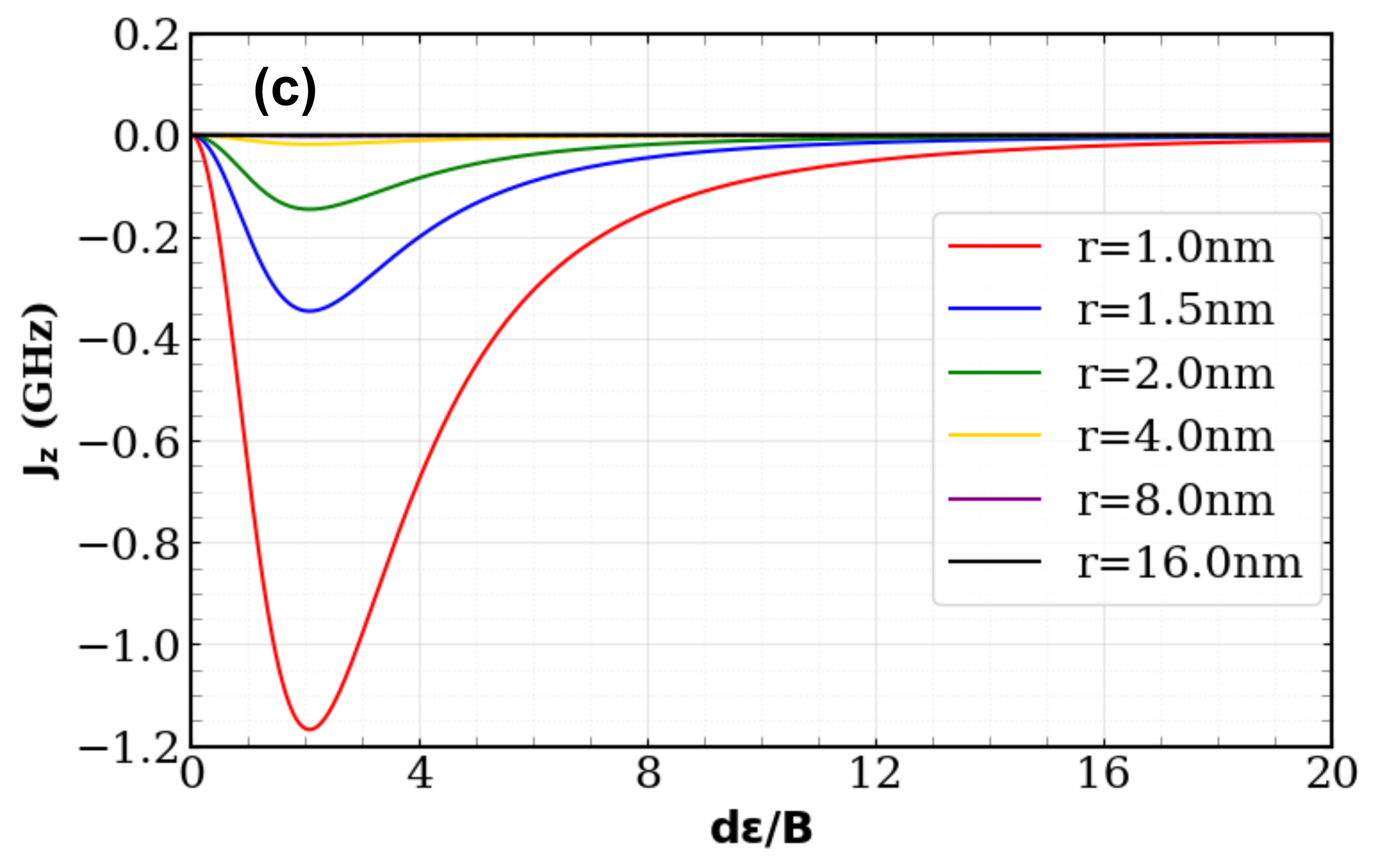} % Update filename
  \end{minipage}\hfill
  % Bottom-Right Plot
  \begin{minipage}[t]{0.48\textwidth}
    \centering
    \includegraphics[width=\linewidth]{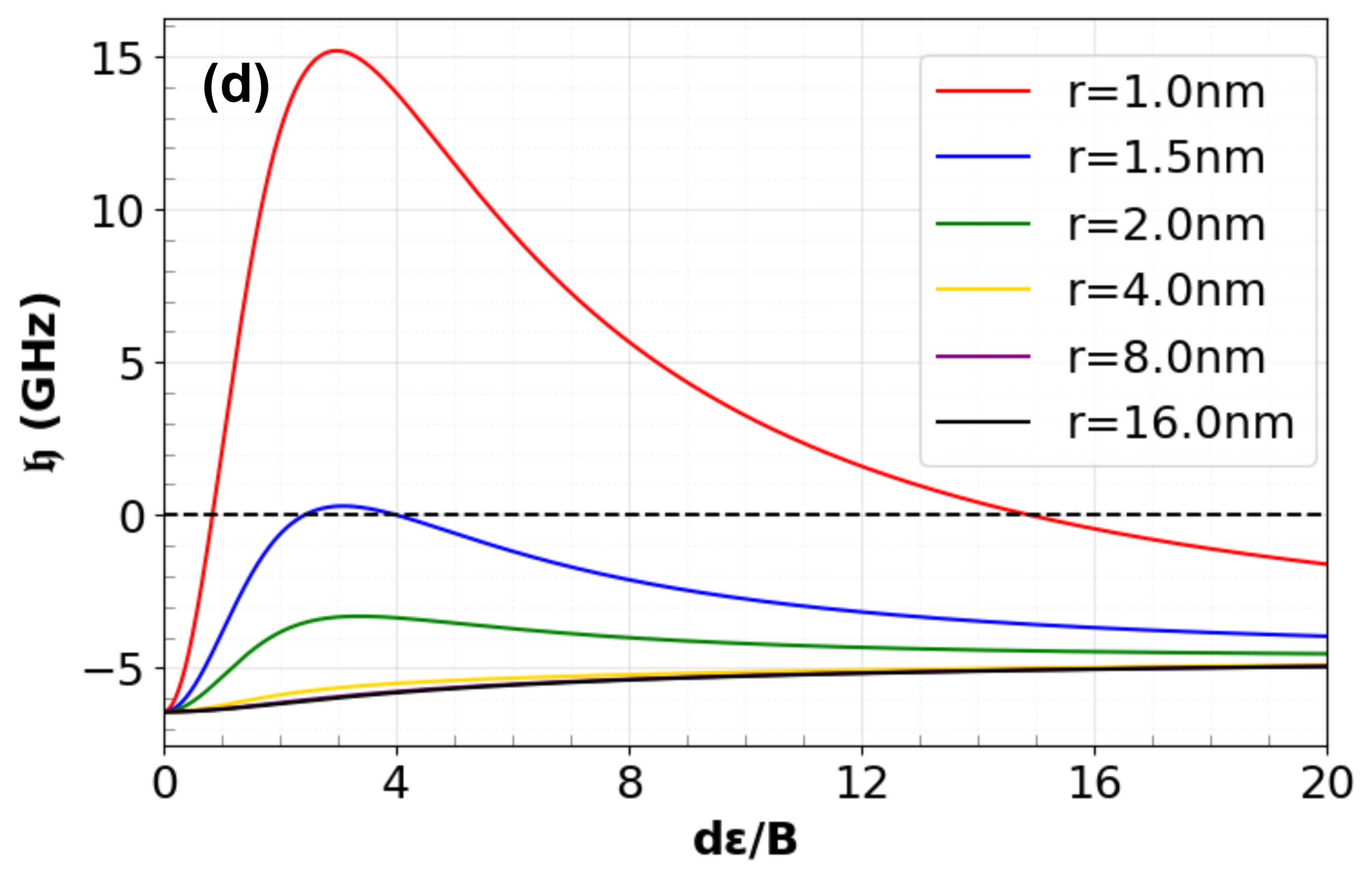} % Update filename
  \end{minipage}

  \caption{Effective coupling parameters of an asymmetric-top molecular Hamiltonian mapped onto a spin-chain model, illustrating their dependence on the intermolecular separation $r$ between $L$- and $R$-enantiomers. (a) Symmetric exchange coupling $J_{xy}$ and (b) Dzyaloshinskii--Moriya interaction $D$, arising from the left- and right-handed enantiomers, decrease monotonically as the electric-field parameter $d\varepsilon/B$ is increased. (c) The Ising interaction $J_z$ remains negative throughout the entire parameter range, indicating a robust ferromagnetic bias. (d) The effective magnetic field $\mathfrak{h}$ displays a strong nonmonotonic behavior, including a sign reversal at short intermolecular distances ($r=1.0\,\mathrm{nm}$), followed by an asymptotic decay.}
  \label{fig:07}
\end{figure*}

Because the unitary transformation $U$ leaves the energy spectrum invariant while rotating the eigenstates, the phase boundaries of the chiral Luttinger liquid in the laboratory frame are identical to those of the standard Luttinger liquid in the rotated frame. To characterize chirality, we examine spin–spin correlation functions, which quantify how a spin at site $i$ couples to a spin at site $j$. In a standard Luttinger liquid (LL), the transverse correlations decay algebraically as \cite{giamarchi2003quantum}
\[
\bigl\langle \hat{\sigma}_i^{-}\hat{\sigma}_j^{+} \bigr\rangle_{\mathrm{LL}}
\sim \frac{1}{|i-j|^{\frac{1}{2K}}}
\]
Where $K=1$ for a Luttinger liquid. The physical spin operators in the laboratory frame are related to the effective spin operators in the rotated frame by a site-dependent twist,
\[
\hat{\sigma}^{+}_{j,\mathrm{lab}} = e^{-i\theta j}\,\hat{\sigma}^{+}_{j,\mathrm{eff}} .
\]
As a result, when evaluating correlation functions in the laboratory frame, an additional phase factor emerges \cite{oshikawa1997field}. Using the relation between laboratory-frame and rotated-frame spin operators,
\[
\hat{\sigma}^{+}_{j,\mathrm{lab}} = e^{-i\theta j}\,\hat{\sigma}^{+}_{j,\mathrm{eff}},
\qquad
\hat{\sigma}^{-}_{j,\mathrm{lab}} = e^{+i\theta j}\,\hat{\sigma}^{-}_{j,\mathrm{eff}},
\]
the transverse spin–spin correlation function in the laboratory frame becomes
\[
\Bigl\langle \hat{\sigma}_{i}^{+}\hat{\sigma}_{j}^{-} \Bigr\rangle_{\mathrm{lab}}
=
\Bigl\langle
e^{-i\theta i}\hat{\sigma}_{i,\mathrm{eff}}^{+}
\,e^{+i\theta j}\hat{\sigma}_{j,\mathrm{eff}}^{-}
\Bigr\rangle .
\]
Factoring out the phase yields
\[
\Bigl\langle \hat{\sigma}_{i}^{+}\hat{\sigma}_{j}^{-} \Bigr\rangle_{\mathrm{lab}}
=
e^{i\theta (j-i)}
\Bigl\langle
\hat{\sigma}_{i,\mathrm{eff}}^{+}\hat{\sigma}_{j,\mathrm{eff}}^{-}
\Bigr\rangle .
\]
The correlator
\(
\bigl\langle
\hat{\sigma}_{i,\mathrm{eff}}^{+}\hat{\sigma}_{j,\mathrm{eff}}^{-}
\bigr\rangle
\)
is the standard Luttinger-liquid correlation, which decays algebraically with distance. The additional phase factor \( e^{i\theta(j-i)} \) induces an oscillatory spatial modulation, causing the laboratory-frame correlations to acquire a spiral structure—characteristic of a chiral Luttinger liquid.\\
Fig. \ref{fig:06}(a) shows the correlation function in the interacting limit for a system size of \(N = 100\). 
The dotted black line represents the reference power-law decay \(r^{-0.5}\) \cite{giamarchi2003quantum}, while the solid blue line 
with filled circles illustrates the decay of correlations for a standard Luttinger liquid when 
\(\tilde{J}_{xy} = \sqrt{J_{xy}^{2} + D^{2}}\).
The coupling constants are fixed at \(\tilde{J}_{xy} \approx 6.4\,\mathrm{GHz}\), 
\(J_{z} \approx -0.2454\), and \(\mathfrak{h} \approx 9.550\), and the objective is to study 
the response of the quantum state to this specific set of parameters. The coupling \(J_{xy}\) is responsible for generating quantum correlations that propagate along the spin chain, while the longitudinal field \(\mathfrak{h}\) competes with this process by favoring spin polarization, thereby tending to orient all spins either up or down. 
At an intermolecular separation of \(r = 1.5\,\mathrm{nm}\), the ratio \(\mathfrak{h}/J_{xy} \approx 1.5\). 
Therefore, for this specific molecular geometry, the system remains well within the Luttinger-liquid regime. 
This is clearly evidenced in the correlation function: for a long chain (\(N = 100\)), the correlations exhibit a power-law decay of the form
\[
\langle S_i^+ S_j^- \rangle \sim \frac{1}{|i - j|^{0.5}} .
\]
In Fig. \ref{fig:06}(b), we plot the correlation function 
\(\bigl\langle \hat{\sigma}_i^{+}\hat{\sigma}_j^{-} \bigr\rangle_{\mathrm{lab}}\), 
measured in the laboratory frame, which explicitly includes the physical twist induced by the Dzyaloshinskii–Moriya interaction. 
The red curve represents the effective-frame correlation 
\(\bigl\langle \hat{\sigma}_{i,\mathrm{eff}}^{+}\hat{\sigma}_{j,\mathrm{eff}}^{-} \bigr\rangle\), 
previously shown in Fig. \ref{fig:06}(a), plotted here on a linear scale. 
This curve acts as an envelope for the laboratory-frame correlations, since the oscillatory spiral structure (blue curve) is bounded by the condition 
\(|e^{i\theta(i-j)}| = 1\) and therefore cannot exceed the effective-frame amplitude. The blue curve exhibits a clear wave-like pattern, confirming the presence of the phase factor 
\(e^{i\theta(i-j)}\). 
The period of these oscillations is directly determined by the twist angle 
\(\theta = \tan^{-1}(D/J_{xy})\).

\section{Reslut and Discussion}
\label{sec 3}

\subsection{ Effective Coupling Constants
\texorpdfstring{$(J_{xy}, J_z, D,\text{ and } \mathfrak{h})$}{(Jxy, Jz, D, h)} }

Fig. \ref{fig:07}(a--d) shows the evolution of the effective coupling constants
$(J_{xy}, J_z, D,\text{ and } \mathfrak{h})$ as functions of the dimensionless dc electric-field strength
$d\varepsilon/B$ for the $L$- and $R$-enantiomers of 1,2-propanediol at different intermolecular separations. The symmetric exchange coupling $J_{xy}$ varies nonmonotonically with the electric-field strength $d\varepsilon/B$, as shown in the Fig. \ref{fig:07}(a)
The nonmonotonic behavior of $J_{xy}$ originates from Stark-induced mixing of the molecular rotational eigenstates, as $k$ is not a good quantum number for asymmetric-top molecules and the electric field therefore mixes different $k$ components.

In the weak-field regime, the coupling increases rapidly and attains a maximum near
$d\varepsilon/B \approx 0.5$, indicating that the rotational states remain only weakly
polarized. In this regime, the applied electric field breaks parity symmetry and induces
strong mixing between states of opposite parity. As a result, the transition dipole
matrix elements are significantly enhanced, giving rise to a pronounced coherent
spin-exchange process described by $\left( \hat{\sigma}_1^{+}\hat{\sigma}_2^{-}
      + \hat{\sigma}_1^{-}\hat{\sigma}_2^{+} \right).$
In contrast, in the strong-field regime, \((d\varepsilon/B \gg 1)\), the off-diagonal transition amplitudes that mediate spin-flip processes are progressively suppressed, leading to an asymptotic reduction of the
exchange coupling $J_{xy}$ toward zero. This behavior arises because the molecular
states become strongly Stark localized and enter the pendular regime, where the
wave functions acquire a dominant magnetic quantum number ($m$) character, which
remains a good quantum number.

Observing appreciable quantum correlations requires the enantiomers to be separated by an optimal intermolecular distance. As evident from the Fig. \ref{fig:07}(a), $J_{xy}$ is the dominant interaction mechanism at very short separations, i.e., on the nanoscale. At the optimal field strength $d\varepsilon/B \approx 0.5$, the exchange coupling reaches approximately $33\,\mathrm{GHz}$ for an intermolecular distance of $r = 1.0\,\mathrm{nm}$ (red solid line), but is reduced to about $10\,\mathrm{GHz}$ when the separation increases to $r = 1.5\,\mathrm{nm}$ (blue solid line). This strong sensitivity of the coupling to the intermolecular separation underscores the need for precise spatial control in experimental realizations, as subnanometer variations in distance can produce significant changes in the interaction strength. For larger separations, $r \geq 4.0\,\mathrm{nm}$ (yellow, purple, and black solid lines), the exchange coupling becomes essentially negligible ($< 1\,\mathrm{GHz}$) for all values of the dc-field--induced detuning.

Fig. \ref{fig:07}(b) presents the Dzyaloshinskii--Moriya interaction parameter $D$ for the real molecular system
$\mathrm{C_3H_8O_2}$ in a heterochiral right--left (RL) molecular configuration. Notably, the
magnitude of the resulting DMI is comparable to values typically encountered in solid-state
systems containing heavy elements, where strong relativistic spin--orbit coupling is present. In contrast, herein, the chiral interaction is
generated purely by molecular geometry because the
intrinsic chirality of the 1,2-propanediol molecule locally breaks
inversion symmetry. On the other hand, the applied dc electric field breaks inversion symmetry globally by mixing (hybridizing) rotational states
and thereby creating states of indefinite parity, which renders the transition dipole
moment $d_{\pm 1}$ nonzero. When a spin excitation propagates from a right-handed molecule
$(\ket{\downarrow})$ to a left-handed molecule $(\ket{\uparrow})$, the transfer is no longer
trivial; instead, the excitation undergoes a chiral twisting process and acquires a
direction-dependent geometric (Berry) phase. This phase
manifests itself microscopically as the Dzyaloshinskii--Moriya interaction,
favoring orthogonal spin orientations over collinear alignment and
stabilizing chiral spin textures. For the RL configuration, the acquired phase is $-\pi/2$, and the resulting DMI is
antisymmetric under site exchange, $(D_{ij} = -D_{ji})$. Reversing the ordering to an
LR configuration, therefore, flips the sign of $D$ while leaving its magnitude unchanged.
Physically, this sign reversal corresponds to inverting the handedness of the effective
synthetic spin--orbit coupling.  This realization of DMI thus renders the molecular spin--orbit coupling highly tunable, in contrast to
its atomic counterpart, which is intrinsically fixed by the material composition.
Such tunability constitutes the principal advantage of mapping the DMI term onto the
underlying molecular geometry. We note that the handedness of the molecular frame is defined by the scalar triple product $\mathbf{d}_a \cdot (\mathbf{d}_b \times \mathbf{d}_c)$, whose sign changes under inversion between L and R enantiomers \cite{hirota_2012,patterson_2013}. This sign reversal constitutes the microscopic origin of the corresponding sign change of the DMI parameter $D$. Fig. \ref{fig:07}(b) clearly demonstrates that the strength of the chiral interaction can be
continuously tuned from zero up to $\sim 15\,\mathrm{GHz}$ by varying the dc electric field
strength $d\varepsilon/B$. This tunability enables controlled access to phase transitions
between nonchiral and chiral spin-spiral phases.

The Ising coupling $J_z$ for the heterochiral right--left (RL) molecular pair is shown in
Fig. \ref{fig:07}(c). Over the examined range of the applied electric-field parameter $d\varepsilon/B$,
$J_z$ remains consistently negative, indicating a robust ferromagnetic bias
($J_z < 0$). Physically, this implies that the dipole--dipole interaction energy is
minimized when neighboring molecules occupy identical rotational states, either
$\ket{\uparrow\uparrow}$ or $\ket{\downarrow\downarrow}$. Since the Ising coupling arises
from the difference between like-spin and unlike-spin interactions
($C_{\uparrow\uparrow} + C_{\downarrow\downarrow}$ versus
$C_{\uparrow\downarrow} + C_{\downarrow\uparrow}$), a weak applied field only slightly
polarizes the molecules. Consequently, the induced dipole moments
$\langle d_z \rangle$ associated with the two spin states remains small and nearly
identical, rendering the energy difference between oriented and anti-oriented spins
negligible and driving $J_z$ toward zero, a trend that also appears in the strong-field
limit. At intermediate values of $d\varepsilon/B$, however, the Stark effect mixes the
rotational states differently for $\ket{\uparrow}$ and $\ket{\downarrow}$, producing a
maximal contrast in the attainable laboratory orientation. This enhances the energy difference between oriented
and anti-oriented dipole configurations, thereby stabilizing a ferromagnetic regime.
Accessing a quantum phase transition requires the transverse couplings ($J_{xy}$ and $D$)
to become comparable to $J_z$, which can be achieved by operating either at
$d\varepsilon/B < 2.5$ or $d\varepsilon/B > 2.5$.

Fig. \ref{fig:07}(d) presents the effective transverse magnetic field \(\mathfrak{h}\), which
sets the single-site energy cost for creating a spin excitation. The
magnitude and sign of \(\mathfrak{h}\) result from a competition between two
distinct physical contributions: (i) The first is the internal Stark
splitting, given by the energy difference between the dressed rotational
states \(E_{\uparrow}-E_{\downarrow}\), which is strictly positive and
increases monotonically with the applied electric field. (ii) The second is
an external dipolar bias, arising from the difference in electrostatic
environments experienced by the two states and proportional to the
geometric factor \(\Omega(r)\,[C_{1}-C_{4}]\). At large intermolecular separations
(\(r \gtrsim 2.0~\mathrm{nm}\)), the dipolar bias is strongly suppressed
by its \(1/r^{3}\) scaling and becomes negligible. In this regime, \(\mathfrak{h}\)
is dominated by the Stark splitting and remains large and positive,
favoring a paramagnetic configuration in which spins are effectively
frozen into the ground state.

For the smallest separation \(r = 1.0~\mathrm{nm}\) (red solid line),
Fig. \ref{fig:07}(d) reveals a nontrivial re-entrant behavior characterized by two
distinct resonance points. The first crossing occurs at a low field
(\(d\varepsilon/B \approx 0.8\)), where the rising Stark splitting
initially compensates the static dipolar bias. Upon further increasing
the field, the system enters a regime of strong dipolar saturation, and
a second zero crossing appears near \(d\varepsilon/B \approx 15\),
beyond where \(\mathfrak{h}\) becomes negative again. This double-crossing
indicates that at extremely short distances, the geometric dipolar bias
remains sufficiently strong to overwhelm the Stark splitting even at
high fields. While physically rich, this behavior creates a narrow and
potentially unstable window for quantum simulation, as small
fluctuations in the electric field may drive the system out of the
critical regime.

In contrast, the intermediate separation
\(r = 1.5~\mathrm{nm}\) (blue solid line) exhibits a single, well-defined resonance at
\(d\varepsilon/B \approx 2.5\). In this case, the dipolar interaction is strong enough to achieve field cancellation, 
yet sufficiently moderate to avoid the re-entrant zero-field crossing ($\mathfrak{h} \to 0$) 
observed at high fields ($d\varepsilon/B \approx 15$) for shorter distances. For
fields above the resonance, \(h\) remains small and positive, providing
a stable and controllable platform for exploring the phase diagram.
This identifies \(r = 1.5~\mathrm{nm}\) as an optimal distance for
experimental realization, balancing large interaction strengths with a monotonic and predictable tuning
parameter.

\section{Phase Diagram}
\label{sec 4}
\begin{figure}[t!]
  \centering

  % --- Top Plot ---
  \begin{minipage}[b]{1.0\linewidth}
    \centering
    \includegraphics[width=\linewidth]{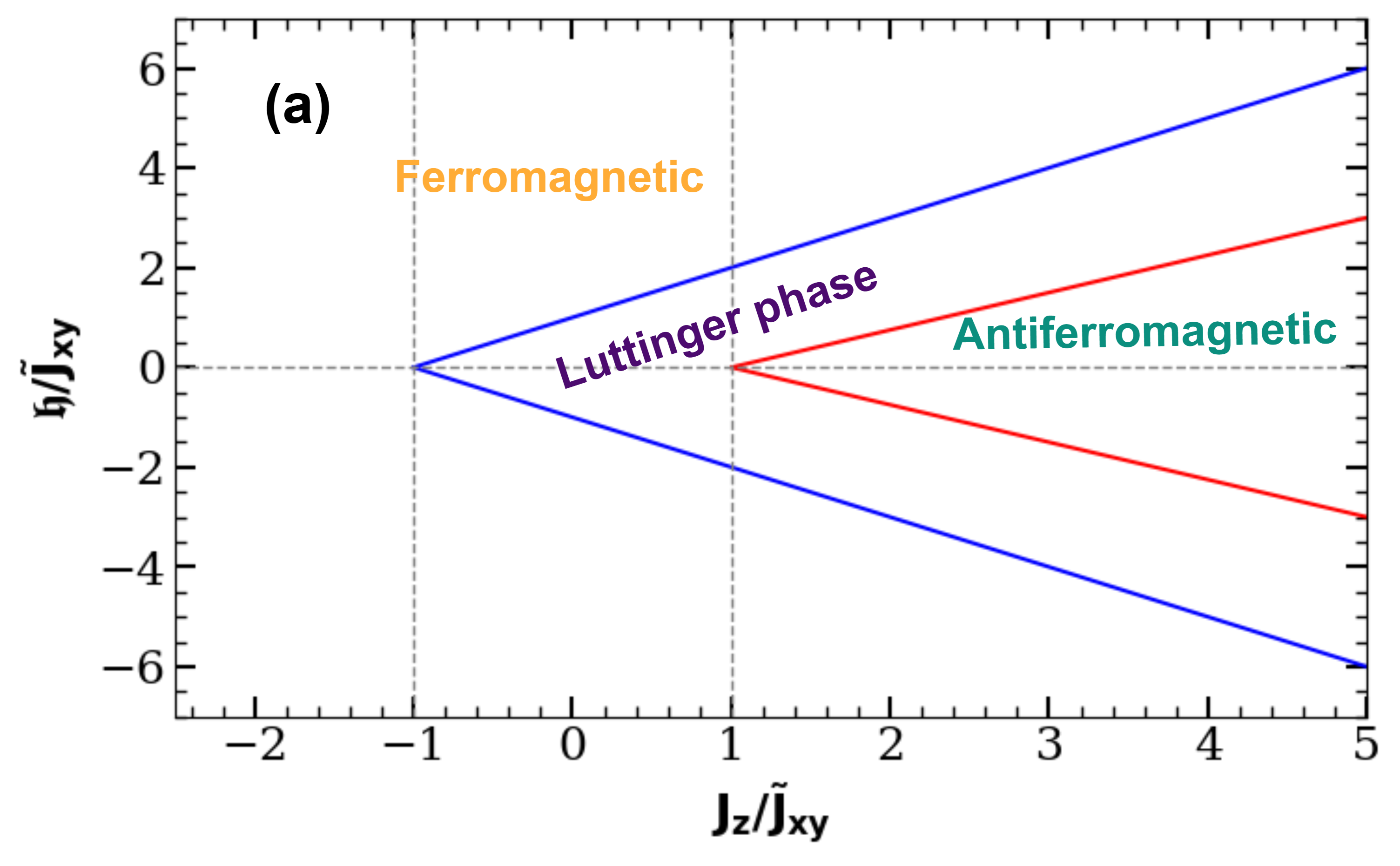}
  \end{minipage}
  \par\vspace{0.2cm}

  % --- Middle Plot (With Horizontal Space) ---
  % Reducing width to 0.9\linewidth adds 5% whitespace on each side
  \hspace{-0.6cm}
  \begin{minipage}[b]{1.05\linewidth} 
    \centering
    \includegraphics[width=\linewidth]{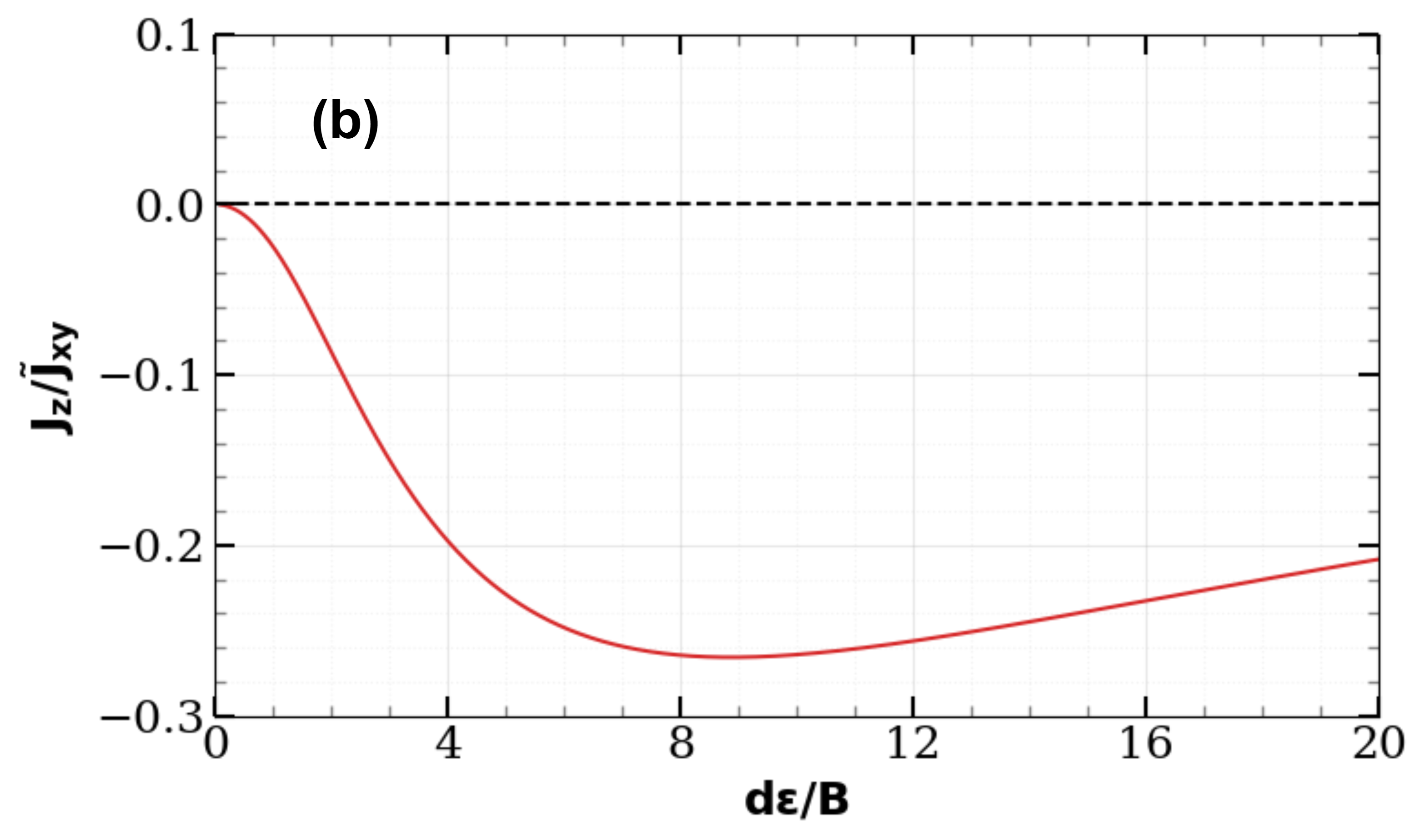}
  \end{minipage}
  \par\vspace{0.2cm}

  % --- Bottom Plot ---
  \hspace{-0.5cm}
  \begin{minipage}[b]{1.03\linewidth}
    \centering
    \includegraphics[width=\linewidth]{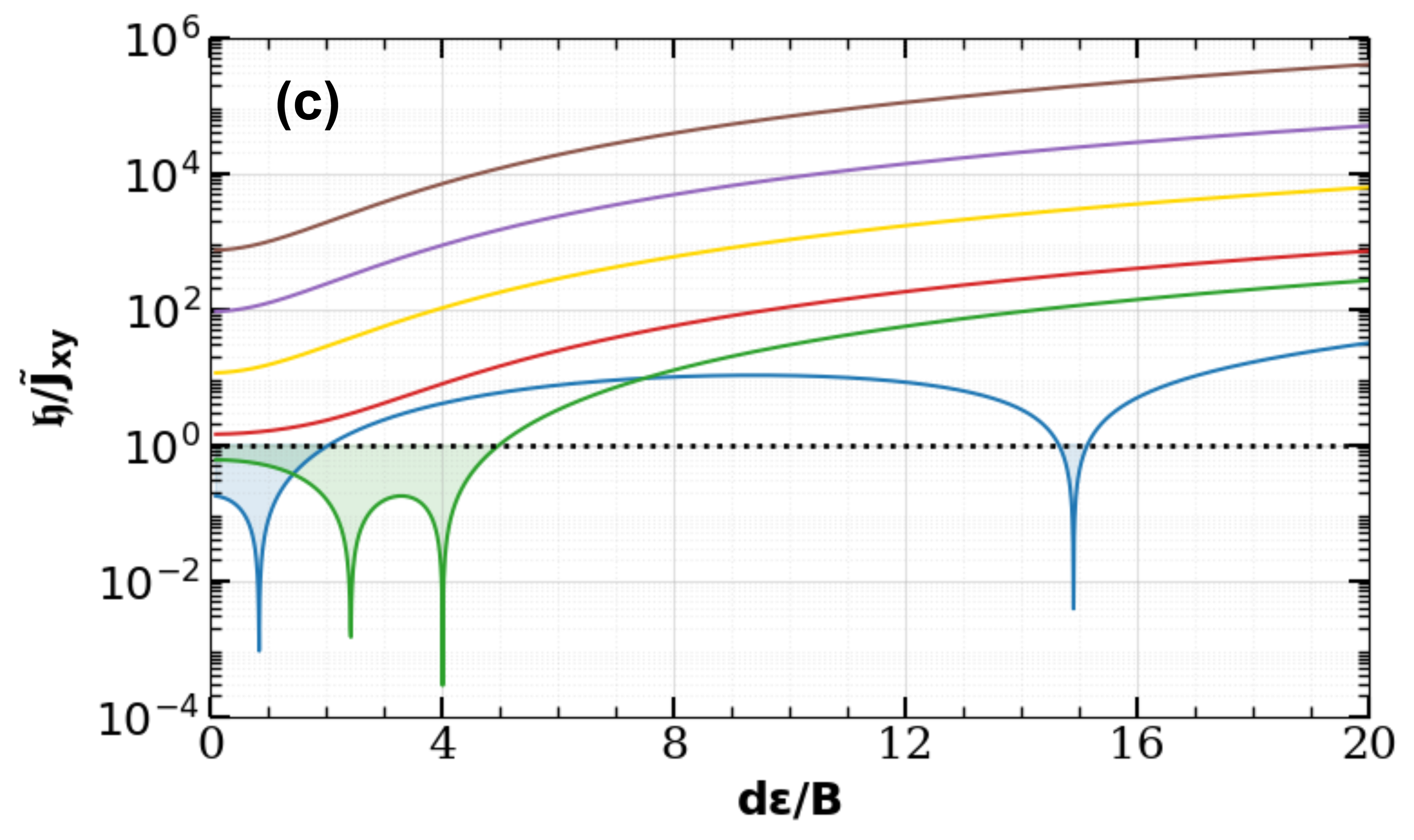}
  \end{minipage}

  \caption{(a) Universal ground-state phase diagram of the $XXZ$ spin chain as a function of the normalized Ising coupling $J_z/\tilde{J}_{xy}$ and field $\mathfrak{h}/\tilde{J}_{xy}$, where the effective transverse coupling is defined as $\tilde{J}_{xy}=\sqrt{J_{xy}^{2}+D^{2}}$. 
  (b) The anisotropy ratio $J_z/\tilde{J}_{xy}$ for Stark-dressed asymmetric top molecules indicates a robust ferromagnetic Ising bias ($J_z < 0$) that persists across the entire field range. (c) Ratio $\mathfrak{h}/\tilde{J}_{xy}$ as a function of the dimensionless electric-field parameter $d\varepsilon/B$ for several intermolecular separations $r$. The dotted horizontal line marks the quantum critical value $\mathfrak{h}/\tilde{J}_{xy} \approx 1$; regions below (above) this line correspond to the Luttinger liquid (ferromagnetic) phase. The shaded region highlights an experimentally accessible window where quantum correlations are stabilized by the interplay between Stark dressing and dipole-dipole interactions.}
  \label{fig:08}
\end{figure}

Fig. \ref{fig:08}(a) presents the ground state phase diagram of the nearest neighbor
$\text{spin-}1/2$ $XXZ$ model \cite{braiorr2016phase,ohanyan2012magnetothermal,cabra1998magnetization}. The horizontal axis is scaled by the
anisotropy parameter $J_{z}/\tilde{J}_{xy}$, where $\tilde{J}_{xy}=\sqrt{J_{xy}^{2}+D^{2}}$ denotes
the effective transverse exchange coupling after the local gauge transformation
and the vertical axis is scaled by the field strength $\mathfrak{h}/\tilde{J}_{xy}$.
The phase diagram exhibits two gapped phases: (i) an antiferromagnetic
phase for $J_{z}/\tilde{J}_{xy}>1$, and (ii) a ferromagnetic phase
for $J_{z}/\tilde{J}_{xy}<-1$. These two phases are separated by a gapless
Luttinger liquid phase \cite{haldane1980general}, which occurs for $-1<J_{z}/\tilde{J}_{xy}<1$.
According to Eq. (\ref{coupling parameters}), for asymmetric-top molecules the ratio $J_{z}/\tilde{J}_{xy}$
depends solely on the dimensionless field strength $d\epsilon/B$.
Accordingly, Fig. \ref{fig:08}(b) illustrates the evolution of
$\ensuremath{J_{z}/\tilde{J}_{xy}}\hspace{0.1cm}\text{as}\hspace{0.1cm}\ensuremath{d\epsilon/B}$
is increased from 0 to 20. The results show that the anisotropy remains
consistently negative but is strictly confined to the interval $-0.3\lesssim J_{z}/\tilde{J}_{xy}\lesssim0$.
The ratio attains a minimum value of approximately $-0.27$ at an
intermediate field strength $d\epsilon/B\approx8$, coinciding with
a pronounced dip in $J_{z}$, where the molecular polarizability is
maximal. Because $-0.27>-1$, the molecular system never enters the
pure Ising ferromagnetic phase and instead remains in the easy-plane
regime in which quantum fluctuations are dominant. To fully characterize
the phase behavior of the asymmetric molecular configuration, it is
therefore necessary to determine the accessible range of $\mathfrak{h}/\tilde{J}_{xy}$.
Fig. \ref{fig:08}(c) shows the dependence of $\mathfrak{h}/\tilde{J}_{xy}$ on $d\epsilon/B$
for several intermolecular separations $r.$ At low electric fields
$(d\epsilon/B\approx0.5-1.0)$, pronounced dips appear in the curves.
These features do not indicate any physical inconsistency of the system;
rather, they are a consequence of the logarithmic scale used, since $\mathfrak{h}/\tilde{J}_{xy}\rightarrow0$
leads to a divergence in the logarithm, producing sharp features in
the Fig. \ref{fig:08}(c). The use of a logarithmic scale on the vertical axis is required
by the strong sensitivity of the dipolar interaction to the intermolecular
distance; a linear scale would compress the critical transition boundary
$(\mathfrak{h}/\tilde{J}_{xy}=1)$ onto the horizontal axis, thereby obscuring
the Luttinger liquid window. We identify these dips as optimal operating
points, where quantum fluctuations are maximized due to the competition
between two mechanisms.

% ===== Figure 06 (Three panels, single column stack) =====

Specifically, at low electric fields, the exchange
interaction $\tilde{J}_{xy}$ is enhanced relative to the Stark splitting
energy $\mathfrak{h}$. Within this narrow regime, the kinetic energy associated
with flip--flop processes dominates over the static energy cost of
spin flips $(\tilde{J}_{xy}\gg \mathfrak{h})$, driving the system well below unity
and into the shaded green and blue regions. These shaded regions correspond
to parameter regimes that predominantly support the formation of a
gapless Luttinger liquid phase. As the electric field is increased
$(d\epsilon/B>5)$, the Stark splitting $h$ continues to grow while
$\tilde{J}_{xy}$ is progressively suppressed, thereby driving the
system into a trivial field-polarized phase. At the shortest intermolecular
separation $(r=1.0 \mathrm{nm})$, the effective field approaches zero $(\mathfrak{h}\rightarrow0)$
at two distinct points: first at very low fields $(d\epsilon/B\approx0.5-1.0)$
and again near a specific higher field $(d\epsilon/B\approx15).$ The
latter point is unstable and highly sensitive to geometric details,
making it a challenging regime for controlled simulations. The green
curve in Fig. \ref{fig:08}(c), corresponding to $r=1.5\,\mathrm{nm}$,
identifies the Goldilocks regime for realizing the Luttinger liquid
phase. This separation is optimal because, unlike $r=1.0\,\mathrm{nm}$,
which exhibits sharp dives toward zero and unstable resonances, or
$r=2.0\,\mathrm{nm}$ (red curve), which only marginally approaches
the phase boundary, the $r=1.5\,\mathrm{nm}$ geometry provides a
broad and stable operational window. The shaded green region further
indicates that the system remains in the Luttinger liquid phase over
a wide range of electric fields $(\ensuremath{d\epsilon/B\approx0.5\rightarrow4.5})$,
providing experimentalists with substantial flexibility to tune the
system without risking the onset of uncontrolled higher-order multipole
interactions.

\begin{figure}[t!]
  \centering 
  % This commands forces the image to exactly fit the width of your text
  \includegraphics[width=1.05\linewidth]{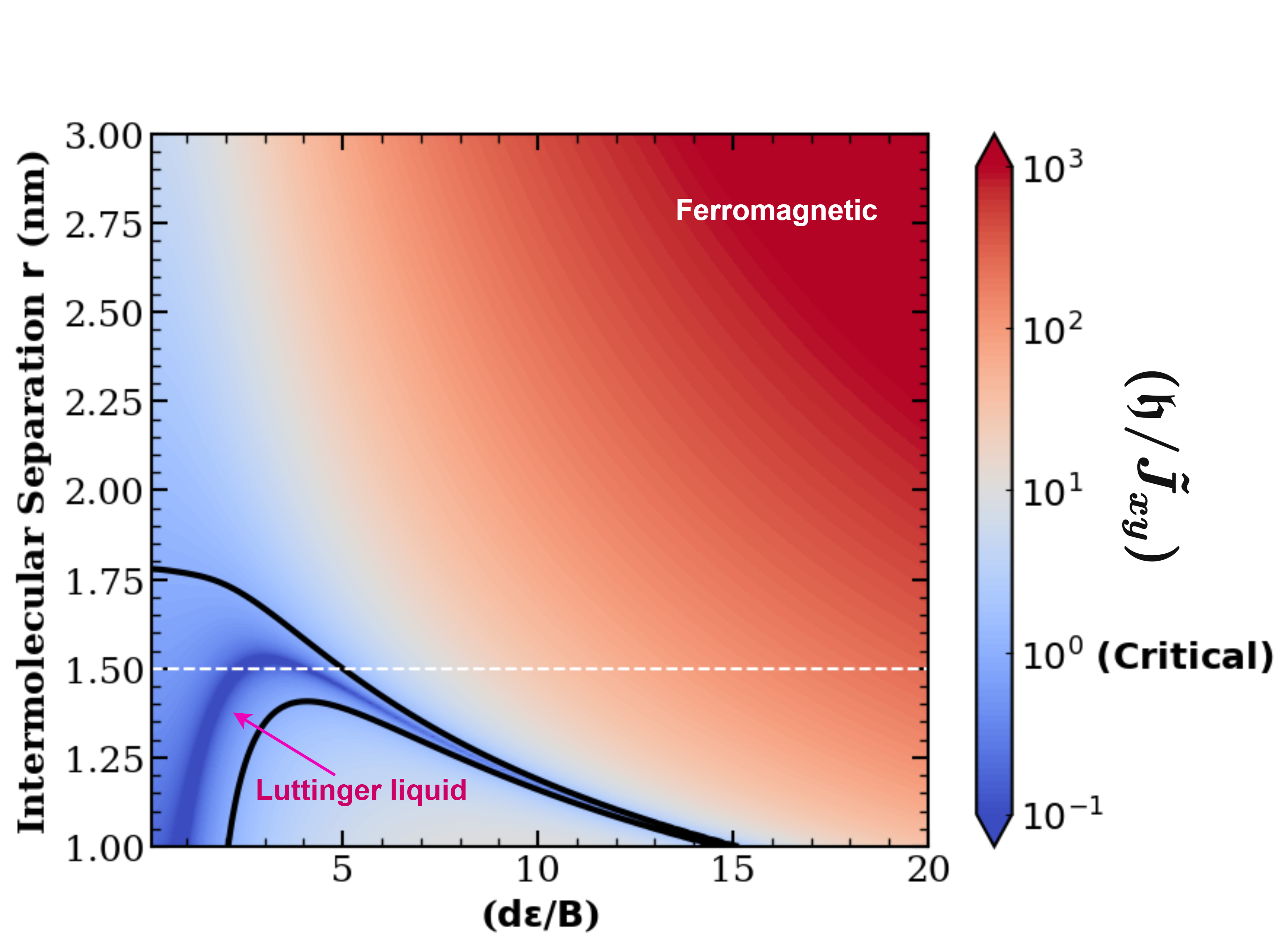} 
  
  \caption{Ground-state phase diagram of the $XXZ$ spin chain realized in a linear array
of asymmetric-top molecules, shown as a function of the field parameter
\( d\epsilon/B \) and the intermolecular separation \( r \).
The color scale represents the ratio \( \mathfrak{h}/\tilde{J}_{xy} \), while the solid
black lines denote the quantum critical boundary
\( \mathfrak{h} \approx \tilde{J}_{xy} \), which separates the Luttinger liquid phase
from the ferromagnetic phase.
}
  \label{fig:09}
\end{figure}

The contour plot in Fig. \ref{fig:09} shows the ground-state phase diagram of a one-dimensional 1,2-propanediol chain with nearest-neighbor interactions, plotted as a function of the dimensionless electric-field strength \( d\epsilon/B \) and the intermolecular separation. The color scale represents the ratio \( \mathfrak{h}/\tilde{J}_{xy} \), which quantifies the competition between electric-field–induced polarization and dipole-mediated quantum exchange processes. The white dashed line corresponds to a realistic experimental scenario in which the \(L\) and \(R\) enantiomers are fixed at an intermolecular separation of \(r = 1.5\,\mathrm{nm}\). Tracing this line from left to right, the system initially resides in the trivial (red) phase, then enters the quantum (blue) phase near \( d\epsilon/B \approx 2.0 \), and remains in this phase over a broad and stable parameter window before exiting at higher fields. This behavior further confirms that \( r = 1.5\,\mathrm{nm} \) represents an optimal lattice spacing, providing a robust operational window for realizing the chiral Luttinger liquid phase.
The solid black lines mark the critical quantum phase boundary defined by \( \mathfrak{h}/\tilde{J}_{xy} \approx 1 \), separating the gapless chiral Luttinger liquid phase from the field-polarized ferromagnetic phase. The blue region enclosed by these boundaries does not form a flat band; instead, it develops distinct lobe-like structures that extend upward at specific electric-field values ( \( d\epsilon/B \approx 2.5 \) ). These lobes correspond to resonance points where \( \mathfrak{h} \approx 0 \), as analyzed in Fig. \ref{fig:08}(c). Within these lobes, the external Stark splitting is nearly perfectly canceled by the geometric dipolar interactions. This cancelation stabilizes the quantum phase at larger intermolecular separations, allowing it to persist up to \( r \approx 1.8\,\mathrm{nm} \), which would otherwise be inaccessible.

\section{Proposed Experimental Implementation}
\label{sec 5}

%\tableofcontents
\begin{figure*}[t] % two-column APS layout
\centering

\setlength{\fboxsep}{0pt} % Removes padding inside the border
    % Subtract 2x the rule thickness and 2x the separation padding
\fbox{\includegraphics[width=0.99\textwidth]{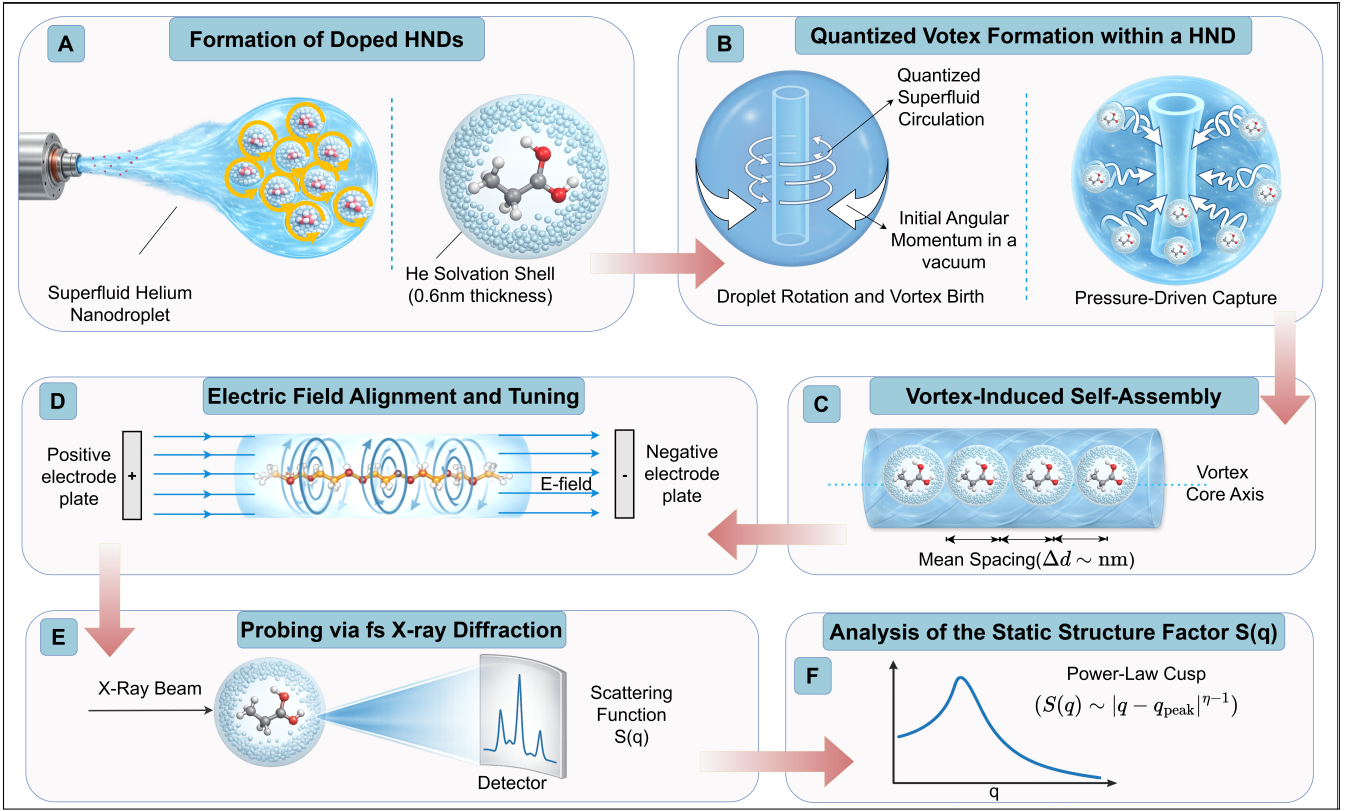}}

\caption{Schematic of the experimental protocol for realizing and detecting a Chiral Luttinger Liquid (CLL) in superfluid helium nanodroplets.
    \textbf{(A)} \emph{Formation of superfluid helium nanodroplets (HND) doped with 1,2 propanediol molecules} via co-expansion of $^4$He gas seeded with 1,2-propanediol with superfluid $^4$He into vacuum through a cryogenic nozzle. The panel depicts the cooling and solvation of the dopant molecules (left) as well as a dopant molecule within a helium solvation shell that forms due to van der Waals attraction between the molecule and helium (right).
    \textbf{(B)} \emph{Quantized vortex formation within a HND.} As the droplet expands into vacuum through the nozzle, it cools and acquires angular momentum. Below the superfluid transition, a quantized vortex core is formed to conserve angular momentum. The resulting pressure gradient (Bernoulli force) drives the solvated molecules radially inward, trapping them in the vortex core.
    \textbf{(C)} \emph{Vortex-induced self-assembly.} The quantized vortex core within a given superfluid droplet guides the solvated molecules doped therein into a 1D linear filament. The mean spacing is defined by the effective diameter of the solvation shell. Based on DFT calculations \cite{toennies2004superfluid,Dalfovo1994}, the shell thickness is $\Delta R_{\text{shell}} \approx 6\,\text{\AA}$. Thus, the spacing is estimated as $\Delta r \approx 2 \times (R_{\text{mol}} + \Delta R_{\text{shell}}) \approx 2 \times (2.5\,\text{\AA} + 6.0\,\text{\AA}) \approx 1.7\,\text{nm}$.
    \textbf{(D)} \emph{Electric field alignment and tuning of the interaction strength.} An external dc electric field $\boldsymbol{\varepsilon}$, applied via parallel plate electrodes, aligns the molecular dipoles along the vortex axis (head-to-tail) stabilizes the chain and allows for tuning the interaction strength.
    \textbf{(E)} \emph{Structural probing via femtosecond X-ray diffraction} to measure the scattering function.
    \textbf{(F)} \emph{Analysis of the static structure factor $S(q)$.} The CLL phase is distinguished from a crystalline solid by the absence of a sharp Bragg peak and the emergence of a power-law singularity, $S(q) \sim |q - q_{\text{peak}}|^{\eta-1}$, characteristic of the algebraic decay of correlations in 1D quantum fluids\cite{giamarchi2003quantum}}.
    \label{fig:experimental_stages}
\end{figure*}

\emph{Conventional Experimental Platforms.} Studying the chiral Luttinger liquid (CLL) phase requires an intermolecular spacing of \( r \approx 1.7~\mathrm{nm} \), which lies in a regime that poses severe challenges for conventional experimental platforms. Although advanced optical trapping techniques, such as structured light, allow precise control over neutral particles \cite{yang2021optical}, a major limitation remains the attainable trap size. Conventional optical lattices are restricted by the diffraction limit of light, leading to minimum lattice site spacings of about \( 300\text{--}500~\mathrm{nm} \) \cite{Gross2017OptLattSim,Schafer2020ToolsOptLatt}. Such separations are too large to accomplish sufficiently strong dipolar interactions %at nanometer-scale intermolecular distances 
and, as a result, preclude the formation of a Luttinger liquid phase.

Turning to nanophotonics, where evanescent electromagnetic fields are exploited to generate traps that are much smaller than the wavelength of light, such as optical nanofiber traps \cite{goban2012demonstration} and metallic plasmonic nanostructures \cite{tame2013quantum}, enabled trapping and manipulating single atoms with nanometer-scale precision. In particular, plasmonic nano-optical tweezers \cite{juan2011plasmon} and nanoplasmonic lattices \cite{gullans2012nanoplasmonic} have been successful in this regard. 
However, close proximity of trapped atoms or molecules to surfaces introduces deleterious decoherence channels, including Johnson--Nyquist noise and Casimir--Polder interactions \cite{lin2003impact}, which are known to destabilize fragile coherent quantum phases. A robust realization of the CLL phase thus requires a confinement platform that provides sub-nanometer positional control while remaining free from surface-induced noise. Such combination of requirements is naturally met by superfluid helium nanodroplets.

\emph{Superfluid Helium Nanodroplets (HNDs)} comprised of $^4$He atoms provide a cold environment ($0.38~\mathrm{K}$) which is largely perturbation-free to the extent that molecules embedded in it can retain their rotational degrees of freedom \cite{toennies2004superfluid,stienkemeier2006spectroscopy,gomez2014shapes}. Imparting angular momentum to the HNDs results in quantized vortex lines that act as effective one-dimensional confinement channels. While doping the droplets via pick up of molecules during passage through a vapor cell  has a limited efficiency (leading to loading of just a few molecules per droplet), a co-expansion with a helium beam seeded with the atoms or molecules of interest (such as 1,2-propanediol in our case or Xe in Ref. \cite{gomez2014shapes}) upstream from the nozzle greatly increases the doping efficency. For Xe, the number of Xe atoms, $N_{Xe}$, loaded into a droplet comprised of $N_{He}\approx 10^6$ helium atoms was shown to be as high as $N_{Xe}\approx 10^{-3} N_{He}$ \cite{gomez2014shapes}.
  
Meanwhile, coherent diffractive imaging experiments showed that heavy dopants preferentially localize along vortex cores \cite{feinberg2025x}, producing characteristic Bragg features in the diffraction signal. These experiments thus provide direct structural evidence of vortex-confined matter -- and a concrete roadmap for achieving the nanometer-scale, quasi-one-dimensional regime, which is key to creating and studying the chiral Luttinger phase. In particular, X-ray diffraction could be used to probe the unique periodicity of the collective density wave of a chiral Luttinger liquid phase and thereby obtain its definitive signature. The diffraction intensity is directly proportional to the static structure factor $S(q)$, defined as the Fourier transform of the density--density correlation function:
\begin{equation}
S(q) = \frac{1}{N} \sum_{j,l} e^{-iq(r_j - r_l)} \langle \hat{n}_j \hat{n}_l \rangle,
\end{equation}
where $\hat{n}_j$ denotes the local density (or effective spin projection) at site $j$. This quantity measures the strength of spatial correlations at a given wavevector $q$, revealing the dominant ordering pattern of the system. Detecting diffraction features at a wavevector $q = 2\pi / (1.7~\mathrm{nm})$, corresponding to intermolecular spacing $r= 1.7~\mathrm{nm}$, would therefore provide direct structural evidence for the formation of the CLL phase. Crucially, the Luttinger liquid state could be distinguished from trivial ordered or fluid phases by analyzing lineshape of the diffraction peak. Unlike the sharp, delta-function-like Bragg peaks of a crystalline solid or the broad, smooth features characteristic of a conventional fluid, a Luttinger liquid exhibits power-law singularities at $2k_F$ \cite{giamarchi2003quantum,Haldane1981,Cazalilla2011}. As a result, the diffraction peaks are expected to display a characteristic cusp-like profile governed by an algebraic decay exponent, thereby providing a rigorous %mathematical 
 signature of the underlying quantum critical state.

More generally, a chiral Luttinger liquid is characterized by a chiral excitation spectrum, a well-defined collective plasmon velocity, and power-law correlation dynamics. In solid-state electronic systems, these properties are typically inferred from transport observables, such as quantized conductance or Hall response. For isolated quantum fluids such as helium nanodroplets (HNDs), however, direct contact-based measurements are not possible. In this case, the appropriate experimental analogue is time-resolved pump--probe spectroscopy. In this experimental framework, a femtosecond pump pulse generates a localized excitation within the vortex-confined molecular filament, and a time-delayed probe pulse subsequently interrogates its propagation and relaxation. If the underlying phase is chiral, the excitation dynamics should display strict directionality, indicative of broken time-reversal symmetry, and follow the non-Fermi-liquid power-law decay associated with Luttinger physics. Although conventional transport measurements are not feasible in this platform, the excitation spectrum and relaxation dynamics offer a rigorous and experimentally meaningful alternative to transport-based evidence.

A key challenge in neutral droplet experiments is the interaction time limited millisecond-scale due to high beam velocities ($200\text{--}400~\mathrm{m/s}$). To enable the extended observation times necessary for pump-probe spectroscopy, we propose utilizing electrostatic trapping of charged nanodroplets, a technique recently demonstrated to store HNDs for up to one minute \cite{veternik2026extending}.
A critical concern is whether the presence of a net charge on the droplet surface would generate electrostatic noise disruptive of the delicate dipole-dipole interactions that drive the CLL phase. The intermolecular dipole-dipole interaction at $r=1.7$ nm is approximately $V_{dd} \approx 1.0$ K. In contrast, the Stark perturbation from a surface charge on a macroscopic droplet ($R \approx 500$ nm) is estimated at $V_{charge} \approx 0.003$ K. Since the signal-to-noise ratio $V_{dd}/V_{charge} \sim 300$, the phase can thus be expected to be robust against electrostatic perturbations. For more detail see Appendix \ref{app:E}. Furthermore, although achieving strong dc electric fields in superfluid droplets can be challenging due to breakdown limits, fields on the order of 10 kV/cm (corresponding to $d\varepsilon/B \approx 2$) are achievable and sufficient to hybridize the rotational states required \cite{Stiles2003PRL_HeDropletDipoles}.

\section{Conclusion}
\label{sec 6}
In this paper, we demonstrate that a hetero-enantiomer (R--L) pair can be mapped onto an effective Heisenberg $XXZ$ spin model, exhibiting an emergent Dzyaloshinskii--Moriya interaction (DMI) in the presence of an external electric field and dipole--dipole coupling. The effective spin-$1/2$ description is realized by projecting onto the two lowest pendular states of the molecules, each formed as a coherent superposition of rotational basis states, and labeled by different values of $\tilde{j}$ and $\tilde{k}$. An external electric field is then used to couple these two states, forming an effective two-level system that provides a new platform for studying the Heisenberg model based on chiral molecules. In order to characterize the model comprehensively, we investigate a wide parameter regime involving the rotational constants, the strength and orientation of the external electric field, and the strength of the dipole–dipole coupling between the molecules. The external electric field is essential for controlling and engineering the effective spin interactions: it is required for hybridizing the rotational states as well as generating nonzero transition dipole moments $C_{d_1}$ and $C_{d_2}$ that make the requisite one-photon transition $\lvert\downarrow\rangle \leftrightarrow \lvert\uparrow\rangle$ fully allowed.

In previous studies, linear, spherical, and symmetric-top molecules were employed, with the pendular states $\lvert00\rangle$ and $\lvert10\rangle$ chosen to represent an effective two-level system. However, because these systems possess inversion symmetry, they did not allow for the study of hetero-enantiomer (R--L) interactions. In our model, the Dzyaloshinskii--Moriya interaction arises naturally from the interference between the transition dipole moments of heterochiral (R--L) enantiomers. This establishes a clear connection between molecular chirality and the resulting spin–spin interactions, thus providing a distinct advantage for controlling and engineering effective spin couplings. Through a rigorous analysis of the quantum phase diagrams shown in Figs.~\ref{fig:08} and Fig. \ref{fig:09}, we identify an optimal experimental regime characterized by an intermolecular separation of $r \approx 1.5~\mathrm{nm}$ and intermediate electric-field strengths $d\varepsilon/B$, where the chiral Luttinger liquid phase emerges and remains protected against trivial phases. A further advantage of the model presented herein is that, in contrast to solid-state platforms where spin--orbit coupling is fixed by material properties, the chiral Dzyaloshinskii--Moriya interaction can be continuously tuned using an external field. In their sum, these properties establish arrays of 1,2-propanediol molecules as a promising platform for quantum simulation, as this work bridges the gap between single-molecule rotational spectroscopy and many-body condensed-matter physics.

Moreover, utilizing an array of asymmetric top molecules has the folloiwng virtues: 

\emph{Topological Switching.} By replacing left-handed (L) enantiomers
with right-handed (R) enantiomers in the molecular array, the sign of
$D$ is instantaneously reversed. As a consequence, the chirality of the resulting Luttinger liquid is inverted, switching the spiral ordering from clockwise to counterclockwise (or vice versa).

\emph{Engineered Domain Walls.} If a heterogeneous molecular
chain is constructed, for example
, · · · L–L–L–R–R–R · · ·,
the sign of the DMI parameter $D$ necessarily changes at the interface between the two enantiomeric regions. This enforced sign change generates a topological domain wall, which can host exotic fractionalized excitations or localized zero modes. In this way, quantum defects can be engineered deterministically through chemical substitution alone.\\

As a future extension, the present model can be engineered to simulate the Su–Schrieffer–Heeger (SSH) model \cite{su1979solitons} by introducing lattice dimerization, thereby giving rise to topologically protected zero-energy modes localized at the chain boundaries \cite{asboth2016short}.
This suggests that molecular arrays may serve as decoherence-free quantum memories \cite{rabl2007molecular}, in which quantum information is stored in topologically protected edge states and remains immune to thermal noise in the interior of the chain. The present study thus contributes to the exploration of complex topological phases, which may yield insights into the microscopic mechanisms responsible for chirality-induced spin selectivity (CISS) in biological and chemical systems \cite{gohler2011spin,akbar2025chiral}. Electron transfer in biological systems, such as photosynthesis, occurs over long distances with surprisingly low scattering. While highly speculative, if such systems were to exhibit a topological phase akin to the SSH model \cite{yuen2014topological}, one might expect transport pathways involving edge-like channels that are comparatively resilient to biological environmental noise.

\section{Acknowledgments}
We  would like to acknowledge the financial support from the Quantum Science Center, a National Quantum Information Science Research Center of the U.S. Department of Energy (DOE), operated at Oak Ridge National Laboratory
(ORNL).

\clearpage
\bibliographystyle{apsrev4-2}
\bibliography{bibliography}   % assuming bibliography.bib is the one you want

\clearpage
\appendix
\onecolumngrid

\setcounter{equation}{0}
\makeatletter
\renewcommand{\theequation}{A\arabic{equation}}
\makeatother

\refstepcounter{section}
\section*{APPENDIX \Alph{section}: Derivation of the Interacting Hamiltonian}
\phantomsection
\label{app:A}

In quantum mechanics and rotational spectroscopy, spherical tensors are preferred over Cartesian tensors because they can transform naturally between the laboratory and molecule frames. In spherical representation, spatial directions are labeled by angular momentum quantum numbers rather than Cartesian components ($x,y,z$). Consequently, to describe two interacting asymmetric-top molecules such as 1,2-propanediol, it is natural to work in the basis $\lvert jkm\rangle$, which follows directly from the spherical-tensor formalism and allows a symmetric transformation of vector operators between frames.

A spherical tensor of rank $l$ represents an object with angular momentum $l$. Since an angular momentum $l$ has $2l+1$ possible projections along a quantization axis, labeled by $m=-l,-l+1,\dots,l$, a spherical tensor of rank $l$ consists of $2l+1$ components $T_{m}^{(l)}$.

Consider a two-dimensional vector $\boldsymbol{v}$, such as the dipole moment of a molecule, rigidly attached to the molecular frame. The same vector may be described from two complementary perspectives: (i) in the molecule-fixed frame, where the vector remains fixed relative to the molecular body and has components $\boldsymbol{v}=(v_{x},v_{y})$; and (ii) in the laboratory frame, where the observer is fixed and the molecule rotates, yielding rotated components $\boldsymbol{v}=(v_{X},v_{Y})$. If the molecule is rotated by an angle $\theta$, the transformation between these components follows directly from elementary trigonometry:
\begin{align}
v_{X} &= v_{x}\cos\theta - v_{y}\sin\theta, \\
v_{Y} &= v_{x}\cos\theta + v_{y}\sin\theta.
\end{align}
We can write this as a rotation matrix:
\begin{equation}
\begin{pmatrix}
v_{X}\\
v_{Y}
\end{pmatrix}
=
\begin{pmatrix}
\cos\theta & -\sin\theta\\
\sin\theta & \cos\theta
\end{pmatrix}
\begin{pmatrix}
v_{x}\\
v_{y}
\end{pmatrix}.
\end{equation}

In three dimensions, transformations of spherical components are described using Wigner D-matrices rather than ordinary rotation matrices. Thus, the molecule-to-laboratory frame transformation can be expressed as:
\begin{equation}
T_{q}^{(l)} = \sum_{r=-l}^{l} D_{q,r}^{l*}(\Omega) T_{r}^{(l)}.
\end{equation}
Here, $T_{r}^{(l)}$ denotes the tensor components in the molecule-fixed frame, $T_{q}^{(l)}$ denotes the corresponding components in the laboratory frame, and $D_{q,r}^{l*}(\Omega)$ is the Wigner D-matrix element that specifies how the $r$-th molecular-frame component contributes to the $q$-th laboratory-frame component. The inverse transformation, which maps laboratory-frame measurements back to the molecule’s intrinsic properties, is given by:
\begin{equation}
T_{r}^{(l)} = \sum_{q=-l}^{l} D_{q,r}^{l*}(\Omega) T_{q}^{(l)}.
\end{equation}

For a vector (rank-1 tensor, $l=1$), we have 3 spherical components: $T_{-1}^{(1)}$, $T_{0}^{(1)}$, and $T_{1}^{(1)}$. Suppose we know the vector in the molecule frame: $(T_{-1}^{(1)},\,T_{0}^{(1)},\,T_{1}^{(1)})$. To find it in the laboratory frame, we perform the summation over $r=-1,0,1$:
\begin{subequations}
\begin{align}
T_{-1}^{(1)} &= D_{-1,-1}^{1*}(\Omega)T_{-1}^{1} + D_{-1,0}^{1*}(\Omega)T_{0}^{1} + D_{-1,1}^{1*}(\Omega)T_{1}^{1}, \\
T_{0}^{(1)}  &= D_{0,-1}^{1*}(\Omega)T_{-1}^{1}  + D_{0,0}^{1*}(\Omega)T_{0}^{1}  + D_{0,1}^{1*}(\Omega)T_{1}^{1}, \\
T_{1}^{(1)}  &= D_{1,-1}^{1*}(\Omega)T_{-1}^{1}  + D_{1,0}^{1*}(\Omega)T_{0}^{1}  + D_{1,1}^{1*}(\Omega)T_{1}^{1}.
\end{align}
\end{subequations}

In this paper, we employ this transformation to study the dipole–dipole interaction:
\begin{equation}
\hat{H}_{\mathrm{dd}} = \frac{\hat{\boldsymbol{d}}_{1}\cdot\hat{\boldsymbol{d}}_{2} - 3(\hat{\boldsymbol{d}}_{1}\cdot\boldsymbol{e}_{r})(\hat{\boldsymbol{d}}_{2}\cdot\boldsymbol{e}_{r})}{r^{3}},
\end{equation}
where $\hat{\boldsymbol{d}}_{1}$ and $\hat{\boldsymbol{d}}_{2}$ represent molecular dipole moments, which are rank-1 tensor operators ($l=1$). Their spherical components in the molecule-fixed frame are denoted by $d_{1,r}^{(1)}$ and $d_{2,r}^{(1)}$, with $r=-1,0,1$. For dipole 1, the laboratory-frame spherical components are related to the molecule-fixed components by:
\begin{equation}
d_{1,q}^{(\mathrm{lab})} = \sum_{r_{1}=-1}^{1} D_{q,r_{1}}^{1*}(\Omega_{1})\,d_{1,r_{1}}^{(\mathrm{mol})},
\end{equation}
where $\Omega_{1}$ specifies the orientation of molecule 1. Similarly, for dipole 2:
\begin{equation}
d_{2,q}^{(\mathrm{lab})} = \sum_{r_{2}=-1}^{1} D_{q,r_{2}}^{1*}(\Omega_{2})\,d_{2,r_{2}}^{(\mathrm{mol})},
\end{equation}
with $\Omega_{2}$ denoting the orientation of molecule 2.

We are primarily interested in the dot product of the two dipole moment operators, which in spherical components is given by:
\begin{equation}
\hat{\boldsymbol{d}}_{1}\cdot\hat{\boldsymbol{d}}_{2} = \sum_{s=-1}^{1}(-1)^{s}\,d_{1,s}^{(\mathrm{lab})}\,d_{2,-s}^{(\mathrm{lab})}.
\end{equation}
Substituting the laboratory-frame transformations yields:
\begin{equation}
\hat{\boldsymbol{d}}_{1}\cdot\hat{\boldsymbol{d}}_{2} = \sum_{s=-1}^{1}(-1)^{s}
\left[\sum_{r_{1}=-1}^{1} D_{s,r_{1}}^{1*}(\Omega_{1})\,d_{1,r_{1}}^{(\mathrm{mol})}\right]
\left[\sum_{r_{2}=-1}^{1} D_{-s,r_{2}}^{1*}(\Omega_{2})\,d_{2,r_{2}}^{(\mathrm{mol})}\right].
\end{equation}

Similarly, the term $(\hat{\boldsymbol{d}}_{1}\cdot\boldsymbol{e}_{r})(\hat{\boldsymbol{d}}_{2}\cdot\boldsymbol{e}_{r})$ requires a transformation of the dipole components along the inter-molecular axis. Since $\boldsymbol{e}_{r}$ is chosen along the laboratory $Z$-axis, its spherical components satisfy $e_{r,0}=1$ and $e_{r,\pm1}=0$, leading to:
\begin{equation}
\hat{\boldsymbol{d}}_{1}\cdot\boldsymbol{e}_{r} = d_{1,0}^{(\mathrm{lab})} = \sum_{r_{1}=-1}^{1} D_{0,r_{1}}^{1*}(\Omega_{1})\,d_{1,r_{1}}^{(\mathrm{mol})},
\end{equation}
with an analogous expression for dipole 2. Consequently, the dipole–dipole Hamiltonian $\hat{H}_{\mathrm{dd}}$ can be written as a double sum over $r_{1}$ and $r_{2}$ involving products of Wigner D-matrices.

To evaluate the matrix elements of the interaction, $\langle j_{1}k_{1}m_{1},\,j_{2}k_{2}m_{2}|\hat{H}_{\mathrm{dd}}| j'_{1}k'_{1}m'_{1},\,j'_{2}k'_{2}m'_{2}\rangle$, one must compute matrix elements of the form:
\begin{equation}
\langle j_{1}k_{1}m_{1}\lvert D_{q,r_{1}}^{1*}(\Omega_{1}) \lvert j'_{1}k'_{1}m'_{1}\rangle
\quad \text{and} \quad
\langle j_{2}k_{2}m_{2}\lvert D_{q,r_{2}}^{1*}(\Omega_{2}) \lvert j'_{2}k'_{2}m'_{2}\rangle.
\end{equation}
These are given by the Wigner-Eckart theorem using Wigner 3-$j$ symbols:
\begin{equation}
\langle j_{1}k_{1}m_{1}|D_{q,r_{1}}^{1*}(\Omega_{1})|j'_{1}k'_{1}m'_{1}\rangle = (-1)^{m_{1}+k_{1}} \sqrt{(2j_{1}+1)(2j'_{1}+1)}
\begin{pmatrix}
j_{1} & 1 & j'_{1}\\
-m_{1} & q & m'_{1}
\end{pmatrix}
\begin{pmatrix}
j_{1} & 1 & j'_{1}\\
-k_{1} & r_{1} & k'_{1}
\end{pmatrix},
\label{EqA14}
\end{equation}
and
\begin{equation}
\langle j_{2}k_{2}m_{2}|D_{q,r_{2}}^{1*}(\Omega_{2})|j'_{2}k'_{2}m'_{2}\rangle = (-1)^{m_{2}+k_{2}} \sqrt{(2j_{2}+1)(2j'_{2}+1)}
\begin{pmatrix}
j_{2} & 1 & j'_{2}\\
-m_{2} & q & m'_{2}
\end{pmatrix}
\begin{pmatrix}
j_{2} & 1 & j'_{2}\\
-k_{2} & r_{2} & k'_{2}
\end{pmatrix}.
\label{EqA15}
\end{equation}
The only non-zero matrix elements are those satisfying the selection rules $m_{1}=q+m'_{1}$, $m_{2}=q+m'_{2}$, $k_{1}=r_{1}+k'_{1}$, and $k_{2}=r_{2}+k'_{2}$.

% --- Appendix B ---
\setcounter{equation}{0}
\setcounter{figure}{0}
\makeatletter
\renewcommand{\theequation}{B\arabic{equation}}
\makeatother
\renewcommand{\thefigure}{B\arabic{figure}} 
\refstepcounter{section}  % create a real section counter step (gives "B")
\section*{APPENDIX \Alph{section}: $\hat{H}=\hat{H}_{rot}+\hat{H}_{dc}$}
\phantomsection           % ensure a hyperref anchor exists
\label{app:B}

We have derived a comprehensive formulation, based on Wigner D-matrices, for transforming operators between the laboratory and molecule-fixed frames. This formalism applies directly to the rotational and Stark terms of the Hamiltonian, $\hat{H}_{\mathrm{rot}}$ and $\hat{H}_{\mathrm{dc}}$.

The dc Stark interaction is given by:
\begin{equation}
\hat{H}_{\mathrm{dc}} = -\boldsymbol{d}\cdot\boldsymbol{\varepsilon},
\end{equation}
where $\boldsymbol{d}$ is the dipole moment operator and $\boldsymbol{\varepsilon}$ is the external dc electric field. Both $\boldsymbol{d}$ and $\boldsymbol{\varepsilon}$ may be expressed as rank-1 spherical tensors, leading to:
\begin{equation}
\boldsymbol{d}\cdot\boldsymbol{\varepsilon} = \sum_{q=-1}^{1}(-1)^{q}d_{q}\,\varepsilon_{-q},
\end{equation}
where $d_{q}$ and $\varepsilon_{-q}$ denote the spherical components in the laboratory frame. Consequently,
\begin{equation}
\hat{H}_{\mathrm{dc}} = -\sum_{q=-1}^{1}(-1)^{q}d_{q}\,\varepsilon_{-q}.
\label{EqB3}
\end{equation}
Since the molecular dipole moment is naturally defined in the molecule-fixed frame through the components $d_{r}^{(\mathrm{mol})}$, whereas $d_{q}$ denotes the corresponding components in the laboratory frame, it is necessary to transform $d_{q}$ into the molecule-fixed frame. This transformation is given by:
\begin{equation}
d_{q} = \sum_{r=-1}^{1} D_{q,r}^{1*}(\Omega)\,d_{r}^{(\mathrm{mol})},
\label{EqB4}
\end{equation}
where $d_{r}^{(\mathrm{mol})}$ are fixed molecular constants and $\Omega$ specifies the molecular orientation.

Substituting Eq. (\ref{EqB3}) into Eq. (\ref{EqB4}) yields:
\begin{equation}
\hat{H}_{\mathrm{dc}} = -\sum_{q=-1}^{1}(-1)^{q} \left[\sum_{r=-1}^{1}D_{q,r}^{1*}(\Omega)\,d_{r}^{(\mathrm{mol})}\right] \varepsilon_{-q},
\end{equation}
which may be written equivalently as:
\begin{equation}
\hat{H}_{\mathrm{dc}} = -\sum_{q=-1}^{1}\sum_{r=-1}^{1}(-1)^{q} D_{q,r}^{1*}(\Omega)\,d_{r}^{(\mathrm{mol})}\,\varepsilon_{-q}.
\end{equation}

The corresponding matrix elements in the $\lvert jkm\rangle$ basis are therefore:
\begin{equation}
\langle jkm|\hat{H}_{\mathrm{dc}}| j'k'm'\rangle = -\sum_{q=-1}^{1}\sum_{r=-1}^{1}(-1)^{q} \langle jkm| D_{q,r}^{1*}(\Omega) | j'k'm'\rangle\,d_{r}^{(\mathrm{mol})}\,\varepsilon_{-q}.
\end{equation}
Using the standard expression for the matrix elements of the Wigner D-matrix, this becomes:
\begin{equation}
\langle jkm|\hat{H}_{\mathrm{dc}}\lvert j'k'm'\rangle = -\sum_{q=-1}^{1}\sum_{r=-1}^{1}(-1)^{q}d_{r}^{(\mathrm{mol})}\,\varepsilon_{-q}(-1)^{m+k}\sqrt{(2j+1)(2j'+1)}
\begin{pmatrix}
j & 1 & j'\\
-m & q & m'
\end{pmatrix}
\begin{pmatrix}
j & 1 & j'\\
-k & r & k'
\end{pmatrix}.
\end{equation}

Since the external electric field has only a single component along the laboratory $z$-axis, $\boldsymbol{\varepsilon}=\varepsilon\,\boldsymbol{e}_{z}$, its spherical components satisfy:
\begin{equation}
\varepsilon_{0} = \varepsilon_{z}, \qquad \varepsilon_{\pm1} = 0.
\end{equation}
As a result, only the $q=0$ term contributes to the dc Stark Hamiltonian. The matrix elements therefore reduce to:
\begin{equation}
\langle jkm\lvert\hat{H}_{\mathrm{dc}}\lvert j'k'm'\rangle = -\sum_{r=-1}^{1}(-1)^{0}d_{r}^{(\mathrm{mol})}\,\varepsilon_{0}(-1)^{m+k}\sqrt{(2j+1)(2j'+1)}
\begin{pmatrix}
j & 1 & j'\\
-m & 0 & m'
\end{pmatrix}
\begin{pmatrix}
j & 1 & j'\\
-k & r & k'
\end{pmatrix}.
\end{equation}
Using $\varepsilon_{0}=\varepsilon_{z}$, this expression simplifies to:
\begin{equation}
\langle jkm\lvert\hat{H}_{\mathrm{dc}}\lvert j'k'm'\rangle = -\varepsilon_{z}\sum_{r=-1}^{1}d_{r}^{(\mathrm{mol})}(-1)^{m+k}\sqrt{(2j+1)(2j'+1)}
\begin{pmatrix}
j & 1 & j'\\
-m & 0 & m'
\end{pmatrix}
\begin{pmatrix}
j & 1 & j'\\
-k & r & k'
\end{pmatrix}.
\end{equation}

From this expression, several selection rules follow immediately. First, the condition $m=m'$ reflects conservation of the laboratory-frame projection quantum number. Second, the relation $k'=k-r$ with $r=-1,0,1$ implies that the molecule-fixed projection can change only according to $\Delta k=0,\pm1$. Finally, the Wigner 3-$j$ symbol:
\begin{equation}
\begin{pmatrix}
j & 1 & j'\\
-m & 0 & m'
\end{pmatrix}
\end{equation}
is nonzero only when $|j-j'|\leq1$. Combined with the condition $m=m'$, this yields the selection rule $\Delta j=0,\pm1$.

When we consider the full system of two interacting molecules, the dc interaction part becomes:
\begin{equation}
\hat{H}_{\mathrm{dc}} = \hat{H}_{\mathrm{dc},1} + \hat{H}_{\mathrm{dc},2} = -\sum_{i=1}^{2}\hat{\boldsymbol{d}}_{i}\cdot\boldsymbol{\varepsilon}.
\end{equation}
As $\hat{H}_{\mathrm{dc},1}$ acts only on molecule 1, the wavefunction of molecule 2 remains unchanged, so:
\begin{equation}
\langle j_{1}k_{1}m_{1},\,j_{2}k_{2}m_{2}|\hat{H}_{\mathrm{dc},1}| j'_{1}k'_{1}m'_{1},\,j'_{2}k'_{2}m'_{2}\rangle = \langle j_{1}k_{1}m_{1}|\hat{H}_{\mathrm{dc},1}| j'_{1}k'_{1}m'_{1}\rangle\,\delta_{j_{2},j'_{2}}\delta_{k_{2},k'_{2}}\delta_{m_{2},m'_{2}}.
\label{EqB14}
\end{equation}
Similarly for $\hat{H}_{\mathrm{dc},2}$:
\begin{equation}
\langle j_{1}k_{1}m_{1},\,j_{2}k_{2}m_{2}|\hat{H}_{\mathrm{dc},2}| j'_{1}k'_{1}m'_{1},\,j'_{2}k'_{2}m'_{2}\rangle = \langle j_{2}k_{2}m_{2}|\hat{H}_{\mathrm{dc},2}| j'_{2}k'_{2}m'_{2}\rangle\,\delta_{j_{1},j'_{1}}\delta_{k_{1},k'_{1}}\delta_{m_{1},m'_{1}}.
\label{EqB15}
\end{equation}
Combining Eq. (\ref{EqB14}) and Eq. (\ref{EqB15}):
\begin{equation}
\begin{split}
\langle j_{1}k_{1}m_{1},\,j_{2}k_{2}m_{2}|\hat{H}_{\mathrm{dc}}| j'_{1}k'_{1}m'_{1},\,j'_{2}k'_{2}m'_{2}\rangle 
&= \langle j_{1}k_{1}m_{1}|\hat{H}_{\mathrm{dc},1}| j'_{1}k'_{1}m'_{1}\rangle\,\delta_{j_{2},j'_{2}}\delta_{k_{2},k'_{2}}\delta_{m_{2},m'_{2}} \\
&\quad + \langle j_{2}k_{2}m_{2}|\hat{H}_{\mathrm{dc},2}| j'_{2}k'_{2}m'_{2}\rangle\,\delta_{j_{1},j'_{1}}\delta_{k_{1},k'_{1}}\delta_{m_{1},m'_{1}}.
\end{split}
\end{equation}

We can apply the same logic to the rotational part, $\hat{H}_{\mathrm{rot}}$:
\begin{equation}
\hat{H}_{\mathrm{rot}} = \hat{H}_{\mathrm{rot},1} + \hat{H}_{\mathrm{rot},2},
\end{equation}
where $\hat{H}_{\mathrm{rot},i} = A_{i}\hat{J}_{i,a_{i}}^{2} + B_{i}\hat{J}_{i,b_{i}}^{2} + C_{i}\hat{J}_{i,c_{i}}^{2}$. The non-zero matrix elements will be:
\begin{equation}
\begin{split}
\langle j_{1}k_{1}m_{1},\,j_{2}k_{2}m_{2}|\hat{H}_{\mathrm{rot}}| j'_{1}k'_{1}m'_{1},\,j'_{2}k'_{2}m'_{2}\rangle 
&= \langle j_{1}k_{1}m_{1}|\hat{H}_{\mathrm{rot},1}| j'_{1}k'_{1}m'_{1}\rangle\,\delta_{j_{2},j'_{2}}\delta_{k_{2},k'_{2}}\delta_{m_{2},m'_{2}} \\
&\quad + \langle j_{2}k_{2}m_{2}|\hat{H}_{\mathrm{rot},2}| j'_{2}k'_{2}m'_{2}\rangle\,\delta_{j_{1},j'_{1}}\delta_{k_{1},k'_{1}}\delta_{m_{1},m'_{1}}.
\end{split}
\end{equation}

% --- Appendix C ---
\setcounter{equation}{0}
\setcounter{figure}{0}
\makeatletter
\renewcommand{\theequation}{C\arabic{equation}}
\makeatother
\refstepcounter{section}
\section*{APPENDIX \Alph{section}: Derivation of the Dipole-Dipole Interaction Term}
\phantomsection
\label{app:C}

The dipole–dipole interaction is described by the Hamiltonian:
\begin{equation}
\hat{H}_{\mathrm{dd}} = -\frac{(\hat{\boldsymbol{d}}_{1}\cdot\hat{\boldsymbol{d}}_{2})-3(\hat{\boldsymbol{d}}_{1}\cdot\boldsymbol{e}_{r})(\hat{\boldsymbol{d}}_{2}\cdot\boldsymbol{e}_{r})}{r^{3}},
\end{equation}
where $\boldsymbol{e}_{r}=\boldsymbol{e}_{Z}$ in our chosen frame. For simplicity in evaluating the matrix element of $\hat{H}_{\mathrm{dd}}$, we set $1/(4\pi\varepsilon_{0})=1$, where $\varepsilon_{0}$ is the vacuum permittivity. We rewrite the dipole–dipole interaction in terms of spherical tensors, which allows $\hat{H}_{\mathrm{dd}}$ to be written as:
\begin{equation}
\hat{H}_{\mathrm{dd}} = -\frac{\sqrt{6}}{r^{3}}\sum_{p=-2}^{2}(-1)^{p}C_{-p}^{(2)}(\Omega_{r})\left[\hat{\boldsymbol{d}}_{1}\otimes\hat{\boldsymbol{d}}_{2}\right]_{p}^{(2)}.
\end{equation}

For molecules separated by a distance $r$ with the electric field aligned along the laboratory $Z$-axis, the laboratory-frame orientation is $\Omega_{r}=(\theta_{r},\phi_{r})=(0,0)$, and only the $p=0$ term contribution remains. Using the property of spherical harmonics $C_{m}^{(l)}(0,0)=\delta_{m,0}$, we find that $C_{-p}^{(2)}(0,0)=1$ only for $p=0$, and zero otherwise. Thus, the sum collapses to a single term:
\begin{equation}
\begin{split}
\hat{H}_{\mathrm{dd}} &= -\frac{\sqrt{6}}{r^{3}}(-1)^{0}C_{0}^{(2)}(0,0)\left[\hat{\boldsymbol{d}}_{1}\otimes\hat{\boldsymbol{d}}_{2}\right]_{0}^{(2)} \\
&= -\frac{\sqrt{6}}{r^{3}}\sum_{p'=-1}^{1}\langle1\,p',\,1\,p-p'|2\,p\rangle\;d_{1,p'}\,d_{2,p-p'}.
\end{split}
\label{EqC3}
\end{equation}

According to Eq.(\ref{EqC3}), three distinct contributions arise:

(i) Term with $p'=0$:
\begin{equation}
\langle1\,0,\,1\,0|2\,0\rangle=\sqrt{\frac{2}{3}} \implies \sqrt{\frac{2}{3}}\,d_{1,0}\,d_{2,0}.
\end{equation}

(ii) Term with $p'=+1$:
\begin{equation}
\langle1\,1,\,1\,-1|2\,0\rangle=\sqrt{\frac{1}{6}} \implies \sqrt{\frac{1}{6}}\,d_{1,1}\,d_{2,-1}.
\end{equation}

(iii) Term with $p'=-1$:
\begin{equation}
\langle1\,-1,\,1\,1|2\,0\rangle=\sqrt{\frac{1}{6}} \implies \sqrt{\frac{1}{6}}\,d_{1,-1}\,d_{2,1}.
\end{equation}

Inserting these values into Eq. (\ref{EqC3}) gives:
\begin{equation}
\hat{H}_{\mathrm{dd}} = -\frac{\sqrt{6}}{r^{3}}\left(\sqrt{\frac{2}{3}}\,d_{1,0}d_{2,0} + \sqrt{\frac{1}{6}}\,d_{1,1}d_{2,-1} + \sqrt{\frac{1}{6}}\,d_{1,-1}d_{2,1}\right).
\end{equation}
Hence, the expression reduces to:
\begin{equation}
\hat{H}_{\mathrm{dd}} = -\frac{1}{r^{3}}\left(2\,d_{1,0}d_{2,0} + d_{1,-1}d_{2,1} + d_{1,1}d_{2,-1}\right).
\label{EqC8}
\end{equation}

In Eq. (\ref{EqC8}), the dipole components are expressed in the laboratory frame and must be transformed to the molecule-fixed frame. Accordingly,
\begin{equation}
d_{i,q} = \sum_{r_{i}=-1}^{1}D_{q,r_{i}}^{1*}(\Omega_{i})\,d_{i,r_{i}}^{(\mathrm{mol})},
\end{equation}
where $q=-1,0,1$ and $r=-1,0,1$ label the laboratory-frame and molecule-fixed spherical components, respectively; $d_{i,r_{i}}^{(\mathrm{mol})}$ are the known dipole components of molecule $i$, and $D_{q,r_{i}}^{1*}(\Omega_{i})$ is the Wigner D-matrix associated with the orientation $\Omega_{i}$.

The first term in the equation is given by $2\,d_{1,0}d_{2,0}$:
\begin{equation}
\begin{split}
d_{1,0} &= \sum_{r_{1}=-1}^{1}D_{0,r_{1}}^{1*}(\Omega_{1})\,d_{1,r_{1}}^{(\mathrm{mol})}, \\
d_{2,0} &= \sum_{r_{2}=-1}^{1}D_{0,r_{2}}^{1*}(\Omega_{2})\,d_{2,r_{2}}^{(\mathrm{mol})}, \\
\implies 2\,d_{1,0}d_{2,0} &= 2\sum_{r_{1}=-1}^{1}\sum_{r_{2}=-1}^{1}D_{0,r_{1}}^{1*}(\Omega_{1})\,d_{1,r_{1}}^{(\mathrm{mol})}D_{0,r_{2}}^{1*}(\Omega_{2})\,d_{2,r_{2}}^{(\mathrm{mol})}.
\end{split}
\end{equation}

The second term $d_{1,-1}d_{2,1}$:
\begin{equation}
d_{1,-1}d_{2,1} = \sum_{r_{1}=-1}^{1}\sum_{r_{2}=-1}^{1}D_{-1,r_{1}}^{1*}(\Omega_{1})\,d_{1,r_{1}}^{(\mathrm{mol})}D_{1,r_{2}}^{1*}(\Omega_{2})\,d_{2,r_{2}}^{(\mathrm{mol})}.
\end{equation}

Similarly, the third term $d_{1,1}d_{2,-1}$:
\begin{equation}
d_{1,1}d_{2,-1} = \sum_{r_{1}=-1}^{1}\sum_{r_{2}=-1}^{1}D_{1,r_{1}}^{1*}(\Omega_{1})\,d_{1,r_{1}}^{(\mathrm{mol})}D_{-1,r_{2}}^{1*}(\Omega_{2})\,d_{2,r_{2}}^{(\mathrm{mol})}.
\end{equation}

Collecting all terms, one obtains:
\begin{equation}
\begin{split}
\hat{H}_{\mathrm{dd}} = -\frac{1}{r^{3}}\sum_{r_{1}=-1}^{1}\sum_{r_{2}=-1}^{1} \Big[ & 2D_{0,r_{1}}^{1*}(\Omega_{1})\,d_{1,r_{1}}^{(\mathrm{mol})}D_{0,r_{2}}^{1*}(\Omega_{2})\,d_{2,r_{2}}^{(\mathrm{mol})} \\
&+ D_{-1,r_{1}}^{1*}(\Omega_{1})\,d_{1,r_{1}}^{(\mathrm{mol})}D_{1,r_{2}}^{1*}(\Omega_{2})\,d_{2,r_{2}}^{(\mathrm{mol})} \\
&+ D_{1,r_{1}}^{1*}(\Omega_{1})\,d_{1,r_{1}}^{(\mathrm{mol})}D_{-1,r_{2}}^{1*}(\Omega_{2})\,d_{2,r_{2}}^{(\mathrm{mol})} \Big].
\end{split}
\label{EqC13}
\end{equation}
By substituting Eqs.(\ref{EqA14}) and Eq. (\ref{EqA15}) into Eq. (\ref{EqC13}), we obtain a general expression for the matrix elements of the dipole--dipole interaction,

\begin{equation}
\begin{split}
& \langle j_{1}k_{1}m_{1},\,j_{2}k_{2}m_{2}|\hat{H}_{\mathrm{dd}}| j'_{1}k'_{1}m'_{1},\,j'_{2}k'_{2}m'_{2}\rangle \\
&\quad = -\frac{1}{r^{3}} \Bigg[ 2\,\langle j_{1}k_{1}m_{1}|D_{0,\,k_{1}-k'_{1}}^{1*}|j'_{1}k'_{1}m'_{1}\rangle\,\langle j_{2}k_{2}m_{2}|D_{0,\,k_{2}-k'_{2}}^{1*}|j'_{2}k'_{2}m'_{2}\rangle\,\delta_{0,\,m_{1}-m'_{1}}\,\delta_{0,\,m_{2}-m'_{2}} \Bigg] \\
&\quad\quad -\frac{1}{r^{3}} \Bigg[ \langle j_{1}k_{1}m_{1}|D_{-1,\,k_{1}-k'_{1}}^{1*}|j'_{1}k'_{1}m'_{1}\rangle\,\langle j_{2}k_{2}m_{2}|D_{1,\,k_{2}-k'_{2}}^{1*}|j'_{2}k'_{2}m'_{2}\rangle\,\delta_{-1,\,m_{1}-m'_{1}}\,\delta_{1,\,m_{2}-m'_{2}} \Bigg] \\
&\quad\quad -\frac{1}{r^{3}} \Bigg[ \langle j_{1}k_{1}m_{1}|D_{1,\,k_{1}-k'_{1}}^{1*}|j'_{1}k'_{1}m'_{1}\rangle\,\langle j_{2}k_{2}m_{2}|D_{-1,\,k_{2}-k'_{2}}^{1*}|j'_{2}k'_{2}m'_{2}\rangle\,\delta_{1,\,m_{1}-m'_{1}}\,\delta_{-1,\,m_{2}-m'_{2}} \Bigg] \\
&\quad\qquad \times d_{1,\,k_{1}-k'_{1}}\,d_{2,\,k_{2}-k'_{2}}.
\end{split}
\label{EqC14}
\end{equation}

% --- Appendix D ---
\setcounter{equation}{0}
\setcounter{figure}{0}
\makeatletter
\renewcommand{\theequation}{D\arabic{equation}}
\makeatother
\refstepcounter{section}
\section*{APPENDIX \Alph{section}}

\phantomsection
\label{app:D}
\subsection{Matrix elements of $\hat{H}_{\mathrm{dd}}$}
We now construct the matrix elements of the dipole--dipole interaction
in the dressed-state basis
\begin{align}
|\uparrow\rangle & =\sum_{J,K}c_{J,K}^{(\uparrow)}(x)|J,K,M=0\rangle\\
|\downarrow\rangle & =\sum_{J,K}c_{J,K}^{(\downarrow)}(x)|J,K,M=1\rangle
\end{align}

Basis:$\{\lvert\downarrow\downarrow\rangle,|\downarrow\uparrow\rangle,\lvert\uparrow\downarrow\rangle,\lvert\uparrow\uparrow\rangle\}$, using the selection rules $\Delta m_{1}=m_{1}-m_{1}',\qquad\Delta m_{2}=m_{2}-m_{2}'.$ There
are three rule which gives non-zero element according to Eq. (\ref{EqC14}):

\begin{enumerate}[label=(\roman*)]
\item \(\Delta m_{1}=0,\ \Delta m_{2}=0\), implying
      \(m_{1}=m_{1}'\) and \(m_{2}=m_{2}'\).
\item \(m_{1}-m_{1}'=-1,\quad m_{2}-m_{2}'=+1\).
\item \(m_{1}-m_{1}'=+1,\quad m_{2}-m_{2}'=-1\).
\end{enumerate}

\noindent (i)\quad{}$\langle\downarrow\downarrow|H_{\mathrm{dd}}|\downarrow\downarrow\rangle$

\[
m_{1}=1,\;m_{1}'=1,\qquad m_{2}=1,\;m_{2}'=1
\]
\[
\Delta m_{1}=0,\qquad\Delta m_{2}=0
\]

From Eq. (\ref{EqC14}), we define
\[
C_{1}=-\sum_{\substack{J_{1},K_{1}\\
J'_{1},K'_{1}
}
}\sum_{\substack{J_{2},K_{2}\\
J'_{2},K'_{2}
}
}c_{J_{1},K_{1}}^{(\downarrow)*}c_{J'_{1},K'_{1}}^{(\downarrow)}\,c_{J_{2},K_{2}}^{(\downarrow)*}c_{J'_{2},K'_{2}}^{(\downarrow)}\left[2\,\langle j_{1}k_{1}1|D_{0,k_{1}-k_{1}'}^{1}|j_{1}'k_{1}'1\rangle\langle j_{2}k_{2}1|D_{0,k_{2}-k_{2}'}^{1}|j_{2}'k_{2}'1\rangle\right]\times d_{1,\,k_{1}-k'_{1}}\,d_{2,\,k_{2}-k'_{2}}
\]

\noindent (ii)\quad{}$\langle\downarrow\downarrow|H_{\mathrm{dd}}|\downarrow\uparrow\rangle$

\[
m_{1}=1,\;m_{1}'=1,\qquad m_{2}=1,\;m_{2}'=0
\]
\[
\Delta m_{1}=0,\qquad\Delta m_{2}=1-0=1
\]

Zero.

\noindent (iii)\quad{}$\langle\downarrow\downarrow|H_{\mathrm{dd}}|\uparrow\downarrow\rangle$

\[
m_{1}=1,\;m_{1}'=0,\qquad m_{2}=1,\;m_{2}'=1
\]
\[
\Delta m_{1}=1,\qquad\Delta m_{2}=0
\]

Zero.

\noindent (iv)\quad{}$\langle\downarrow\downarrow|H_{\mathrm{dd}}|\uparrow\uparrow\rangle$

\[
m_{1}=1,\;m_{1}'=0,\qquad m_{2}=1,\;m_{2}'=0
\]
\[
\Delta m_{1}=1,\qquad\Delta m_{2}=1
\]

Zero.

\noindent (v)\quad{}$\langle\downarrow\uparrow|H_{\mathrm{dd}}|\downarrow\downarrow\rangle$

\[
m_{1}=1,\;m_{1}'=1,\qquad m_{2}=0,\;m_{2}'=1
\]
\[
\Delta m_{1}=0,\qquad\Delta m_{2}=-1
\]

Zero.

\noindent (vi)\quad{}$\langle\downarrow\uparrow|H_{\mathrm{dd}}|\downarrow\uparrow\rangle$

\[
m_{1}=1,\;m_{1}'=1,\qquad m_{2}=0,\;m_{2}'=0
\]
\[
\Delta m_{1}=0,\qquad\Delta m_{2}=0
\]

Non-zero, defining from Eq. (\ref{EqC14})
\[
C_{2}=-\sum_{\substack{J_{1},K_{1}\\
J'_{1},K'_{1}
}
}\sum_{\substack{J_{2},K_{2}\\
J'_{2},K'_{2}
}
}c_{J_{1},K_{1}}^{(\downarrow)*}c_{J'_{1},K'_{1}}^{(\downarrow)}\,c_{J_{2},K_{2}}^{(\uparrow)*}c_{J'_{2},K'_{2}}^{(\uparrow)}\left[2\,\langle j_{1}k_{1}1|D_{0,k_{1}-k_{1}'}^{1}|j_{1}'k_{1}'1\rangle\langle j_{2}k_{2}0|D_{0,k_{2}-k_{2}'}^{1}|j_{2}'k_{2}'0\rangle\right]\times d_{1,\,k_{1}-k'_{1}}\,d_{2,\,k_{2}-k'_{2}}
\]

\noindent (vii)\quad{}$\langle\downarrow\uparrow|H_{\mathrm{dd}}|\uparrow\downarrow\rangle$

\[
m_{1}=1,\;m_{1}'=0,\qquad m_{2}=0,\;m_{2}'=1
\]
\[
\Delta m_{1}=1,\qquad\Delta m_{2}=-1
\]

From Eq. (\ref{EqC14}), non-zero (exchange-type term) given as
\begin{align*}
C_{d_{1}} & =-\sum_{\substack{J_{1},K_{1}\\
J'_{1},K'_{1}
}
}\sum_{\substack{J_{2},K_{2}\\
J'_{2},K'_{2}
}
}c_{J_{1},K_{1}}^{(\downarrow)*}c_{J'_{1},K'_{1}}^{(\uparrow)}\,c_{J_{2},K_{2}}^{(\uparrow)*}c_{J'_{2},K'_{2}}^{(\downarrow)}\\
 & \qquad\times\langle j_{1}k_{1}1|D_{+1,k_{1}-k_{1}'}^{1}|j_{1}'k_{1}'0\rangle\langle j_{2}k_{2}0|D_{-1,k_{2}-k_{2}'}^{1}|j_{2}'k_{2}'1\rangle\times d_{1,\,k_{1}-k'_{1}}\,d_{2,\,k_{2}-k'_{2}}
\end{align*}

\noindent (viii)\quad{}$\langle\uparrow\downarrow|H_{\mathrm{dd}}|\uparrow\uparrow\rangle$

\[
m_{1}=0,\;m_{1}'=1,\qquad m_{2}=1,\;m_{2}'=1
\]
\[
\Delta m_{1}=-1,\qquad\Delta m_{2}=0
\]

Zero.

\noindent (ix)\textbf{\quad{}$\langle\uparrow\downarrow|H_{\mathrm{dd}}|\downarrow\downarrow\rangle$}

\[
m_{1}=0,\;m_{1}'=1,\qquad m_{2}=1,\;m_{2}'=1
\]
\[
\Delta m_{1}=-1,\qquad\Delta m_{2}=0
\]

zero.

\noindent (x)\quad{}$\langle\uparrow\downarrow|H_{\mathrm{dd}}|\uparrow\downarrow\rangle$

\[
m_{1}=0,\;m_{1}'=0,\qquad m_{2}=1,\;m_{2}'=1
\]
\[
\Delta m_{1}=0,\qquad\Delta m_{2}=0
\]

Non-zero, giving the Hermitian-conjugate exchange term $C_{d_{2}}$ from Eq. (\ref{EqC14})
\begin{align*}
C_{d_{2}} & =-\sum_{\substack{J_{1},K_{1}\\
J'_{1},K'_{1}
}
}\sum_{\substack{J_{2},K_{2}\\
J'_{2},K'_{2}
}
}c_{J_{1},K_{1}}^{(\uparrow)*}c_{J'_{1},K'_{1}}^{(\downarrow)}\,c_{J_{2},K_{2}}^{(\downarrow)*}c_{J'_{2},K'_{2}}^{(\uparrow)}\\
 & \qquad\times\langle j_{1}k_{1}0|D_{-1,k_{1}-k_{1}'}^{1}|j_{1}'k_{1}'1\rangle\langle j_{2}k_{2}1|D_{+1,k_{2}-k_{2}'}^{1}|j_{2}'k_{2}'0\rangle\times d_{1,\,k_{1}-k'_{1}}\,d_{2,\,k_{2}-k'_{2}}
\end{align*}

\noindent (xi)\quad{}$\langle\uparrow\downarrow|H_{\mathrm{dd}}|\uparrow\downarrow\rangle$

\[
m_{1}=0,\;m_{1}'=0,\qquad m_{2}=1,\;m_{2}'=1
\]
\[
\Delta m_{1}=0,\qquad\Delta m_{2}=0
\]

We define 
\[
C_{3}=-\sum_{\substack{J_{1},K_{1}\\
J'_{1},K'_{1}
}
}\sum_{\substack{J_{2},K_{2}\\
J'_{2},K'_{2}
}
}c_{J_{1},K_{1}}^{(\uparrow)*}c_{J'_{1},K'_{1}}^{(\uparrow)}\,c_{J_{2},K_{2}}^{(\downarrow)*}c_{J'_{2},K'_{2}}^{(\downarrow)}\left[2\,\langle j_{1}k_{1}0|D_{0,k_{1}-k_{1}'}^{1}|j_{1}'k_{1}'0\rangle\langle j_{2}k_{2}1|D_{0,k_{2}-k_{2}'}^{1}|j_{2}'k_{2}'1\rangle\right]\times d_{1,\,k_{1}-k'_{1}}\,d_{2,\,k_{2}-k'_{2}}
\]

\noindent (xii)\quad{}$\langle\uparrow\uparrow|H_{\mathrm{dd}}|\uparrow\downarrow\rangle$

\[
m_{1}=0,\;m_{1}'=0,\qquad m_{2}=1,\;m_{2}'=0
\]
\[
\Delta m_{1}=0,\qquad\Delta m_{2}=1
\]

Zero.

\noindent (xiii)\quad{}$\langle\uparrow\uparrow|H_{\mathrm{dd}}|\downarrow\downarrow\rangle$

\[
m_{1}=0,\;m_{1}'=1,\qquad m_{2}=1,\;m_{2}'=1
\]
\[
\Delta m_{1}=-1,\qquad\Delta m_{2}=0
\]

Zero.

\noindent (xiv)\quad{}$\langle\uparrow\uparrow|H_{\mathrm{dd}}|\downarrow\uparrow\rangle$

\[
m_{1}=0,\;m_{1}'=1,\qquad m_{2}=0,\;m_{2}'=0
\]
\[
\Delta m_{1}=-1,\qquad\Delta m_{2}=0
\]

Zero.

\noindent (xv)\quad{}$\langle\uparrow\uparrow|H_{\mathrm{dd}}|\uparrow\downarrow\rangle$

\[
m_{1}=0,\;m_{1}'=0,\qquad m_{2}=0,\;m_{2}'=1
\]
\[
\Delta m_{1}=0,\qquad\Delta m_{2}=-1
\]

Zero.

\noindent (xvi)\quad{}$\langle\uparrow\uparrow|H_{\mathrm{dd}}|\uparrow\uparrow\rangle$

\[
m_{1}=0,\;m_{1}'=0,\qquad m_{2}=0,\;m_{2}'=0
\]
\[
\Delta m_{1}=0,\qquad\Delta m_{2}=0
\]

From Eq. (\ref{EqC14}) we define

$C_{4}=-\sum_{\substack{J_{1},K_{1}\\
J'_{1},K'_{1}
}
}\sum_{\substack{J_{2},K_{2}\\
J'_{2},K'_{2}
}
}c_{J_{1},K_{1}}^{(\uparrow)*}c_{J'_{1},K'_{1}}^{(\uparrow)}\,c_{J_{2},K_{2}}^{(\uparrow)*}c_{J'_{2},K'_{2}}^{(\uparrow)}\left[2\,\langle j_{1}k_{1}0|D_{0,\,k_{1}-k'_{1}}^{1}|j'_{1}k'_{1}0\rangle\,\langle j_{2}k_{2}0|D_{0,\,k_{2}-k'_{2}}^{1}|j'_{2}k'_{2}0\rangle\right]\times d_{1,\,k_{1}-k'_{1}}\,d_{2,\,k_{2}-k'_{2}}$
\vspace{0.3cm}
Collecting all nonzero contributions, the dipole--dipole Hamiltonian
in the dressed basis $\{|\downarrow\downarrow\rangle,|\downarrow\uparrow\rangle,|\uparrow\downarrow\rangle,|\uparrow\uparrow\rangle\}$
takes form
\[
H_{\mathrm{dd}}=-\frac{1}{r^{3}}\begin{pmatrix}C_{1} & 0 & 0 & 0\\
0 & C_{2} & C_{d_{1}} & 0\\
0 & C_{d_{2}} & C_{3} & 0\\
0 & 0 & 0 & C_{4}
\end{pmatrix}
\]

Here, the off-diagonal terms $C_{d_{1}}$ and $C_{d_{2}}$ are Hermitian
conjugates: 
\[
C_{d_{2}}=C_{d_{1}}^{*}
\]

Because the molecules under consideration are chiral, the dipole components
$d_{\pm1}$ carry complex phase $\mp\frac{d_{b}\pm id_{c}}{\sqrt{2}}$,
and the resulting matrix elements are therefore complex. We decompose
these matrix elements into their real and imaginary parts as

\begin{align*}
C_{d_{1}} & =J_{xy}+i\mathcal{D}\\
C_{d_{2}} & =J_{xy}-i\mathcal{D}
\end{align*}

To map this systems to a spin-$1/2$ model, we define Pauli operators
acting on the dressed states:
\begin{itemize}
\item $\lvert\uparrow\rangle$ (state $M=0$) corresponds to spin ``up'' ($+z$), 
\item $\lvert\downarrow\rangle$ (state $M=1$) corresponds to spin ``down''
($-z$). 
\end{itemize}
The ladder operators are defined as 
\begin{align*}
\hat{\sigma}^{+} & =\lvert\uparrow\rangle\langle\downarrow \lvert &  & \text{(flips \ensuremath{\downarrow\to\uparrow})}\\
\hat{\sigma}^{-} & =\lvert\downarrow\rangle\langle\uparrow \lvert &  & \text{(flips \ensuremath{\uparrow\to\downarrow})}
\end{align*}
with the Pauli matrices: 
\begin{align}
\hat{\sigma}^{x} & =\hat{\sigma}^{+}+\hat{\sigma}^{-}\\
\hat{\sigma}^{y} & =-i(\hat{\sigma}^{+}-\hat{\sigma}^{-})\quad\Rightarrow\quad\hat{\sigma}^{\pm}=\frac{1}{2}\left(\hat{\sigma}^{x}\pm i\hat{\sigma}^{y}\right)\\
\hat{\sigma}^{z} & =\lvert\uparrow\rangle\langle\downarrow\lvert-\lvert\downarrow\rangle\langle\uparrow \lvert
\end{align}
The off-diagonal matrix elements connect $\lvert\downarrow\uparrow\rangle$
and $\lvert\uparrow\downarrow\rangle$. We analyze the transition corresponding
to $C_{d_{1}}$.
\begin{itemize}
\item Initial state (ket): $\lvert\downarrow\uparrow\rangle$ (Mol~1 is $\uparrow$,
Mol~2 is $\downarrow$). 
\item Final state (bra): $\langle\downarrow\uparrow\lvert$ (Mol~1 is $\downarrow$,
Mol~2 is $\uparrow$). 
\item Action: Mol~1 flips $\uparrow\to\downarrow$ ($\hat{\sigma}_{1}^{-}$)
and Mol~2 flips $\downarrow\to\uparrow$ ($\hat{\sigma}_{2}^{+}$). 
\end{itemize}
This term is represented by the operator
\begin{equation}
H_{\mathrm{off}}
=
- C_{d_{1}}\,\hat{\sigma}_{1}^{-}\hat{\sigma}_{2}^{+}
- C_{d_{2}}\,\hat{\sigma}_{1}^{+}\hat{\sigma}_{2}^{-},
\label{EqD6}
\end{equation}
Substitute the Pauli-matrix forms
\begin{align*}
\hat{\sigma}_{1}^{-}\hat{\sigma}_{2}^{+} & =\left[\frac{1}{2}(\hat{\sigma}_{1}^{x}-i\hat{\sigma}_{1}^{y})\right]\left[\frac{1}{2}(\hat{\sigma}_{2}^{x}+i\hat{\sigma}_{2}^{y})\right]\\
 & =\frac{1}{4}\bigl[\hat{\sigma}_{1}^{x}\hat{\sigma}_{2}^{x}+\hat{\sigma}_{1}^{y}\hat{\sigma}_{2}^{y}+i(\hat{\sigma}_{1}^{x}\hat{\sigma}_{2}^{y}-\hat{\sigma}_{1}^{y}\hat{\sigma}_{2}^{x})\bigr]
\end{align*}
and similarly for the conjugate term 
\begin{align*}
\hat{\sigma}_{1}^{+}\hat{\sigma}_{2}^{-} & =\left[\frac{1}{2}(\hat{\sigma}_{1}^{x}+i\hat{\sigma}_{1}^{y})\right]\left[\frac{1}{2}(\hat{\sigma}_{2}^{x}-i\hat{\sigma}_{2}^{y})\right]\\
 & =\frac{1}{4}\bigl[\hat{\sigma}_{1}^{x}\hat{\sigma}_{2}^{x}+\hat{\sigma}_{1}^{y}\hat{\sigma}_{2}^{y}-i(\hat{\sigma}_{1}^{x}\hat{\sigma}_{2}^{y}-\hat{\sigma}_{1}^{y}\hat{\sigma}_{2}^{x})\bigr]
\end{align*}
Substitute $C_{d_{1}}=J_{xy}+i\mathcal{D}$ and $C_{d_{2}}=J_{xy}-i\mathcal{D}$
into Eq \ref{EqD6}): 
\begin{align*}
H_{\mathrm{off}} & =-(J_{xy}+i\mathcal{D})\frac{1}{4}\bigl[A+iB\bigr]-(J_{xy}-i\mathcal{D})\frac{1}{4}\bigl[A-iB\bigr]\\
 & =-\frac{J_{xy}+i\mathcal{D}}{4}[A+iB]-\frac{J_{xy}-i\mathcal{D}}{4}[A-iB]
\end{align*}
where, for convenience, 
\begin{align*}
A & =\hat{\sigma}_{1}^{x}\hat{\sigma}_{2}^{x}+\hat{\sigma}_{1}^{y}\hat{\sigma}_{2}^{y}\\
B & =\hat{\sigma}_{1}^{x}\hat{\sigma}_{2}^{y}-\hat{\sigma}_{1}^{y}\hat{\sigma}_{2}^{x}
\end{align*}
Collecting the real contributions ($J_{xy}$), we obtain
\begin{align*}
-\frac{J_{xy}}{4}\bigl[(A+iB)+(A-iB)\bigr] & =-\frac{J_{xy}}{4}(2A)=-\frac{J_{xy}}{2}A\\
 & =-\frac{J_{xy}}{2}\bigl(\hat{\sigma}_{1}^{x}\hat{\sigma}_{2}^{x}+\hat{\sigma}_{1}^{y}\hat{\sigma}_{2}^{y}\bigr)
\end{align*}
which is the XY exchange interaction.\\

\noindent The Imaginary Terms contribution ($\mathcal{D}$) is given by
\begin{align*}
-\frac{i\mathcal{D}}{4}\bigl[(A+iB)-(A-iB)\bigr] & =-\frac{i\mathcal{D}}{4}(2iB)=+\frac{\mathcal{D}}{2}B\\
 & =+\frac{\mathcal{D}}{2}\bigl(\hat{\sigma}_{1}^{x}\hat{\sigma}_{2}^{y}-\hat{\sigma}_{1}^{y}\hat{\sigma}_{2}^{x}\bigr).
\end{align*}
which is the Dzyaloshinskii--Moriya interaction (DMI).

\subsection{Mapping of  Hamiltonian}
\label{D2}

The total Hamiltonian for the two-molecule system is
\begin{equation}
\hat{H}_{\text{total}}=(\hat{H}_{\text{rot}}^{(1)}+\hat{H}_{\text{dc}}^{(1)})+(\hat{H}_{\text{rot}}^{(2)}+\hat{H}_{\text{dc}}^{(2)})+\hat{H}_{\text{dd}}
\end{equation}

Since the single-molecule Hamiltonian $(\hat{H}_{rot}+\hat{H}_{dc})$
has already been solved, its eigenvalues are: (i) Spin Up: $\lvert\uparrow\rangle$
(State $M=0$) has energy $E_{\uparrow}$(ii) Spin Down: $\lvert\downarrow\rangle$
(State $M=1$) has energy $E_{\downarrow}$. 

The full diagonal energies of the product states are obtained by adding
the single-molecule energies to the diagonal dipole terms ($C_{1},C_{2},C_{3},C_{4}$):

\begin{enumerate}
\item State $\lvert\downarrow\downarrow\rangle$:
\begin{equation}
E_{\text{tot}}(\downarrow\downarrow)
= E_{\downarrow}+E_{\downarrow}-C_{1}
= 2E_{\downarrow}-C_{1}
\label{eq:Etot_dd}
\end{equation}

\item State $\lvert\downarrow\uparrow\rangle$:
\begin{equation}
E_{\text{tot}}(\downarrow\uparrow)
= E_{\downarrow}+E_{\uparrow}-C_{2}
\label{eq:Etot_du}
\end{equation}

\item State $\lvert\uparrow\downarrow\rangle$:
\begin{equation}
E_{\text{tot}}(\uparrow\downarrow)
= E_{\uparrow}+E_{\downarrow}-C_{3}
\label{eq:Etot_ud}
\end{equation}

\item State $\lvert\uparrow\uparrow\rangle$:
\begin{equation}
E_{\text{tot}}(\uparrow\uparrow)
= E_{\uparrow}+E_{\uparrow}-C_{4}
= 2E_{\uparrow}-C_{4}
\label{eq:Etot_uu}
\end{equation}
\end{enumerate}

\begin{equation}
\hat{H}_{\text{diag}}
= J_{z}\hat{\sigma}_{1}^{z}\hat{\sigma}_{2}^{z}
+ \mathfrak{h}_{1}\hat{\sigma}_{1}^{z}
+ \mathfrak{h}_{2}\hat{\sigma}_{2}^{z}
+ E_{0}
\label{eq:Hdiag}
\end{equation}

\begin{itemize}
\item \textbf{Eq 1} ($\lvert\downarrow\downarrow\rangle$, spins $-1,-1$):
\begin{equation}
J_{z}(-1)(-1)+\mathfrak{h}_{1}(-1)+\mathfrak{h}_{2}(-1)+E_{0}
= 2E_{\downarrow}-C_{1}
\label{EqD13}
\end{equation}
\end{itemize}

\begin{equation}
J_{z}-\mathfrak{h}_{1}-\mathfrak{h}_{2}+E_{0}
= 2E_{\downarrow}-C_{1}
\label{eq:match_dd}
\end{equation}

\begin{itemize}
\item \textbf{Eq 2} ($\lvert\downarrow\uparrow\rangle$, spins $-1,+1$):
\end{itemize}

\begin{equation}
J_{z}(-1)(+1)+\mathfrak{h}_{1}(-1)+\mathfrak{h}_{2}(+1)+E_{0}
= E_{\downarrow}+E_{\uparrow}-C_{2}
\label{EqD15}
\end{equation}

\begin{equation}
- J_{z}-\mathfrak{h}_{1}+\mathfrak{h}_{2}+E_{0}
= E_{\downarrow}+E_{\uparrow}-C_{2}
\label{eq:match_du}
\end{equation}

\begin{itemize}
\item \textbf{Eq 3} ($\lvert\uparrow\downarrow\rangle$, spins $+1,-1$):
\begin{equation}
- J_{z}(+1)(-1)+\mathfrak{h}_{1}(+1)+\mathfrak{h}_{2}(-1)+E_{0}
= E_{\uparrow}+E_{\downarrow}-C_{3}
\label{EqD17}
\end{equation}

\begin{equation}
- J_{z}+\mathfrak{h}_{1}-\mathfrak{h}_{2}+E_{0}
= E_{\uparrow}+E_{\downarrow}-C_{3}
\label{eq:match_ud}
\end{equation}

\item \textbf{Eq 4} ($\lvert\uparrow\uparrow\rangle$, spins $+1,+1$):
\begin{equation}
J_{z}(+1)(+1)+\mathfrak{h}_{1}(+1)+\mathfrak{h}_{2}(+1)+E_{0}
= 2E_{\uparrow}-C_{4}
\label{EqD19}
\end{equation}

\begin{equation}
J_{z}+\mathfrak{h}_{1}+\mathfrak{h}_{2}+E_{0}
= 2E_{\uparrow}-C_{4}
\label{eq:match_uu}
\end{equation}
\end{itemize}

A. Solving for $J_{z}$

We isolate $J_{z}$ by computing 
\[
(\text{Eq. (\ref{EqD13})}+\text{Eq. (\ref{EqD19})})-(\text{Eq. (\ref{EqD15})}+\text{Eq. (\ref{EqD17})})
\]

LHS: 
\begin{align*}
[(J_{z}-2\mathfrak{h}+E_{0})+(J_{z}+2\mathfrak{h}+E_{0})]-[(-J_{z}+E_{0})+(-J_{z}+E_{0})] & =[2J_{z}+2E_{0}]-[-2J_{z}+2E_{0}]\\
 & =4J_{z}
\end{align*}

RHS: 
\begin{align*}
[(2E_{\downarrow}-C_{1})+(2E_{\uparrow}-C_{4})] & -[(E_{\downarrow}+E_{\uparrow}-C_{2})+(E_{\uparrow}+E_{\downarrow}-C_{3})]\\
 & =(2E_{\downarrow}+2E_{\uparrow}-C_{1}-C_{4})-(2E_{\downarrow}+2E_{\uparrow}-C_{2}-C_{3})\\
 & =-C_{1}-C_{4}+C_{2}+C_{3}.
\end{align*}

Thus 
\begin{equation}
J_{z}=\frac{1}{4}(C_{2}+C_{3}-C_{1}-C_{4})
\end{equation}

B. Solving for Fields

Compute Eq. (\ref{EqD19}) - Eq. (\ref{EqD13}): 
\begin{equation}
(J_{z}+\mathfrak{h}_{1}+\mathfrak{h}_{2}+E_{0})-(J_{z}-\mathfrak{h}_{1}-\mathfrak{h}_{2}+E_{0})=2(\mathfrak{h}_{1}+\mathfrak{h}_{2})
\end{equation}

\begin{equation}
2(\mathfrak{h}_{1}+\mathfrak{h}_{2})=2(E_{\uparrow}-E_{\downarrow})+(C_{1}-C_{4})
\end{equation}

\begin{equation}
\mathfrak{h}_{1}+\mathfrak{h}_{2}=(E_{\uparrow}-E_{\downarrow})+\frac{1}{2}(C_{1}-C_{4})
\label{EqD24}
\end{equation}

Compute Eq. (\ref{EqD17}) - Eq. (\ref{EqD15}): 
\begin{equation}
(-J_{z}+\mathfrak{h}_{1}-\mathfrak{h}_{2}+E_{0})-(-J_{z}-\mathfrak{h}_{1}+\mathfrak{h}_{2}+E_{0})=2\mathfrak{h}_{1}-2\mathfrak{h}_{2}
\end{equation}

\begin{equation}
2(\mathfrak{h}_{1}-\mathfrak{h}_{2})=(E_{\uparrow}+E_{\downarrow}-C_{3})-(E_{\downarrow}+E_{\uparrow}-C_{2})
\end{equation}

\begin{equation}
2(\mathfrak{h}_{1}-\mathfrak{h}_{2})=C_{2}-C_{3}
\end{equation}

\begin{equation}
\mathfrak{h}_{1}-\mathfrak{h}_{2}=\frac{1}{2}(C_{2}-C_{3})
\label{EqD28}
\end{equation}

By adding and subtracting the results of Eq. (\ref{EqD24}) and Eq. (\ref{EqD28}), we obtain 
\begin{align}
\mathfrak{h}_{1} & =\frac{1}{2}(\mathfrak{h}_{1}+\mathfrak{h}_{2})+\frac{1}{2}(\mathfrak{h}_{1}-\mathfrak{h}_{2})\\
 & =\frac{1}{2}\left[(E_{\uparrow}-E_{\downarrow})+\frac{1}{2}(C_{1}-C_{4})\right]+\frac{1}{2}\left[\frac{1}{2}(C_{2}-C_{3})\right]\\
 & =\frac{1}{2}\left[(E_{\uparrow}-E_{\downarrow})+\frac{1}{2}(C_{1}-C_{4})\right]+\frac{1}{4}\left[(C_{2}-C_{3})\right]
\end{align}
and similarly 
\begin{align}
\mathfrak{h}_{2} & =\frac{1}{2}(\mathfrak{h}_{1}+\mathfrak{h}_{2})-\frac{1}{2}(\mathfrak{h}_{1}-\mathfrak{h}_{2})\\
 & =\frac{1}{2}\left[(E_{\uparrow}-E_{\downarrow})+\frac{1}{2}(C_{1}-C_{4})\right]-\frac{1}{2}\left[\frac{1}{2}(C_{2}-C_{3})\right]\\
 & =\frac{1}{2}\left[(E_{\uparrow}-E_{\downarrow})+\frac{1}{2}(C_{1}-C_{4})\right]-\frac{1}{4}\left[(C_{2}-C_{3})\right]
\end{align}

As from the Fig. \ref{fig:04} it is confirm that $C_{2}=C_{3}$then $\mathfrak{h}_{1}=\mathfrak{h}_{2}=\mathfrak{h}=\frac{1}{2}(E_{\uparrow}-E_{\downarrow})+\frac{1}{4r^{3}}(C_{1}-C_{4})$. Collecting all terms gives the exact Spin-1/2 Hamiltonian of the system
\begin{equation}
\hat{H}_{\text{spin}}=\underbrace{J_{xy}\left(\hat{\sigma}_{1}^{x}\hat{\sigma}_{2}^{x}+\hat{\sigma}_{1}^{y}\hat{\sigma}_{2}^{y}\right)}_{\text{Symmetric Exchange}}-\underbrace{D\left(\hat{\sigma}_{1}^{x}\hat{\sigma}_{2}^{y}-\hat{\sigma}_{1}^{y}\hat{\sigma}_{2}^{x}\right)}_{\text{Chiral DMI}}+\underbrace{J_{z}\,\hat{\sigma}_{1}^{z}\hat{\sigma}_{2}^{z}}_{\text{Ising Interaction}}+\mathfrak{h}(\hat{\sigma}_{1}^{z}+\hat{\sigma}_{2}^{z}).
\end{equation}

Where 
\[
J_{xy}=-\frac{1}{2r^{3}}\Re(C_{d_{1}}),\qquad D=\frac{1}{2r^{3}}\Im(C_{d_{1}})
\]
\begin{equation}
J_{z}=\frac{1}{4r^{3}}(C_{2}+C_{3}-C_{1}-C_{4})
\end{equation}

\[
\mathfrak{h}=\frac{1}{2}(E_{\uparrow}-E_{\downarrow})+\frac{1}{4r^{3}}(C_{1}-C_{4})
\]

\subsection{Guage Transformation}
\label{D3}

From Eq.~(\ref{EqD6}), the real and imaginary parts of the off-diagonal Hamiltonian
$H_{\text{off}}$ described as

\begin{equation}
H_{\text{off}}
=
J_{xy}\left[2\left(\hat{\sigma}_{i}^{+}\hat{\sigma}_{i+1}^{-}
+\hat{\sigma}_{i}^{-}\hat{\sigma}_{i+1}^{+}\right)\right]
-
D\left[2i\left(\hat{\sigma}_{i}^{+}\hat{\sigma}_{i+1}^{-}
-\hat{\sigma}_{i}^{-}\hat{\sigma}_{i+1}^{+}\right)\right].
\label{EqD37}
\end{equation}

Collecting terms, this expression can be rewritten as

\begin{equation}
H_{\text{off}}
=
2\left(J_{xy}-iD\right)\hat{\sigma}_{i}^{+}\hat{\sigma}_{i+1}^{-}
+
2\left(J_{xy}+iD\right)\hat{\sigma}_{i}^{-}\hat{\sigma}_{i+1}^{+}.
\end{equation}

It is convenient to express the complex coefficients
$\left(J_{xy}\pm iD\right)$ in polar form as $R e^{\pm i\theta}$.
Defining the effective transverse coupling
$\tilde{J}_{xy}=\sqrt{J_{xy}^{2}+D^{2}}$
and the phase
$\theta=\tan^{-1}\!\left(D/J_{xy}\right)$,
we may write

\begin{equation}
J_{xy} \mp iD = \tilde{J}_{xy} e^{\mp i\theta}.
\label{EqD39}
\end{equation}

Substituting Eq. (\ref{EqD39}) into Eq. (\ref{EqD37}), the Hamiltonian yields the compact form

\begin{equation}
H_{\text{off}}
=
2\tilde{J}_{xy}
\left(
e^{-i\theta}\hat{\sigma}_{i}^{+}\hat{\sigma}_{i+1}^{-}
+
e^{+i\theta}\hat{\sigma}_{i}^{-}\hat{\sigma}_{i+1}^{+}
\right).
\label{EqD40}
\end{equation}

We perform a local unitary transformation about the $z$-axis

\begin{equation}
U=\prod_{i=1}^{n}e^{-i\frac{\phi_{i}}{2}\sigma_{i}^{z}} .
\end{equation}

This transforms the ladder operators as

\begin{equation}
U^{\dagger}\hat{\sigma}_{i}^{+}U
=
e^{i\phi_{i}}\hat{\sigma}_{i}^{+},
\end{equation}

\begin{equation}
U^{\dagger}\hat{\sigma}_{i}^{-}U
=
e^{-i\phi_{i}}\hat{\sigma}_{i}^{-}.
\end{equation}

The term $\hat{\sigma}_{i}^{+}\hat{\sigma}_{i+1}^{-}$ becomes

\begin{equation}
\hat{\sigma}_{i}^{+}\hat{\sigma}_{i+1}^{-}
=
\left(e^{i\phi_{i}}\hat{\sigma}_{i}^{+}\right)
\left(e^{-i\phi_{i+1}}\hat{\sigma}_{i+1}^{-}\right)
=
e^{-i(\phi_{i+1}-\phi_{i})}
\hat{\sigma}_{i}^{+}\hat{\sigma}_{i+1}^{-}.
\end{equation}

To eliminate the complex phase appearing in
$\left(J_{xy}-iD\right)=\tilde{J}_{xy}e^{-i\theta}$ and
$\left(J_{xy}+iD\right)=\tilde{J}_{xy}e^{+i\theta}$,
we require the rotation phases to satisfy

\begin{equation}
e^{-i(\phi_{i+1}-\phi_i)} = e^{i\theta}
\quad \Longrightarrow \quad
\phi_{i+1}-\phi_i = -\theta .
\end{equation}

This condition is satisfied by choosing the rotation angle to vary linearly
along the chain,

\begin{equation}
\phi_i = - i\,\theta,
\qquad
\theta = \tan^{-1}\!\left(\frac{D}{J_{xy}}\right).
\end{equation}

Substituting values of $\phi_i = - i\,\theta$ into Eq. (\ref{EqD40}) the transformed off-diagonal Hamiltonian becomes

\begin{equation}
\tilde{H}_{\text{off}}
=
2\tilde{J}_{xy}
\left(
e^{-i\theta} e^{i(\phi_{i+1}-\phi_i)}
\hat{\sigma}_i^{+}\hat{\sigma}_{i+1}^{-}
+
e^{+i\theta} e^{-i(\phi_{i+1}-\phi_i)}
\hat{\sigma}_i^{-}\hat{\sigma}_{i+1}^{+}
\right),
\end{equation}

and using $\phi_{i+1}-\phi_i=-\theta$, we obtain

\begin{equation}
e^{-i\theta} e^{i(\phi_{i+1}-\phi_i)}
=
e^{-i\theta} e^{i\theta}
=
1,
\end{equation}

which yields the simplified form

\begin{equation}
\tilde{H}_{\text{off}}
=
2\tilde{J}_{xy}
\left(
\hat{\sigma}_i^{+}\hat{\sigma}_{i+1}^{-}
+
\hat{\sigma}_i^{-}\hat{\sigma}_{i+1}^{+}
\right).
\end{equation}

Converting back to Pauli matrices,

\begin{equation}
\hat{H}_{\mathrm{spin}}
=
\sum_{j=1}^{N-1}
\Big[
\tilde{J}_{xy}
\left(
\hat{\sigma}_{j}^{x}\hat{\sigma}_{j+1}^{x}
+
\hat{\sigma}_{j}^{y}\hat{\sigma}_{j+1}^{y}
\right)
+
J_{z}\,
\hat{\sigma}_{j}^{z}\hat{\sigma}_{j+1}^{z}
\Big]
+
\mathfrak{h} \sum_{j=1}^{N} \hat{\sigma}_{j}^{z}.
\label{eq:general_hamiltonian}
\end{equation}

% --- Appendix E ---
\setcounter{equation}{0}
\setcounter{figure}{0}
\makeatletter
\renewcommand{\theequation}{E\arabic{equation}}
\makeatother
\refstepcounter{section}
\section*{APPENDIX \Alph{section}}

\phantomsection
\label{app:E}

We assume a macroscopic droplet doped with 1,2-propanediol (PDO) forming a vortex filament.
\begin{itemize}
    \item Dipole Moment of PDO ($\mu$): $\approx 2.5$ Debye $\approx 8.34 \times 10^{-30}$ C$\cdot$m.
    \item Intermolecular Spacing ($r$): $1.7$ nm (Critical density for CLL phase).
    \item Droplet Radius ($R_{drop}$): $500$ nm (Assuming $N \approx 10^8$ He atoms).
    \item Droplet Charge ($q$): $+1e \approx 1.602 \times 10^{-19}$ C (Localized on the surface).
\end{itemize}

The stabilization energy of the CLL phase is governed by the interaction between nearest neighbors in the filament. Assuming a head-to-tail alignment along the vortex core:
\begin{equation}
    V_{dd} = \frac{1}{4\pi\epsilon_0} \frac{\mu^2}{r^3}
\end{equation}
Substituting the values:
\begin{align*}
    V_{dd} &\approx \frac{(8.34 \times 10^{-30})^2}{4\pi (8.85 \times 10^{-12}) (1.7 \times 10^{-9})^3} \\
    V_{dd} &\approx 1.35 \times 10^{-23} \text{ Joules}
\end{align*}
Converting to temperature units (Kelvin) by dividing by the Boltzmann constant ($k_B$):
\begin{equation}
    {V_{dd} \approx 0.98 \text{ K} \approx 1.0 \text{ K}}
\end{equation}

In a charged helium droplet, the charge localizes on the surface due to the electrostriction of the helium. The vortex filament is located at the center of the droplet. The electric field $E$ generated by the surface charge at the center is:
\begin{equation}
    E_{surf} = \frac{1}{4\pi\epsilon_0} \frac{q}{R_{drop}^2}
\end{equation}
The interaction energy with a molecular dipole is $V_{charge} = -\vec{\mu} \cdot \vec{E}_{surf}$. The maximum perturbation magnitude is:
\begin{align*}
    |V_{charge}| &\approx \mu \times \left( \frac{1}{4\pi\epsilon_0} \frac{q}{R_{drop}^2} \right) \\
    |V_{charge}| &\approx (8.34 \times 10^{-30}) \times \frac{1.602 \times 10^{-19}}{4\pi (8.85 \times 10^{-12}) (500 \times 10^{-9})^2} \\
    |V_{charge}| &\approx 4.8 \times 10^{-26} \text{ Joules}
\end{align*}
Converting to Kelvin:
\begin{equation}
    {V_{charge} \approx 0.0035 \text{ K}}
\end{equation}

Comparing the two energy scales:
\begin{equation}
    \frac{V_{dd} \text{ (Signal)}}{V_{charge} \text{ (Noise)}} = \frac{1.0 \text{ K}}{0.0035 \text{ K}} \approx 285
\end{equation}
The intermolecular interaction is nearly 300 times stronger than the electrostatic perturbation. Furthermore, because the source of the electric field ($R=500$ nm) is distant compared to the molecular spacing ($r=1.7$ nm), the field appears as a uniform background that may define a quantization axis without disrupting the internal Luttinger liquid dynamics.

\end{document}